\documentclass[twocolumn]{aastex6}
\usepackage{natbib}

\usepackage{amsmath}
\def \bea {\begin{eqnarray}}     
\def \ena {\end{eqnarray}}          
\def \bee {\begin{equation}}
\def \ene {\end{equation}}
\def \gas {{\rm gas}}
\def	\th		{\rm {th}}
\def	\H		{\rm {H}}
\def	\K		{~\rm {K}}
\def	\cm		{\rm {cm}}
\def	\yr		{\rm {yr}}
\def	\B		{\rm {B}}
\def	\erg	{\, \rm {erg}}
\def	\s		{\, \rm {s}}

\def	\mum	{\,{\mu \rm{m}}}

\def    \apjl  		{\rm {ApJL}}
\def    \apj  		{\rm {ApJ}}
\def    \mnras  	{\rm {MNRAS}}

\def    \apjl  		{\rm {ApJL}}

\begin{document}

\title{Magnetic Properties of Dust Grains, Effect of Precession and Radiative Torque Alignment}
\author{A. Lazarian}
\affil{Department of Astronomy, University of Wisconsin-Madison, USA; \href{mailto:alazarian@facstaff.wisc.edu}{alazarian@facstaff.wisc.edu}}
\author{Thiem Hoang}
\affil{Korea Astronomy and Space Science Institute, Daejeon 34055, South Korea; \href{mailto:thiemhoang@kasi.re.kr}{thiemhoang@kasi.re.kr}}
\affil{Korea University of Science and Technology, 217 Gajeong-ro, Yuseong-gu, Daejeon, 34113, South Korea}

\date{Draft version \today}            

\begin{abstract}

Alignment of dust grains in astrophysical environments results in the polarization of starlight as well as the polarization of radiation emitted by dust. We demonstrate the advances in grain alignment theory allow the use of linear and circular polarization to probe not only the magnetic field, but also dust composition, the dust environment, etc. We revisit the process of grain alignment by Radiative Torques (RATs) and focus on constraining magnetic susceptibility of grains via observations. We discuss the possibility of observational testing of the magnetic properties of grains as the alignment changes from being in respect to the magnetic field to being in respect to the radiation direction. This opens both a possibility of constraining the uncertain parameters of the RATs theory and provides a new way of measuring magnetic fields in interstellar medium and circumstellar regions. We provide a detailed discussion of the precession induced both by the magnetic field and the anisotropic radiation and revisit a number of key processes related to magnetic response of the grains. We consider various effects that increase the rate of magnetic relaxation both in silicate and carbonaceous grains. In particular, we find a new relaxation process related to the change of the amplitude of internal magnetization within a wobbling triaxial grain and identify a range of grain sizes in which this effect can dominate the internal alignment of angular momentum within grain axes. We show that these relaxation processes significantly change the dynamics of grains in the presence of RATs. We apply our analysis for observed grain alignment in special environments to put constraints on the enhanced magnetic properties of dust grains in the cloud near supernovae, in cometary coma, and protoplanetary disks.

\end{abstract}

\keywords{
ISM- Magnetic fields- polarization, ISM: dust-extinction}

\section{Introduction}

Magnetic fields play a crucial role in many astrophysical processes, e.g., star formation, accretion of matter, transport processes, including heat conduction and propagation of cosmic rays. One of the easiest ways to study magnetic field topology is via polarization of radiation arising from extinction or/and emission by aligned dust grains (see \citealt{Andersson:2015bq}). 

For decades, the mechanism of paramagnetic alignment (\citealt{1951ApJ...114..206D}) and its modifications (see \citealt{2003JQSRT..79..881L} for the review of the further developments related to the mechanism) were considered the most important means of the alignment of interstellar grains. However, later studies revealed
the inability of the paramagnetic mechanism to explain observational data (see \citealt{2003JQSRT..79..881L}). In this situation, the mechanism based on radiative torques (RATs)
became the most likely one for explaining interstellar polarimetric data. The radiative torques were introduced by \cite{1976Ap&SS..43..291D} and were numerically
studied by \cite{{1996ApJ...470..551D},{1997ApJ...480..633D}} and \cite{2003ApJ...589..289W}. The analytical model of the RAT alignment was suggested in \cite{2007MNRAS.378..910L} (henceforth LH07) and 
significantly extended and elaborated in the subsequent publications (e.g. \citealt{HoangLazarian:2008}; \citealt{2009ApJ...695.1457H}; \citealt{2009ApJ...697.1316H}). For RATs to be important, the grains should have helicity, which was shown to be typical for irregular grain shapes. This is in contrast to the earlier grain alignment studies that were dealing with spherical or ellipsoidal grains (see \citealt{{Purcell:1969p3641},{Purcell:1979}}; \citealt{Spitzer:1979p2708}).

For grains with normal paramagnetic susceptibility, the effects of paramagnetic dissipation are negligible compared to the effects of RATs in typical interstellar environments. However, if grains have strong magnetic response, e.g., superparamagnetic, ferromagnetic or ferrimagnetic, the alignment measure of the RAT+magnetic (MRAT) alignment is enhanced (\citealt{Lazarian:2008fw}, henceforth LH08; \citealt{2016ApJ...831..159H}, henceforth HL16). 

Incidentally, the MRAT alignment should not be confused with the reincarnation of the paramagnetic alignment. Within the MRAT process, it is the RATs that produce the alignment, while the role of the magnetic field is to stabilize the grain rotation with the high angular momentum. As it was shown in LH07, for a given grain shape, there is a particular range of angles between the direction of radiation and the magnetic field that results in a grain getting into the state of high angular momentum (high-J). This state of high-J corresponds to fast, i.e. suprathermal, rotation and to the perfect alignment with longer grain axes perpendicular to magnetic field. For other directions, the RATs aligned grains, but not perfectly and, in fact, not spin up, but {\it slow down the grain}, tending to bring them into the state of slow or subthermal (low-J) rotation. The enhanced magnetic relaxation increases the range of angles for which the high-J alignment is possible. Note, that the high-J alignment is the perfect alignment, while the low-J alignment demonstrates a significantly lower alignment degree (LH07; \citealt{HoangLazarian:2008}; HL16). Therefore, the magnetic inclusions tip the balance between the two modes of the RAT alignment.
More theoretical discussion of the RAT alignment and its modifications can be found in the reviews (e.g., \citealt{2007JQSRT.106..225L}; \citealt{LAH15}). The observational evidence in favor of the RAT alignment processes is summarized in \cite{Andersson:2015bq}. Recent study of RATs for many grain shapes in \cite{Herranen:2019kj} supports the analytical theory in LH07 and confirms that the RAT alignment is a really universal process for irregular grains. 

This paper addresses the particular aspect of the RAT alignment related to the grain precession. Grain precession arising from the interaction of grain magnetic moment with the ambient magnetic field has been discussed extensively in the grain alignment literature (see \citealt{1999MNRAS.305..615R} and ref. therein). In LH07, another type of precession has been introduced and quantified. It was shown there that, in the presence of anisotropic radiation, all grains, ever non-helical ellipsoidal grains, precess about the direction of radiation. The effect was shown to be particularly important in the vicinity of the bright stars where the rapid precession is induced about the radiation direction, i.e., the direction of incoming photons given by the wave vector ${\bf k}$.  If the corresponding precession that we term k-precession is faster than the Larmor precession, or B-precession, the alignment happens in respect to the radiation. Following the suggestion of our colleague B-G Andersson, we shall term the alignment in respect to radiation the {\it k-RAT} alignment, while the traditional alignment in respect to magnetic field we will call {\it B-RAT} alignment.

The k-RAT alignment complicates the magnetic field tracing in the vicinity of bright stars (\citealt{2007MNRAS.378..910L}). This type of alignment is becoming more important as with high spatial resolutions astronomers are exploring the polarization arising from dust in the close vicinity of radiation sources. For instance, the k-RAT  alignment was shown to be very important for accretion disks (see \citealt{2017ApJ...839...56T}). This serves as a motivation for us to provide a detailed study of when the k-precession is faster compared to the B-precession and what observational signatures are expected for the RAT alignment for grains of different sizes. This calls for revisiting a number of key processes, e.g., the internal relaxation of energy within wobbling grains. 

In this paper, we do not question the dust theory paradigm that separates grains into two classes: silicate and carbonaceous one. This paradigm faces an apparent problem of how to avoid mixing up the grain composition as grains coagulate in the dense part of the interstellar medium (ISM). For most of the paper, we assume that this does not happen in diffuse media and consider the alignment of silicate and carbonaceous grains separately.
A recent studies in \cite{Hoangetal:2019} and \cite{Hoang:2019} suggest the importance of centrifugal disruption of grains spun-up by RATs. As predicted by \cite{Hoang:2019}, the composite silicate-carbonaceous grains are being destroyed in diffuse media by centrifugal stress arising from their fast rotation induced by the RATs. This is an interesting possibility that requires further studies. At the same time, we discuss a possibility that in molecular clouds the silicate and carbonaceous grains do coagulate and form grains composed both of silicates and carbonaceous components. This can be the consequence of decreased influence of RATs and also of the effects of ices that can glue to different components of the grain together. If this happens the  dust in molecular clouds is different from that in diffuse gas, not only in terms of the size of dust grains, but also in terms of the grain composition. Therefore we expect that multi-frequency polarization observations to reveal features corresponding to the aligned carbonaceous grains.

For the silicate grains, we consider both pure silicate grains and the grains with magnetic inclusions. Such clusters can induce strong paramagnetic relaxation enabling MRAT alignment (LH08, HL16). The clusters may be made of iron or other strongly magnetic materials.  Incidentally, iron inclusions in interstellar silicate grains are suggested by recent studies (see \citealt{2018ApJ...857...94Z} and ref. therein).

In what follows, in \S \ref{sec:KBRAT}, we summarize the observational evidence for B-RAT and k-RAT alignment, provide the quantitative discussion of the RAT alignment in the presence of paramagnetic relaxation in \S \ref{sec:magnetic}, revisit the issue of internal Barnett and nuclear relaxation within wobbling grains in \S \ref{sec:relaxation}, discuss the processes of grain randomization in \S \ref{sec:random}, and address the precession of grains induced by magnetic field and by radiation in \S \ref{sec:precession}. In \S \ref{sec:theory}, we discuss how the values of the magnetic field and grain magnetic susceptibility can be obtained from observations, and in \S \ref{sec:kBRAT} we calculate the alignment of grains in terms in the presence of for both the state of high and low rate of grain rotation. Possible ways of probing grain magnetic susceptibilities from observations are discussed in \S \ref{sec:obs}, while the discussion of the results and the summary of the paper are presented in \S \ref{sec:discuss} and \S \ref{sec:summary}, respectively. 

\section{Properties of B-RAT and k-RAT alignment}\label{sec:KBRAT}

\begin{figure}
\includegraphics[width=0.5\textwidth]{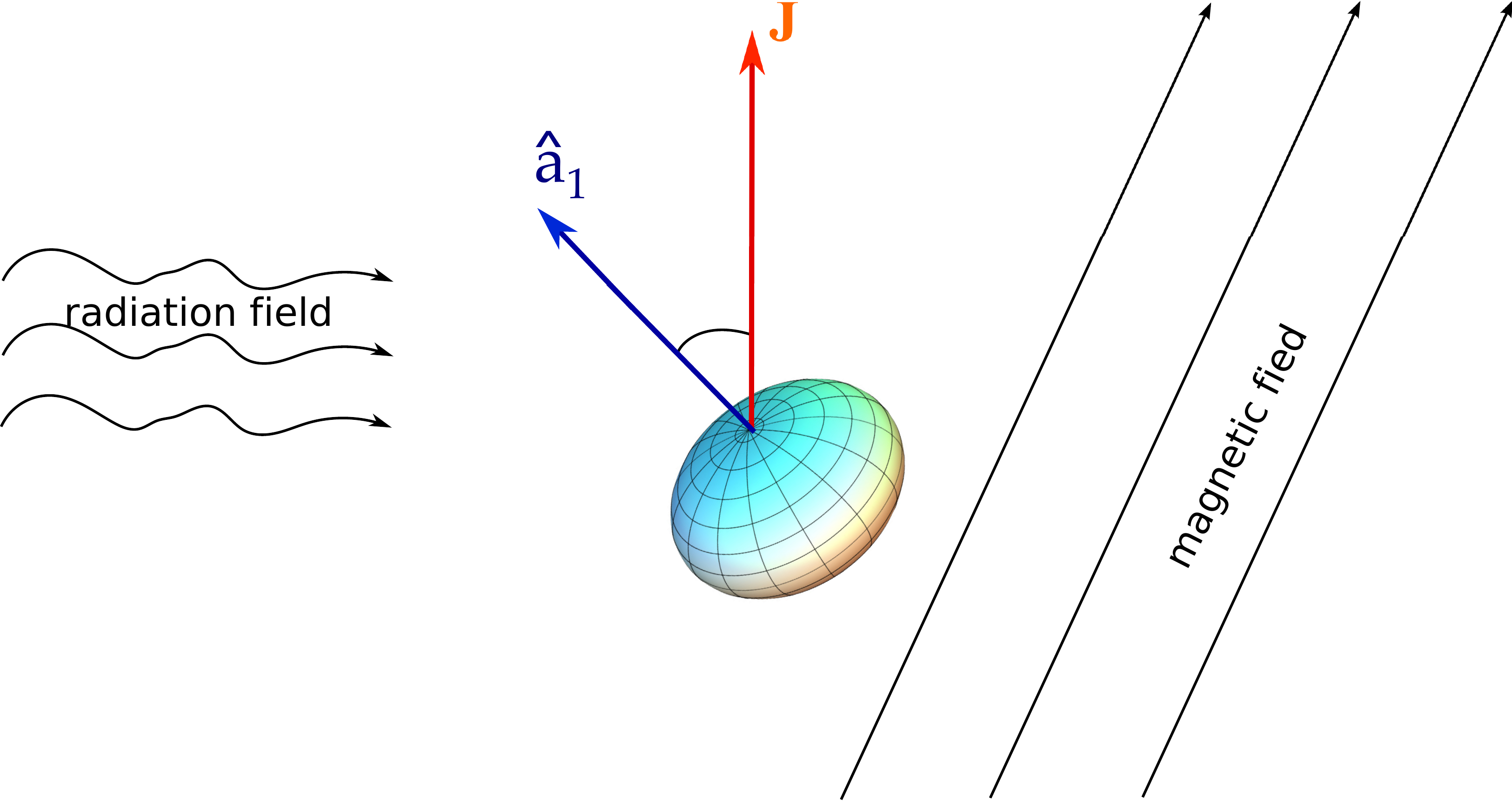}
\caption{Schematic illustration of grain alignment. The radiation direction and magnetic field directions provide two directions that are, in general, do not coincide. If Larmor precession is faster than the grain precession induced by RATs we deal with B-alignment. In the opposite case the alignment in respect to the direction of radiation, i.e. k-alignment, takes place. }
\label{fig:k-B_align}
\end{figure}

Figure \ref{fig:k-B_align} shows the schematic illustration of a grain with angular momentum ${\bf J}$ at an angle to the grain axis ${\bf a}_{1}$ of the maximal moment of inertia. The grain interacts with the radiation flux {\bf k} and the ambient magnetic field {\bf B}. As a result, the grain is subject both to RATs and magnetic torques. It is important that both types of torques induce the precession. According LH07 whether the grain gets aligned with the direction of {\bf B} or {\bf k} depends on whether the Larmor precession rate is larger or smaller than the precession rate induced by RATs. The former rate, as well as magnetic dissipation, depend on grain magnetic properties. In addition, both the direction and the degree of grain alignment depends on whether the grain angular momentum ${\bf J}$ is aligned with the axis of the grain maximal moment of inertia and, if it is aligned, what is the degree of alignment. The internal alignment of ${\bf J}$ is controlled by the internal relaxation processes and those for silicate grains are also related to magnetic dissipation.\footnote{The internal relaxation for carbonaceous grains can be due to inelastic deformations of wobbling grains (see Lazarian \& Efroimsky 1999).} The importance of magnetic phenomena within astrophysical dust grains motivate our revisiting this subject in our present paper.

The basic predictions of the RAT theory have been tested extensively, in particular, by B-G Anderson and his collaborators (see \citealt{Andersson:2015bq}). Nevertheless, the issue of the magnetic response of the interstellar and circumstellar grains has not been completely resolved yet. For instance, the analysis of interstellar grains captured in the Earth upper atmosphere revealed that these grains contained magnetic inclusions capable to render superparamagnetic response to dust particles (\citealt{Bradley:1994p6379}; \citealt{1995ApJ...455L.181G}; \citealt{Altobelli:2016dl}). 
The interpretation of the interstellar polarization data requires high degree of grain alignment in diffuse gas (\citealt{2015A&A...576A.104P}; \citealt{Guillet:2017hg}). At the same time the RAT theory predicts that the alignment of ordinary paramagnetic grains is not guaranteed to be perfect. Indeed, it was demonstrated in LH07 that depending on the grain shape, in particular, on $q^{\rm max}$ parameter introduced in that paper, there is  a range of angles between the magnetic field and the direction of radiation for which the alignment is reduced. The degree of alignment can still be increased in the presence of the imbalanced torques, e.g. the Purcell's (\citealt{Purcell:1979}) torques, in particular, those arising from H$_2$ formation (\citealt{HoangLazarian:2008}) or mechanical uncompensated torques arising as an irregular moves through the ambient gas (\cite{LazarianHoang:2007b}).\footnote{The difference between these two types of torques is that Purcell's torques are torques acting in the grain reference frame, while those introduced in \cite{LazarianHoang:2007b}) act in the lab frame. As a result, the former do not induce the alignment of their own, while the latter act similar to RATs and induce their own alignment. However, if the Larmor precession is faster than both the precession induces by RATs or mechanical torques, the resulting grain alignment happens in respect to magnetic field.} The magnitude of these torques for typical ISM conditions are not well constrained. In the absence of the aforementioned uncompensated torques the numerical study of the distribution of $q^{\rm max}$ parameter for a distribution of grain shapes in \cite{Herranen:2019kj} indicates that perfect alignment of grains with long axes perpendicular to the magnetic field is unlikely in the diffuse ISM. Instead, the values of space averaged alignment the order of 50\% or less are expected. We do not think that existing polarization measurements can rule out this possibility, but it is likely that the realistic interstellar grains with size $10^{-5}$ cm or larger do contain magnetic inclusions.   

The alignment can be significantly improved if grains demonstrate the enhanced magnetic relaxation typical for grains with magnetic inclusions (LH08; HL16). Such a response stabilizes grain rotation with high value of angular momentum, and this makes the alignment perfect for the grains, irrespectively of their shape and the direction of illumination. The only requirement for the perfect alignment of grains with strong magnetic response is that RATs are strong enough to overcome the randomization processes. In this paper, we claim that there are observational ways to
determine the magnetic response of grains by studying the transfer from the alignment in respect to magnetic field, i.e., B-alignment to the alignment in respect to radiation, i.e., the k-alignment.  In particular, we discuss how the actual value of the grain magnetic response can be obtained from the analysis of the high resolution studies of linear and circular polarization. Indeed, it is easy to see that the transition of one type of alignment to another depends on the magnetic susceptibility of grains. 

We should mention that the recent ALMA polarization data from accretion disks showed the pattern that deviates from the expectations based on the the alignment of grains with very large magnetic susceptibilities. Indeed, such grains are expected to be aligned with long axes perpendicular to the mostly toroidal magnetic field that is expected in the circumstellar disks. This is not what is
seen (\citealt{Kataoka:2017fq}; \citealt{Stephens:2017ik}; \citealt{2018ApJ...854...56L}; \citealt{2018ApJ...854...56L}; \citealt{Alves:2018vl}; \citealt{2018ApJ...854...56L}), suggesting that only moderate enhancements of the magnetic susceptibilities are present. In the paper we discuss in detail magnetic effects on grains.

The complication that exist for detecting the point of the transition from one type of alignment to another stems from the changes of the direction of alignment that are present if the internal relaxation of energy within the grains is slow. It was demonstrated in \cite{2009ApJ...697.1316H} that, in the absence of fast internal relaxation of energy within wobbling grains, the alignment of grains can happen with the long grain axes parallel to ${\bf k}$ for k-RATs and ${\bf B}$ for B-RATs. Note, that if the internal relaxation is fast, the alignment takes place with long grain axes perpendicular to the two aforementioned axes. This complication motivates us to revisit the processes of internal alignment. 

The alignment of silicate grains is better understood compared to that of  carbonaceous grains. In fact, the alignment of the latter presents longstanding puzzle. Observations suggest that carbonaceous grains are marginally aligned with the magnetic field for most of the diffuse ISM (see \citealt{Andersson:2015bq} and ref. therein). The reason for that was attributed in \cite{LAH15} to the difference of the magnetic moment of carbonaceous and silicate grains, which makes the k-RAT alignment for the former the preferred one compared to the B-RAT alignment. There has not been a systematic study of the k-alignment of carbonaceous grains and we feel it is important to search for the signatures of the k-RAT alignment of these grains. For them, however, the theoretical predictions of the degree of k-alignment are not trivial in view of the anomalous randomization process introduced in \cite{2009MNRAS.400..536J}. The process 
 is related to charged grains being accelerated by MHD turbulence (\citealt{2002ApJ...566L.105L}; \citealt{2003ApJ...592L..33Y}; \citealt{Yan:2004ko}; \citealt{Hoang:2012cx}). The key idea is that fast moving grains experience additional electric field which causes the randomization of their angular momentum when the grain electric dipole moment fluctuates. The corresponding randomization can be orders of magnitude faster than the randomization that arises from gaseous collisions.\footnote{ In addition, the anomalous randomization can speed up the relaxation towards a new direction of alignment when the ambient radiation flux changes its direction or its amplitude, i.e., in the latter case it can help to the transfer from B-alignment to k-alignment.} This motivates us to revisit the anomalous randomization process.

In this paper we present both theoretical calculations of the basic processes relevant to k-RAT and B-RAT alignment as well as provide suggestions for observational testing of our predictions. 

\section{Grain precession}\label{sec:precession}
\subsection{Grain precession about magnetic field}

The rate of precession determines whether B-RAT or k-RAT alignment takes place. Below we discuss how magnetic properties affect grain precession. 

A rotating grain with a magnetic moment $\mu$ precesses in the ambient magnetic field $B$ with the Larmor rate 
\begin{equation}
\Omega_{\rm L}=\frac{\mu B}{I_{\|}\omega}, 
\label{eq:tl}
\end{equation}
where $I_{\|}$ is the grain maximal moment of inertia. For an oblate spheroid
\begin{equation}
I_{\|}=\frac{8\pi}{15}\rho_{\rm s}a_1a_2^4,\ I_{\perp}=\frac{4\pi}{15}\rho_{\rm s}a_1a_2^2(a_1^2+a_2^2),
\end{equation}
where $\rho_s$ is the grain density, $a_1,\ a_2$ denotes the minor and major grain axes. The implicit assumption that we made is that the grain is rotating about its maximal axis of inertia, which corresponds
to the conditions that the internal relaxation within grains (see \citealt{Purcell:1979}; \citealt{LazEfroim:1999}; \citealt{1999ApJ...520L..67L}) is fast and the grain rotation is much faster that its rotation corresponding to the grain temperature (\citealt{1995MNRAS.277.1235L}; \citealt{LazarianRoberge:1997}). If one of these conditions is not satisfied, the grain is wobbling.

 For paramagnetic interstellar grains, the Barnett effect was identified as the major source of magnetic moment in \cite{1976Ap&SS..43..291D}.\footnote{As we discussed earlier, the Barnett effect is a reciprocal phenomenon of the Einstein--de Haas effect in which magnetization of a body induces mechanical rotation (see e.g., \citealt{1960ecm..book.....L}), a subtle effect related to the orienting electron spins as a paramagnetic body shares its angular momentum with the electron spin system. Grains also get magnetized due to the Barnett effect arising from nuclear spins. The magnetization then is $\sim 1000$ times reduced.}  Due to the Barnett effect more spins get oriented in the rotation direction as compared to the opposite direction. As a result, the rotating body gets magnetized. Within the wobbling triaxial body at low frequencies of rotation the alignment of spins happens in relation to grain angular velocity.  Thus, we assume the magnetic moment induced by the Barnett process (\citealt{1960ecm..book.....L}) is
\begin{equation}
\mu_{\rm Bar}=\frac{\chi(0)V\hbar}{g\mu_{\rm B}}\omega \label{eq:mubar}
\end{equation}
where $V$ is the volume of the dust grain, $\hbar$ is the Planck constant divided by $2\pi$, and $\chi(0)$ is given by Equation (\ref{chi_p}). The factor $\mu_{\rm B}=e\hbar/2m_ec\approx 9.274\times 10^{-21}$ erg G$^{-1}$ in Eq. (\ref{eq:mubar}) is the Bohr magneton, where $m_e$ is the mass of the electron, $c$ is the speed of light, and $g$ is the $g$-factor, which is $g\approx 2$ for electrons.

The Larmor rate can be obtained combining Eqs. (\ref{eq:tl}) and (\ref{eq:mubar}), one can obtain the timescale of Larmor precession of a paramagnetic grain
\begin{eqnarray}
t_{\rm L}&=&\frac{4\pi}{5} \frac{g_e\mu_{\rm B}}{\hbar}\rho_s s^{-2/3}a_{\rm eff}^{2}B^{-1}\chi(0)^{-1}\\ \nonumber
&\approx& 1.3\ \hat{\rho}\hat{s}^{-2/3}a_{-5}^2\hat{B}^{-1}\hat{\chi}^{-1}\ {\rm yr}
\label{larmor_p}
\end{eqnarray}
where the normalized value of magnetic field is $\hat{B}=B/5\ \mu$G and $\hat{\chi}=\chi(0)/10^{-4}$. Larger values of magnetic field strength can be present in accretion disks. For grains with iron inclusions, the enhancement of $\chi(0)$ (see Appendix, Eq. \ref{eq:chi_sp}) entails the corresponding enhancement of the rate of the Larmor precession (see Eq. \ref{larmor_p}). 

For paramagnetic grains the magnetization induced by grain charging is, as a rule, a subdominant process. However, if the carbonaceous grains are diamagnetic, then for such grains the magnetic moment induced by the charge on the surface of the surface of a rotating grain can be important. The corresponding magnetic moment  for a grain with charge $q$ (see \citealt{1950clme.book.....G}) is
\begin{equation}
\mu_{q}=\alpha \frac{qJ r^2}{3cI_{\|}},
\end{equation}
where $J$ is grain angular momentum and $r$ is the effective radius of the grain. The coefficient $\alpha$ depends on the spatial distribution of the grain charges. Within our order of magnitude approach it is safe to assume that $\alpha=1$. The corresponding precession rate is
\begin{equation}
\omega_{q}\approx \frac{q r^2 B}{3c I_{\|}}
\end{equation}
is significantly smaller compared to the rate induced by Barnett effect for typical paramagnetic interstellar dust grains.

We would like to mention that the carbonaceous grains can also have unpaid electrons due to the existence of unpaid bonds and free radicals. For instance, the paramagnetic response was seen with coal samples. In LD00, it was assumed that the number of unpaired electrons is $10^{21}\cm^{-3}$, which is somewhat larger than the number of unpaired electrons in coals, but less than the concentration of free radicals suggested in \cite{1982come.coll..131G}. Compared to the estimate for the paramagnetic grains (see Eq. \ref{larmor_p}), this provides an increase of the Larmor period by a factor 10, which is still significantly shorter than the precession period arising from the typical interstellar grain charging. 

\subsection{Grain precession induced by radiative torques}

Our discussion of grain precession in the magnetic field shows that the rate of precession depends sensitively on the grain magnetic properties. Thus, by measuring this rate, one can get insight into the grain composition. Such a measurement can be accomplished by comparing this rate with that arising from RATs. LH07 found that the precession arising from RATs is universal, and it takes place even for ellipsoidal grains. If the resulting precession is faster than the Larmor one, the radiation direction rather than that of magnetic field becomes the axis of alignment. 

Unlike the Larmor precession, the RAT-induced one is much less sensitive to the composition of grains.  The estimate of RAT-induced precession can be obtained using the known expressions for the radiative torques (DW97; LH07):
\begin{equation}
\mathbf{\Gamma}_{\rm rad}=\frac{u_{\rm rad}a_{\rm eff}^2\overline{\lambda}}{2}\overline{\gamma}\overline{\mathbf{Q_{\Gamma}}},
\end{equation}
where a bar indicates the spectrum-averaged quantities, 
\begin{eqnarray}
&& \overline{\mathbf{Q_{\Gamma}}}=\frac{\int \mathbf{Q_{\Gamma}}u_{\lambda} d\lambda}{u_{\rm rad}}, \
\overline{\lambda}=\frac{\int u_{\lambda}\lambda d\lambda}{u_{\rm rad}},\\
&& \overline{\gamma}=\frac{\int u_{\lambda}\gamma_{\lambda} d\lambda}{u_{\rm rad}},\
 u_{\rm rad}={\int u_{\lambda}d\lambda},
\end{eqnarray}
where $u_{\rm \lambda}$ is the energy spectrum of the radiation field, $\lambda$ is the radiation wavelength, $\gamma_\lambda$ is the anisotropy parameter, and $\mathbf{Q_{\Gamma}}$ is the RAT efficiency. The amplitude\footnote{Further in the paper we discuss the dependence of  $\mathbf{Q_{\Gamma}}$ on the direction between the grain rotational axis and the direction of radiation (see LH07). This dependence is important for understanding many processes of RAT alignment.} of $\mathbf{Q_{\Gamma}}$ is estimated by the DDA calculation (LH07), 
\begin{eqnarray}
|\mathbf{Q_{\Gamma}}|&\approx& 2.3\left(\frac{\lambda}{a_{\rm eff}}\right)^{-3}\ \ {\rm for\ } \lambda>1.8 a_{\rm eff}\\
&\approx& 0.4\ \ \ \ \ \ \ \ \ \ \ \ \ {\rm for\ } \lambda\leq1.8 a_{\rm eff} \label{eq:QRATs}.
\end{eqnarray}

As a result LH07 defined the precession time scale:
\begin{eqnarray}
t_{rad,\ p}&=&\frac{2\pi}{\Omega_{\rm p}}, \label{eq:tradp} \\
\Omega_{\rm p}&=& \frac{u_{\rm rad}\bar{\lambda} a_{\rm eff}^2}{I_{\|}\omega}\gamma\overline{|\mathbf{Q_{\Gamma}}|}
\end{eqnarray}
where $I_{\|}$ is the maximum moment of inertia, and $\omega$ is the angular velocity of a grain.

Plugging typical numerical parameters, one obtains
\begin{eqnarray}
t_{rad, p}&\approx&1.1\times10^{2}\ \hat{\rho}^{1/2}\hat{s}^{-1/3}a_{-5}^{1/2}\hat{T_d}^{1/2}\\
&\times&\left(\frac{u_{\rm rad}}{u_{\rm ISRF}}\right)^{-1}\left(\frac{\lambda}{1.2\ \mu{\rm m}}\right)^{-1}\left(\frac{\gamma \overline{|\mathbf{Q_{\Gamma}}|}}{0.01}\right)^{-1}\ {\rm yr},\nonumber  
\label{rad_p}
\end{eqnarray}
where we have assumed $\omega$ equal to the thermal angular velocity $\omega_{d}=\sqrt{2k T_{\rm d}/I_{\|}}$.

\subsection{Other types of precession and the comparison of the rates}

Apart from analytically describing the RAT alignment  \citealt{LazarianHoang:2007b} introduced the alignment by Mechanical Torques or MT alignment. The nature of MTs is similar to RATs, but they arise due to irregular grains interacting not with photons but with gaseous atoms as the grain moves in respect to gas.  It was found that grains moving in the gas precess about the direction of their motion (\citealt{LazarianHoang:2007b}). The expression for torque component inducing precession of an ellipsoidal grain  subject to impinging particles was obtained in LH07.  Assuming the reflective interactions of hydrogen atoms with the surface of the ellipsoid is we can rewrite the expression in LH07 as:
\begin{equation}
    \Gamma_{MAT}\approx m_{H} V_{grain}^2 s^{-2}a^3 e(e^2-1) K(\Theta, e) \sin 2\Theta,
    \label{MechQ3}
\end{equation}
where $s=a/b<1$ is the axial ratio of oblate grains, $V_{grain}$ is the velocity of the grain relative the gas of atoms with mass $m$, $\Theta$ is an angle between the grain symmetry axis and the vector ${\bf V_{grain}}$,  $e$ is the eccentricity of the oblate grain and $K(\Theta, e)$ is a function tabulated in LH07. This function for most of the range of $\Theta$ is of the order of unity and, taking into account that real grains are not ideal oblate spheroids, it can be put to be equal to unity. As a result, repeating the arguments for the precession induced by RATS (see \S 3.2) one gets
\bea
    t_{mech,p}&=& \frac{2\pi}{\Omega_{p}}\sim\frac{2\pi I_{\|}\omega}{\pi a^{3}s^{-2}n_{\rm H}m_{\rm H}V_{grain}^{2}e(e^{2}-1)K(\Theta,e)\sin 2\Theta}\nonumber\\
    &\simeq& 36C\left(\frac{\omega}{\omega_{th}}\right)\left(\frac{v_{th}}{V_{grain}}\right)^{2}\left(\frac{s}{0.5}\right)^{2}\frac{1}{\sin 2\Theta} ~\rm yr,
    \label{t_mex}
\ena
where $C$ is a factor of order unity arising with our inability to describe the grain shapes exactly ( see \citealt{Hoangetal:2018} for numerical modeling of the MTs), $v_{th}$ is the thermal velocity of hydrogen, and $\omega$ is taken to be the grain thermal angular velocity at $T=T_{\rm gas}$.

Comparing Eq. (\ref{t_mex}) and Eq. (\ref{rad_p}) we see that for trans-sonic grain motions the precession rate arising from interstellar radiative torques and from mechanical torque precession can be of the same order. Both rates are significantly smaller than the rate of Larmor precession $t_L^{-1}$ given by Eq. (\ref{larmor_p}). This ensures that over most of the volume of the interstellar medium the alignment is happening in respect to magnetic field irrespectively of whether the alignment is caused by RATs or MTs. 

The dependence on $\sin 2\Theta$ in Eq.(\ref{MechQ3}) is important. Indeed, the acceleration by turbulence mostly induces relative motion of grain perpendicular to magnetic field. If grains are perfectly aligned with the magnetic field the factor $\sin 2\Theta$ is close to zero and therefore the rate of precession induce by the mechanical torques is reduced. Similarly, if mechanical precession dominates the alignment with the direction of the relative grain-gas motion cannot be perfect, as in this case $\Theta$ equals to zero and the mechanical precession vanish. 

The Analytical Model (AMO) in LH07 predicts\footnote{In Eq. (\ref{eq:QRATs}) this dependence is hidden in ${\bf Q_{\Gamma}}$.} of  the same dependence on $\Theta$ as we have for the mechanical torques. This dependence is confirmed with the numerical modeling of the RATs. As a result, we do not expect the perfect alignment of grains in respect to radiation in typical interstellar conditions, unless the direction of magnetic field and radiation coincide. Indeed, as the degree of k-RAT alignment increases, the effect of Larmor precession gets more important, causing the deviating of the direction of grain rotation axis from the direction of ${\bf k}$. 

We note, that while the values of two other components of mechanical torques vary significantly from one grain shape to another (see \citealt{Hoangetal:2018}), the component of torque that causes grain precession does not depend on the resulting helicity of the grain and its value has the uncertainty of the order of unity. In terms of the precession the properties of the mechanical torques are identical to those of RATs. However, our estimate in Eq.(\ref{t_mex}) shows that rather special conditions are necessary for mechanical torques to induce the precession faster than the Larmor precession in the typical interstellar conditions. This is in contrast to the radiation torques the amplitude of which easily increases hundred times and more as the radiation flux increases with the decrease of the distance to the star or as the result of a novae or supernovae outburst. As a result, we expect that the change of the direction of alignment happens mostly due to RATs rather than MTs.    

One difference between the two processes that we noted above is the dependence of the Larmor precession on the grain composition. This means that in the presence of both the Larmor and the competing RAT-induced precession
depending on their composition the grains are going to be aligned in respect to the the magnetic field or the radiation. For instance, the diamagnetic grains may be mostly aligned by in respect to the radiation anisotropy, while superparamagnetic grains can be aligned in respect to magnetic field. This is an important effect with the astrophysical consequences that we discuss further in the paper. 

Another difference between the Larmor precession and that induced by RATs is related to the rate of grain rotation. The Larmor one does not depend on the angular velocity of grain rotation, as the higher the velocity of a grain, the larger the magnetic moment of the grain. Thus, the Larmor frequency is the same for the fast and slowly rotating grains. On the contrary, the faster grain rotates, the smaller the RAT-induced precession rate. As we discuss below this can induce non-trivial dynamics of grains and potentially provides a way to distinguish fast and slow rotating grains.
\section{Magnetic relaxation in an external field}\label{sec:magnetic}

\subsection{Relaxation parameter}

Paramagnetic relaxation was proposed by \cite{1951ApJ...114..206D} as a way to align interstellar grains with the long axes perpendicular to the ambient magnetic field and thus to explain the observed interstellar polarization. It is known that variations of magnetic fields induce paramagnetic relaxation. As a result the authors suggested that the changes of the magnetic field in the grain frame, as the grain rotates with angular momentum ${\bf J}$ at an angle with magnetic field, should result in the dissipation of grain kinetic energy.\footnote{This analogy is not exact as was shown in \cite{2000ApJ...536L..15L}. In fact, it does matter whether the grain or magnetic field is rotating. The grain rotation induces an effective magnetization which changes the rate of relaxation. The corresponding effect was termed by \cite{2000ApJ...536L..15L} {\it resonance relaxation} as the magnetization induced by grain rotation results in the resonance paramagnetic relaxation. The difference between the resonance paramagnetic relaxation and the classical \citep{1951ApJ...114..206D} relaxation is important for very small fast rotating grains, e.g., PAH molecules or silicate nanoparticles, and it is not important for the $>10^{-5}$~cm grains that are traditionally discussed in the context of grain alignment in the interstellar medium. } If the corresponding relaxation time is $\tau_{mag, relax}$, the ratio
\begin{equation}
\delta_m=\frac{\tau_{\rm damp}}{\tau_{mag, relax}},
\end{equation}
where $\tau_{\rm damp}$ is the rotational damping time scale,  is a key parameter that characterizes the relaxation efficiency. Note, that we define $\tau_{\rm damp}$ as the timescale for grain randomization (see Eq. (\ref{eq:tdamp})) by all processes apart from the anomalous randomization that we discuss further. One can easily see that if $\delta_m\ll 1$ the magnetic relaxation is negligible. On the contrary, if $\delta_m\gg 1$, the relaxation should dominate the grain dynamics. In reality, thermodynamics suppresses the effects of relaxation for $\delta_m\gg 1$ when the grains rotate with angular velocities $\omega$ comparable to thermal angular velocity, $\omega_{th}$, corresponding to the temperature of the grain material (see \citealt{Jones:1967p2924}). Therefore, fast grain rotation is another requirement for paramagnetic relaxation to be efficient. 

\cite{Jones:1967p2924} suggested a way to increase $\delta_m$ by increasing grain magnetic susceptibilities. This is possible if grains have magnetic inclusions that we discuss further in the paper. However, even with the enhanced magnetic response, the theory based on magnetic relaxation could not explain the properties of polarization in molecular clouds where grain rotation gets close to thermal (see \citealt{2007JQSRT.106..225L}). Instead, these properties, as many other observed features of interstellar polarization, can be well explained by the RATs (see \citealt{Andersson:2015bq}).  

By itself the alignment based on magnetic relaxation has limited applicability. For instance, at present, it is generally agreed that RATs rather than magnetic relaxation are responsible for the alignment of the classical $10^{-5}$ cm grains. However, as we mentioned above, the magnetic relaxation can enhance the RAT alignment. As a result, the parameter $\delta_m$ is an important parameter for the Magnetic-RAT, or MRAT, theory where magnetic relaxation acts together with RATs torques (LH08, HL16). Rather counter-intuitively, the magnetic relaxation induces more grains to be more efficiently spun-up by RATs as more grains find themselves at the high-J attractor (LH08). The higher degree of grain alignment achievable with the MRAT mechanism makes it more favorable for explaining observations of regions with high degree of polarization. In this section, we show why $\delta_m$ for both silicate and carbonaceous grains may be larger than it is customary accepted.

To understand the processes of magnetic relaxation that affect $\delta_m$, we consider both spin-spin and spin-lattice relaxation within grains. The former characterizes the exchange of energy within the spin system, while the latter ensures the transfer of energy to the lattice. Both processes are important for magnetic relaxation process within astrophysical dust. In grain alignment studies, the effects of spin-lattice relaxation are frequently disregarded. However, we show that they can be important. In particular, we discuss the importance of spin-lattice relaxation that arises when the grain magnetization is induced by fast rotation of grains.

The value of $\delta_m$ should be constrained from observations. It is directly related to the grain magnetic susceptibilities that are currently poorly known. Our paper suggests a way to evaluate the magnetic response of grains from polarization measurements. 

\subsection{Ordinary paramagnetic materials and enhanced magnetic relaxation}

The time scale for magnetic relaxation of a grain of size $a$ rotating with angular velocity $\omega$ in the external magnetic field $H_{\rm ext}$ is defined for a spherical grain as (see \citealt{1999MNRAS.305..615R})
\begin{equation}
\tau_{mag, relax}= \frac{2 \rho a^2}{5 K(\omega) H_{\rm ext}^2},
\label{tau_mag}
\end{equation}
where $\rho$ is the grain density and 
\begin{equation}
K(\omega)= \frac{Im (\chi(\omega))}{\omega},
\end{equation}
where $Im (\chi(\omega))$ is the imaginary part of the complex magnetic susceptibility $\chi (\omega)$.

We discuss the properties of this type of relaxation in Appendix \ref{apdx:enhanced} where we show that it can be enhanced by the presence of
inclusions with strong magnetic response. If the inclusions are sufficiently small, their magnetization fluctuates. Such inclusions are called superparamagnetic (see \citealt{Morrish:2001vp}). The enhancement of relaxation increases with the size of the inclusion. However, the frequency at which the enhanced superparamagnetic  response is cut off decreases with the size of the inclusion.

\begin{figure}
\includegraphics[width=0.5\textwidth]{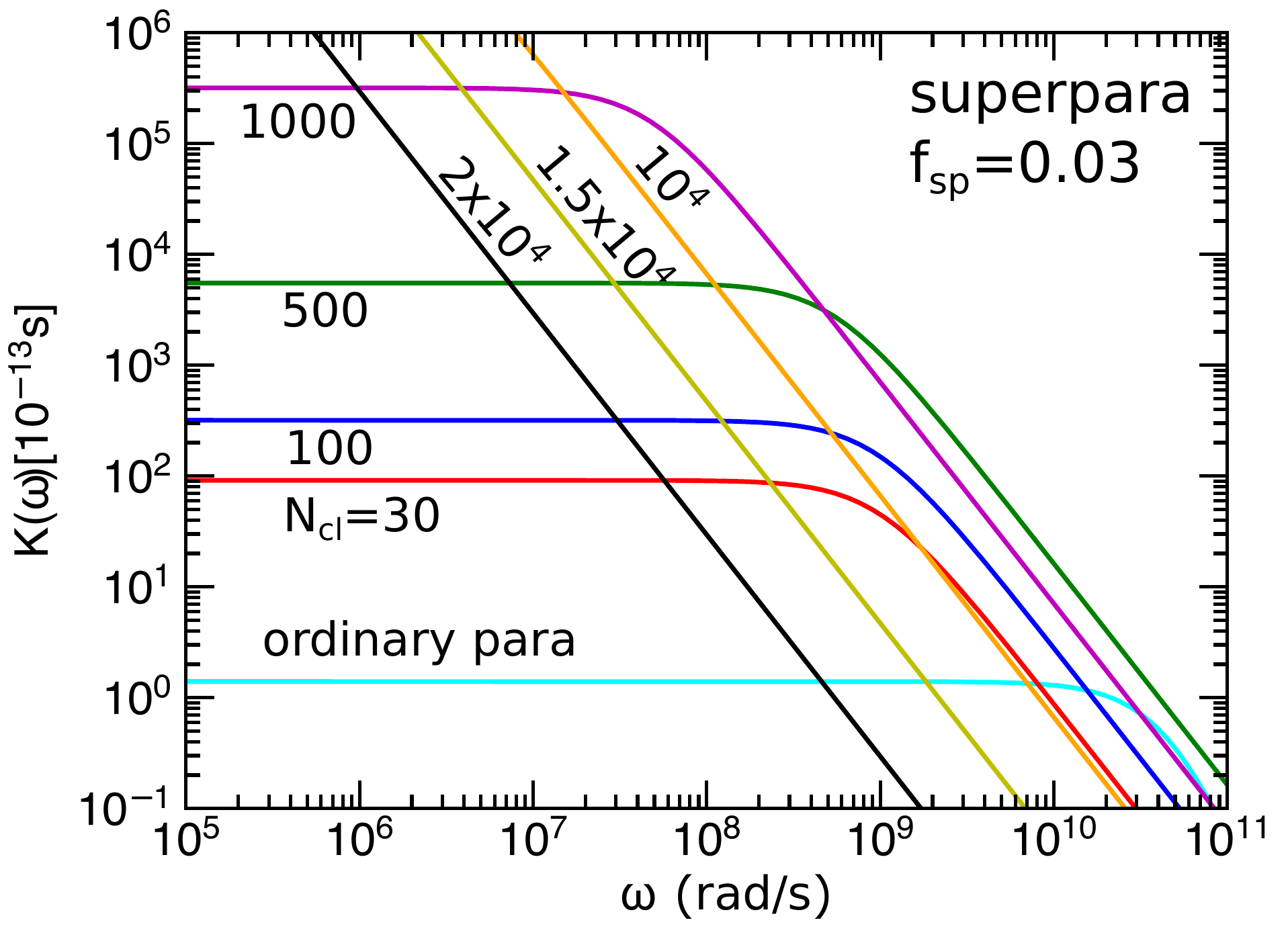}
\caption{Magnetic susceptibility parameter vs. grain angular velocity for ordinary paramagnetic grains and grains with iron inclusions of varying the number of iron atoms per cluster $N_{cl}$. The calculations are done assuming that the fraction of atoms in superparamagnetic clusters is $f_{sp}=0.03$.}
\label{fig:kappa_omega}
\end{figure}

Figure \ref{fig:kappa_omega} illustrates the dependence of magnetic dissipation within a grain as a function of grain angular velocity $\omega$. It is evident that the magnetic dissipation in ordinary paramagnetic silicate grains is valid for the entire range of angular velocities typical for "classical" $10^{-5}$ grains in the ISM. At the same time, the grains with superparamagnetic inclusions should not have too large size. For instance, thermal angular velocity of grains with inclusions containing more than $10^5$ atoms is larger that the critical angular velocity that is determined by the rate of thermal re-magnetization within the grain material. Nevertheless, such large inclusions still increase the magnetic dissipation within rotating grains. However, the mechanism of the relaxation is different, it involves spin-lattice relaxation. 

To understand the difference between the spin-spin and spin-lattice relaxation, we note, that when the magnetization in a solid body changes its direction but does not change its amplitude, the change can be performed through the spin-spin interactions. This is, in fact, the case of a grain rotating with its angular momentum ${\bf J}$ at an angle to interstellar magnetic field. Therefore the classical Davis-Greenstein process in paramagnetic grains involves spin-spin relaxation. On the contrary, when the magnetic field changes the amplitude, in order to change the value of magnetization the system of spins has to share its momentum with the lattice. The corresponding process is the spin-lattice relaxation. 

\cite{Jones:1967p2924} showed that magnetic inclusions induce the external magnetic field that is stationary in the grain body frame. Thus, as the grain rotates in the external field, the total magnetization in the grain frame changes both in terms of its amplitude and its direction. As a result (see Appendix \ref{apdx:enhanced}), the relaxation efficiency is enhanced through the spin-lattice relaxation that takes place in the presence of the magnetic inclusions. 

LH08 identified the synergy of RATs and enhanced magnetic relaxation and it introduced a new mechanism of Magnetic RAT (MRAT) alignment. The subsequent study in HL16 shows that the enhancement of the relaxation by an order of magnitude is already enough to bring the degree of the MRAT alignment to 100\%. Thus, the picture is very different from the classical paramagnetic alignment where a more significant increase of magnetic relaxation was necessary in order to explain observations (see \citealt{Jones:1967p2924}). According to Figure \ref{fig:kappa_omega}, both large inclusions and small inclusions can provide conditions for the efficient MRAT alignment. In view of the abundance of iron in interstellar grains and the analysis of the interstellar grains captured in the Earth atmosphere (\citealt{Bradley:1994p6379}) and interplanetary medium (\citealt{Altobelli:2016dl}), the iron inclusions are likely to be present within silicate grains. As we discuss in this paper the magnetic properties of grains affect the alignment and this opens ways to study them from polarization measurements.

\subsection{Relaxation related to nuclear spins}

 It was noticed by \cite{Purcell:1979} that a grain with nuclear moments rotating in an external magnetic field will also experience magnetic relaxation. 
The effect is easy to understand while referring to our discussion presented in Appendix \ref{apdx:enhanced}.  Combining Eq. (\ref{K_par}) and Eq. (\ref{eq:SS}), it is possible to 
see that, for rotation at sufficiently small $\omega$, the expression for function $K(\omega)$ that does not depend either on the density of paramagnetic species nor the value of paramagnetic moments. Therefore, for this range of $\omega$, the magnetic relaxation arising from nuclear spins will be similar to the relaxation arising from the electron spins. The corresponding value of $\omega$ should be less than the inverse of spin-spin relaxation time for nuclear spins (see more in \citealt{1999ApJ...520L..67L}). For \cite{Purcell:1979}, this finding was a matter of curiosity as for the effect could induce only a correction by a factor of 2 for slowly rotating grains. Below we discuss an effect that can potentially have much more tangible consequences for the carbonaceous grains. This new effect is related to the magnetization of nuclear spins within rotating grains. 

\subsubsection{Barnett-equivalent field for nuclear spin system}

Barnett effect (\citealt{Barnett:1915p6353}) is the the  effect of magnetization of a sample with free spins in the presence of the rotation of the sample. It is inverse to the better know Einstein-de Haas effect that is the rotation of the sample as it gets magnetized. The Barnett magnetization of interstellar grains was first discussed by \cite{1976Ap&SS..43..291D} as the means of rendering magnetic moment to paramagnetic grains. Later, the Barnett magnetization was associated with the process of internal energy dissipation in wobbling grains by \cite{Purcell:1979}. 

Following \cite{Purcell:1979}, one can introduce the effective Barnett field, i.e., the effective magnetic field of the amplitude that can cause the Barnett magnetization. This field is
\begin{equation}
H_{\rm Barnett}= \frac{\hbar \omega}{g_{e} \mu_{B}}
\approx \frac{1}{2} 10^{-2} \left(\frac{\omega}{10^5 {\rm s}^{-1}}\right)~{\rm Oe}, 
\label{barn_equiv}
\end{equation}
where $g_e$ is the electron g-factor (\citealt{Morrish:2001vp}).

For paramagnetic grains that rotate with $\omega<10^7$ s$^{-1}$, the Barnett-equivalent field is much weaker than the intrinsic magnetic field within the paramagnetic materials $H_i$. The latter is $\approx 10$ Oe according to \cite{1978ApJ...219L.129D}. However, for fast rotating silicate nanoparticles and Polycyclic Aromatic Hydrocarbon (PAH) large molecules that rotate at much larger angular velocities, the Barnett-equivalent field becomes dominant. As a result, the nature of energy dissipation within a grain rotating in the external magnetic field changes. \cite{2000ApJ...536L..15L} identified a new relaxation process, that they termed {\it resonance relaxation}, that takes place for fast rotating grains (see also \citealt{2016ApJ...821...91H}). This process is important for the alignment of nanoparticles that can be responsible for the anomalous microwave emission that interferes with the Cosmic Microwave Background (CMB) studies (\citealt{{1998ApJ...494L..19D},{1998ApJ...508..157D}}; \citealt{2016ApJ...821...91H}). 

Here we show that another process associated with the Barnett effect can be important for magnetic relaxation within slowly rotating grains. It is possible to show the Barnett magnetization induced in the system of nuclear spins can significantly change the relaxation. Indeed, the Barnett-equivalent magnetic field associated with nuclear spins is significantly stronger than the Barnett-equivalent magnetic field associated with electron spins. It can be obtained from Eq. (\ref{barn_equiv}) by changing the electron g-factor and the value of the electron magneton to the nuclear g-factor and the nuclear magneton, respectively. For protons, the expression for the equivalent field is \citealt{1999ApJ...520L..67L}
\begin{equation}
H_{\rm Barnett, nucl}= \frac{\hbar \omega}{g_{n} \mu_{n}},
\label{B_nucl}
\end{equation}
where the nuclear magneton $\mu_{n}\equiv e\hbar/2m_p c$, and the proton mass $m_p$ rather than the electron mass $m_e$ is used.  Therefore, the Barnett equivalent magnetic field of nuclear spin system is $\sim 10^3$ times larger than of the electron spin system. According to Eq. (\ref{barn_equiv}), it can be $\sim 10$ Oe for a grain rotating with $\omega=10^5$ s$^{-1}$, which can be comparable or stronger than the internal magnetic fields within interstellar grains, especially the carbonaceous grains. 

\subsubsection{Magnetic relaxation of nuclear spin system subject to Barnett effect}

In analogy with Eq. (\ref{K_sl}) for relation for the system of electron spins discussed in Appendix \ref{apdx:enhanced}, one can write for the relaxation in the system of nuclear spins:
 \begin{equation}
 K_{n}(\omega) = X \frac{\chi_n (0) \tau_{\rm n, SL}}{1+(\omega \tau_{\rm n,SL})^2} +(1-X) \frac{\chi_n (0) \tau_{\rm n,SS}}{1+(\omega \tau_{n,\rm SS})^2},
 \label{K_sl_nucl}
 \end{equation}
 where $\tau_{n,SS}$ and $\tau_{n,SL}$ are, respectively, the spin-spin and the spin-lattice relaxation times for the nuclear spin system, while $\chi_n(0)$ is the magnetic susceptibility of the nuclear spin system with density $n_{n}$ at zero frequency:
 \begin{equation}
 \chi(0)=\frac{n_{n}\mu_n}{3kT}.
 \end{equation}
 The magnetization arises from the Barnett-equivalent field $H_{Barnett, nucl}$ and the factor in Eq. (\ref{K_sl_nucl}), namely
 \begin{equation}
 X=\frac{H_{\rm Barnett, nucl}^2}{H_{\rm Barnett, nucl}^2 + 0.5 H_i^2},
 \label{X}
 \end{equation}
changes with the ratio of $H_{\rm Barnett, nucl}$, and $H_i$ is the internal rms magnetic field in the sample. 

The second term in Eq. (\ref{K_sl_nucl}) is the usual spin-spin paramagnetic relaxation of the system of nuclear spins which we discussed earlier. The first term in Eq. (\ref{K_sl_nucl}) presents a new effect that we discuss here, namely the spin-lattice relaxation arising from the presence of the Barnett magnetization. For $X$ of order of unity, the possible enhancement of the relaxation is due to the spin-lattice effect.

As we discuss in Appendix \ref{apdx:enhanced}, the internal field $H_i$ varies with the grain composition, and we adopt  $H_i \sim 10$ Oe for paramagnetic materials (see \citealt{1978ApJ...219L.129D}). As a result, for grains rotating with 
$\omega=10^5$ s$^{-1}$, the Barnett-equivalent field for nuclear spins and the internal magnetic field are of similar magnitudes. However, for typical paramagnetic materials, $n_n/n_e$ is likely to be less than unity, and therefore the process of both spin-spin and spin-lattice nuclear relaxation are mediated by electron-nuclear interactions (Appendix \ref{apdx:alignment}). In this case, one should use $\tau_{ne}$ instead of $\tau_{n,SL}$ and $\tau_{n,SS}$ in Eq. (\ref{K_sl_nucl}). Therefore, the spin-lattice relaxation in nuclear spins changes the relaxation by a factor of unity. To get an increase of the relaxation through spin-lattice effects, one should decrease the density of electron spins compared to the density of nuclear spins. This is the case of carbonaceous grains. 

We expect that the carbonaceous grains have a significantly smaller $H_i$. Therefore the Barnett-induced magnetization may be very important for the dynamics of their nuclear moments. To evaluate the relative importance of the first term in Eq. (\ref{K_sl_nucl}), one has to evaluate the characteristic timescale of spin-lattice relaxation for the nuclear spin system. We argue in Appendix \ref{apdx:alignment} that we expect the nuclear spin system to undergo spin-lattice relaxation through the transfer of its momentum to the electron spin system. The latter system then can relax through the spin-lattice interactions. The bottleneck for such a process is given by the timescale of nuclear-electron spin system interactions, and for carbonaceous grains the corresponding time is (see Eq. \ref{nucl_electr})
\bea
\tau_{ne}&=& \frac{\hbar g_e}{3.8 n_e g_{n}^2 \mu_{n}^2} \nonumber\\
&=& 3\times 10^{-3} \left(\frac{2.7}{g_{n}}\right) \left(\frac{10^{21} {\rm cm}^{-3}}{n_e}\right) \s.
\label{nucl_electr1}
\ena
Therefore according to Eq. (\ref{K_sl_nucl}), carbonaceous grains will experience the spin-lattice nuclear relaxation if their rotation frequency $\omega<\tau_{ne}^{-1}$. For thermally rotating grains this means grains of $\sim 10^{-4}$ cm or larger. As it was shown in LH07, grains at low-J attractor points rotate subthermally, which means that formally they will be subject to the MRAT alignment and their phase portraits will get the high-J attractor point. The problem with this is that as the grain starts moving to this attractor point, its $\omega$ increases and the attractor point disappears. This means that the grains should be even larger for it to be suprathermal when it rotates with $\omega<\tau_{ne}^{-1}$. Such large grains are very rare in the typical ISM, but they are present in the accretion disks. The major obstacle for making more solid conclusions is the uncertainty of the internal magnetic field $H_i$ within carbonaceous grains.

For silicate grains, the internal magnetic field within grains $H_i$ is too strong, and the Barnett magnetization of grains rotating with $\omega <\tau_{ne}^{-1}$ is not sufficient to induce any appreciable spin-lattice relaxation.
 Naturally, in the presence of magnetic inclusions, the spin-lattice relaxation of the nuclear spin system is also enhanced, as discussed in Appendix \ref{apdx:alignment}, but the effect is subdominant in most cases.

\subsection{Magnetic relaxation within silicate and carbonaceous grains}

For most astrophysical settings, the radiation field is strong enough to ensure the dominance of RATs. 
Therefore, the paramagnetic relaxation is weak and has a marginal effect on the RAT alignment. Above we discussed a number of possible effects that enhance magnetic relaxation and can make RATs more efficient. We confirmed that the presence of magnetic (both small superparamagnetic and larger ferromagnetic) inclusions in grains enables MRAT alignment. Our analysis of the effects of the relaxation processes in the system of nuclear spins can potentially make MRAT alignment relevant to the carbonaceous grains larger than $10^{-4}$ cm. 

The existing observations indicating the marginal alignment of carbonaceous grains in the diffuse ISM are suggestive of carbonaceous grains not having magnetic inclusions. Therefore we expect the pure RAT alignment of carbonaceous grains in most ISM cases. As we discuss further, the low magnetic moment of carbonaceous grains makes their RAT alignment in respect to magnetic field problematic and therefore we expect k-RAT alignment of carbonaceous grains to be more important compared to the B-RAT alignment. The magnitude of the alignment depends on the rate of grain randomization that we discuss in \S 6. The results there indicate that in the grain randomization only by gaseous collisions we should expect the k-RAT alignment to be widely spread in the interstellar medium. However, the processes of anomalous randomization can significantly reduce the effect the alignment. Therefore studies of k-RAT alignment of carbonaceous grains can bring an important insight into the anomalous randomization (see \S 6.2). 

The aforementioned problem with the alignment of carbonaceous grains, however, are not present for the composite grains containing both silicate and carbonaceous fragments. We expect such grains to be present in molecular clouds and to be aligned there. The difference in the ability of forming composite grains in diffuse media and molecular clouds can arise from the difference in the physical conditions in these two environments. In particular the mechanism of centrifugal disruption of grains introduced in \cite{Hoangetal:2019} and \cite{Hoang:2019} may be relevant.  Notably, we may speculate that the radiation field in diffuse media is strong enough to disrupt composite grains with insufficiently strong bonds between carbonaceous and silicate fragments. The reduced strength of radiation field in molecular clouds, on the contrary, may not allow RATs to destroy composite grains. 

\section{Internal relaxation within wobbling grains}\label{sec:relaxation}

\subsection{Relaxation within grains without magnetic inclusions}

Grain alignment is a complex process which includes not only the alignment of the grain angular momentum ${\bf J}$ in respect to either the ambient magnetic field ${\bf B}$ or the direction of radiation ${\bf k}$ (i.e., external alignment). It also involves the alignment of ${\bf J}$ along the grain axis of maximal moment of inertia ${\bf a}$, i.e., internal alignment (see illustration of this in \citealt{2007JQSRT.106..225L}). The latter alignment corresponds to the minimal kinetic energy of a rotating grain for a given value of its angular momentum. Therefore the internal relaxation of kinetic energy within a wobbling grain induces the ${\bf a}$ and ${\bf J}$ alignment.

The issue of internal alignment has been an important branch of research within the grain alignment theory. The earlier studies of paramagnetic alignment, e.g., \cite{Jones:1967p2924} did not consider the internal dissipation energy. This would decrease the expectations of the alignment degree, but would not change the direction of the alignment. Our study for RATs of the alignment of large grains for which the internal dissipation is negligible in \citealt{2009ApJ...697.1316H} showed that the very nature of the alignment can change, i.e., the alignment with long axes of grains parallel to magnetic field gets possible.\footnote{Some new polarization data (B-G Andersson, private communications) suggests that this theoretically predicted effect of the "wrong" alignment can be actually present for large carbonaceous grains in the vicinity of stars.} This means that the expected far infrared polarization from such grains can be {\it parallel} to magnetic field. This effect that potentially can significantly complicate the interpretation of polarized radiation in terms of the underlying magnetic fields. In the case of $k$-alignment, the grains with inefficient internal relaxation can align with long grain axis parallel to the radiation direction. Naturally, these effects can complicate the interpretation of the results of polarimetry, in particular the ALMA polarimetry of disks. This makes it essential to revisit the issue of internal relaxation. 

The process of internal relaxation was introduced in grain alignment theory by \cite{Purcell:1979}. The inelastic relaxation first quantified by \cite{Purcell:1979} was later elaborated in the subsequent studies (\citealt{LazEfroim:1999}; \citealt{Efroimsky:2000p5384}). The latter process is the most important for large grains, i.e., grains of $10^{-4}$ cm and larger. With the structure and composition of such grains being uncertain, it is rather difficult to quantify this type of alignment. In this paper that deals with the magnetic properties of dust, we consider effects of relaxation related to the electron and nuclear spin systems of the grains. Our estimates show that for silicate grains present in the diffuse interstellar medium, i.e., with $a<10^{-4}$ cm, the internal relaxation related to magnetic dissipation dominates that arising from inelastic effects. The internal dissipation within larger grains, i.e. grains in molecular clouds and accretion disks, can be different, especially if the grains at hand present loose aggregates of fragments. The internal dissipation within such grains can be significantly enhanced. The inelastic relaxation for carbonaceous grains is likely to be the dominant mechanism for their internal alignment. 

In addition, \cite{1999ApJ...516L..37L} demonstrated that fluctuations arising from internal relaxation induce thermal flipping effect that significantly changes grain dynamics. For instance, grains can get "thermally trapped" and not being able to rotate fast even in the presence of the uncompensated torques arising from H$_2$ formation \cite{Purcell:1979}. In other words, "thermal trapping" effectively suppresses the effects of Purcell's torques on grain alignment. In terms of the RAT alignment, the effect of Purcell's torques is the increase of the degree of the RAT alignment to make it nearly perfect, irrespectively of the presence of magnetic inclusions \citep{2009ApJ...695.1457H}. Resolving what is the critical size of grains that are "thermally trapped" is very important for interpreting the polarization studies, and this provides yet another stimulus to quantify the rates of internal relaxation within grains. 

In the previous section, we discussed the Barnett effect and its consequences for magnetic relaxation. For classical paramagnetic grains, \cite{Purcell:1979} introduced a process which he termed {\it Barnett relaxation}. Its idea is easy to understand. As a grain wobbles the angular velocity vector precesses within the grain axes. Thus, the direction of grain magnetization caused by the Barnett effect also changes. This induces paramagnetic dissipation, resulting in the alignment of grain angular momentum ${\bf J}$  with the axis of maximal moment of inertia ${\bf a}$. The characteristic time of Barnett relaxation for an oblate grain with of dimensions $2a\times 2a \times a$ was estimated in \cite{Purcell:1979} as
\begin{equation}
\tau_{\rm BR}=\frac{\rho a^2 \gamma^2}{K(\omega) \omega^2}\approx 4\times 10^{7} \left(\frac{10^5~\s^{-1}}{\omega}\right)^2 {\rm s},
\end{equation}
which is a short time compared to the characteristic timescales involved in the alignment of interstellar grains.\footnote{Because of that for years it was assumed that grains must have ${\bf J}$ and ${\bf a}$ {\it perfectly} aligned. \cite{Lazarian:1994}, however, showed that for thermally rotating grains, the alignment is far from being perfect due to the thermal fluctuations within grain body. Later, \cite{LazarianRoberge:1997} quantified how the alignment changes as the ratio of the grain rotational energy $E_{\rm kin}$ to grain temperature $T$ changes, i.e., how the alignment changes with the ratio $E_{\rm kin}/kT$. Currently, the thermal wobbling of grains in the presence of  internal relaxation is accepted as the essential part of grain dynamics.} 
 
The Barnett relaxation is the effect related to electron spins. {\it Very counter-intuitively}, \cite{1999ApJ...520L..67L} introduced the internal relaxation in the system of nuclear spins and demonstrated that this relaxation process is more efficient that the classical Barnett relaxation. The authors termed the process {\it nuclear relaxation"} to distinguish it from the Barnett relaxation that acts on the electron spin system. The high efficiency of nuclear relaxation arises from the fact that the Barnett equivalent magnetic field (see Eq. (\ref{B_nucl}) is inversely proportional to the magnetic moment $\mu_n$. At the same time, the dissipative paramagnetic response is proportional to $K(\omega)$, which, as we discussed earlier, does not depend on the value of magnetic moment $\mu$. According to Eq. (\ref{tau_mag}), the rate of relaxation is proportional to $K(\omega) \times H^2$, which gets proportional to $\mu^{-2}_n$. As $\mu_n\ll \mu_e$, the nuclear relaxation can be $\sim 10^6$ times more efficient than the Barnett relaxation.

The considerations above are correct for sufficiently small rotational frequencies, i.e. for frequencies small compared to the rate of spin-spin relaxation rate $\tau_{SS}^{-1}$ for nuclei in the solid. In the opposite case when $\omega \tau_{SS}$ is large, one should take the suppression of the relaxation into account. The time scale for the nuclear relaxation for a "brick" with dimensions $a\times a\sqrt{3}\times a\sqrt{3}$ is:
\begin{equation}
u_{\rm NR}= G_{\rm NR}\hat{\rho}^2 a_{-5}^{7},
\label{NR}
\end{equation}
where
\begin{equation}
G_{\rm NR}=610 \left(\frac{n_e}{n_n}\right) \left(\frac{\omega_d}{\omega}\right)^2 \left(\frac{g_n}{3.1}\right) \left(\frac{2.7 \mu_N}{\mu_n}\right) \left[1+\left(\omega \tau_n \right)^2 \right] \s, 
\label{Gn}
\end{equation}
where, in turn, $\hat{\rho}\equiv \rho/(2 {\rm g~cm}^{-3})$, $a_{-5}\equiv a/10^{-5}$ cm, $\omega_d$ is the angular velocity of a grain rotating thermally at $T_d=10$K, i.e. $\approx 3\times 10^4$ s$^{-1}$. The nuclear spins with magnetic moments $\mu_n=g_{n}\mu_{N}$ are normalized by the magnetic moment of the proton $\mu_N\equiv 5.05 \times e\hbar/2m_pc=10^{-24}$ erg G$^{-1}$, $\tau_n$ is the time of spin-spin relaxation within the system of nuclear spins.  The difference of our present representation for the expression derived for nuclear relaxation in \cite{1999ApJ...520L..67L} is that we use a different asymptotic behavior of the relaxation. For instance, at high frequencies such that $\omega\gg \tau_n^{-1}$, the relaxation time gets constant rather than increasing as $\sim \omega^2$. The latter asymptotic was adopted in \cite{1999ApJ...520L..67L}. Our current choice corresponds to the high frequency behavior of the magnetic response that we adopted in the Appendix \ref{apdx:enhanced} and the justification for that choice is provided there.

\subsection{Additional relaxation within triaxial grains}

\begin{figure}
\includegraphics[width=0.5\textwidth]{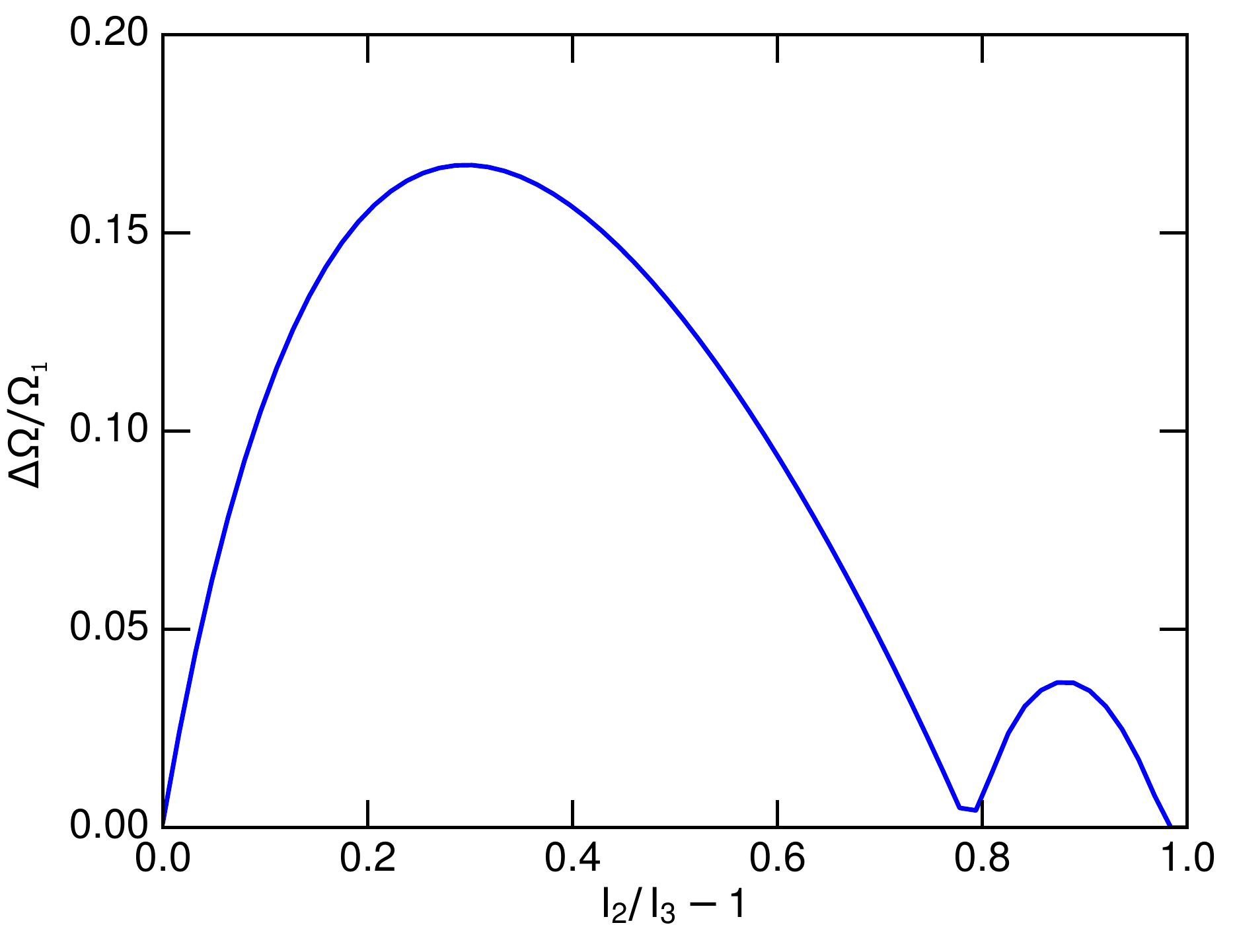}
\caption{Variations of $\Delta \Omega/\Omega_{1}$ as a function $I_{\perp,1}/I_{\perp,2}-1\equiv I_{2}/I_{3}-1$.}
\label{fig:Omega_time}
\end{figure}

Considering triaxial grains without magnetic inclusions, we identify a new way of enhancing  magnetic relaxation that is related to the change of the {\it amplitude} of the Barnett equivalent magnetic field. This new channel of internal relaxation acts both for the relaxation within the electron and nuclear spin systems. Indeed, the amplitude of the ${\bf \Omega}$ is changing for realistic irregular wobbling grains with components of the tensor inertial $I_{\|}>I_{\bot, 1}>I_{\bot, 2}$, where $I_{\|}$ corresponds to the moment of the maximal inertia. The change of the amplitude of angular velocity due to the wobbling of the irregular grain is illustrated in Figure \ref{fig:Omega_time}.  Grain rotation with constant ${\bf J}$ corresponds to the variations of $\Omega$ which are proportional to $1-I_{\bot, 2}/I_{\bot, 1}$. The important consequence of this change is that the variations of the amplitude of $\Omega$ induce the variations of the amplitude of the Barnett magnetization.

Consider first the effect of changing amplitude of the Barnett magnetization on the spin system of electrons. The change of magnetization of the grain is only possible through the spin-lattice relaxation process that has the characteristic time $\tau_{\rm SL}$. Assuming for simplicity that the frequency of fluctuations of the amplitude of ${\bf \Omega}$ vector differs by a factor of unity from the frequency $\Omega$, one can come to the conclusion that for $\Omega \gg \tau_{\rm SL}^{-1}$, the bulk grain magnetization arises from the averaged value of angular velocity $\langle \Omega \rangle$. The direction of ${\bf \Omega}$  precesses in the grain body axes and induces the re-magnetization that involves the spin-spin relaxation with the time scale $\tau_{\rm SS}$. For rotation with $\Omega< \tau_{\rm SL}^{-1}$, the Barnett remagnetization involves changes of the magnetization amplitude. This entails additional spin-lattice relaxation that can be $\sim \tau_{\rm SL}/\tau_{\rm SS}$ faster than the ordinary Barnett relaxation\footnote{This ratio of relaxations times is reduced by the $\Delta \Omega/\Omega$ ratio, which depends on the axis ratio of the triaxial grain (see Figure \ref{fig:Omega_time}). Thus, for an oblate grain, the rate of spin-lattice relaxation is not induced.} For silicate grains with $\tau_{\rm SL}\sim 10^{-6}$ s and $\tau_{\rm SS}\sim 10^{-9}$ s, this can provide an enhancement of factor $\sim 10^3$ for $\Omega<\tau_{\rm SL}^{-1}$. This resulting rate is less than that for nuclear relaxation introduced in LD99b, but the new type of internal relaxation can efficient for larger frequencies compared to the those for nuclear relaxation.

The rate of the relaxation depends on the variation of the amplitude of the Barnett equivalent magnetic field which is proportional to the variations of the amplitude of $\Omega$ as illustrated in Figure (\ref{fig:Omega_time}) for a triaxial grain. To distinguish this type of relaxation from the classical Barnett relaxation that does not involve the change of the amplitude of $\Omega$, we will term this new relaxation Amplitude Barnett Relaxation or, in short, ABR-relaxation. The time ABR relaxations can be written in the form given by Eq. (\ref{NR}), namely
\begin{equation}
\tau_{\rm ABR}= G_{\rm ABR}\hat{\rho}^2 a_{-5}^{7},
\label{abr}
\end{equation}
where the factor 
\begin{equation}
G_{\rm ABR}\approx 2 \times 10^7 \left(\frac{f_1}{0.1}\right)^2 \left(\frac{f_2}{10^3}\right) \left(\frac{\omega_d}{\omega}\right)^2 \left[1+\left(\frac{\omega \tau_{SL}}{2}\right)^2 \right] \s,
\end{equation}
where the factor $f_1$ corresponds to ratio of angular velocity variation $\delta \Omega/\Omega$, factor $f_2$ corresponds to the ratio of the spin-lattice to the spin-spin relaxation times, i.e., to $\tau_{SL}/\tau_{SS}$ (see Appendix \ref{apdx:enhanced}).  Here we choose $f_{2}\sim 10^{3}$ to illustrate the potential effect of ABR for some parameter space of silicate grains.

The Barnett relaxation written in the same form is
\begin{equation}
\tau_{\rm BR}=G_{\rm BR}\hat{\rho}^2 a_{-5}^{7},
\label{br}
\end{equation}
where 
\begin{equation}
G_{\rm BR}\approx 2 \times 10^8 \left(\frac{\omega_d}{\omega}\right)^2 \left[1+\left(\omega \tau_{SS}\right)^2 \right] \s.
\end{equation}

\begin{figure}
\includegraphics[width=0.5\textwidth]{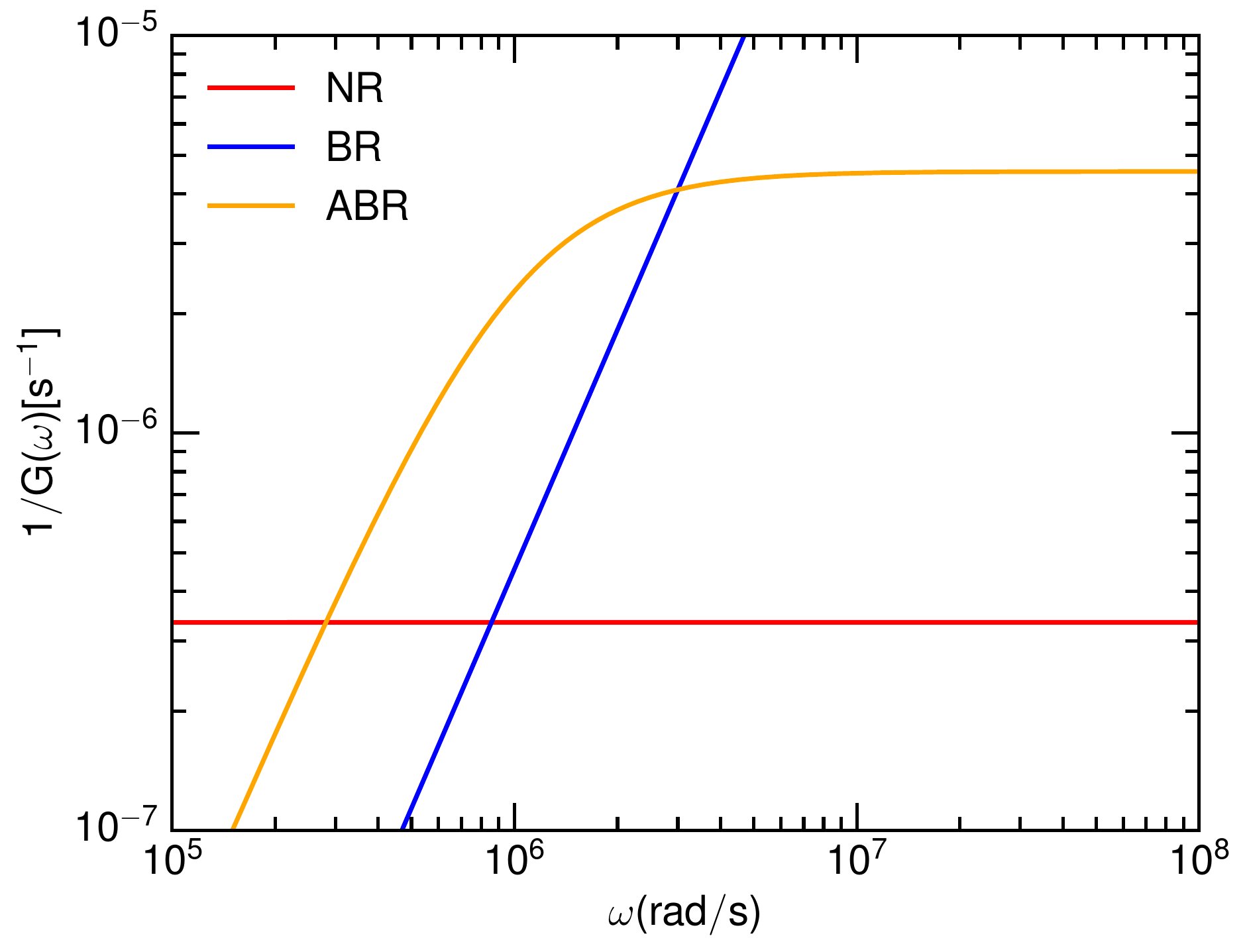}
\caption{The rates of nuclear relaxation (NR), Barnett relaxation (BR), and Amplitude Barnett relaxation (ABR) for silicate grains as a function of grain rotation frequency, assuming the nuclear spin with $g_{n}=1$.}
\label{fig:Gi}
\end{figure}

The thermal velocity of a spherical grain corresponds to
\begin{equation}
\omega_{\rm th}\approx 3.7\times 10^{5} a_{-5}^{-2.5} \hat{\rho} \hat{T}^{1/2} \s^{-1},
\end{equation}
where $\hat{T}\equiv T/100\K$. Figure \ref{fig:Gi} illustrates the frequency behavior $G(i)^{-1}$ for Barnett, ABR and nuclear processes of internal relaxation. The Barnett relaxation is present for high rotation frequencies, while the nuclear relaxation is dominant for the low frequencies. There is a range of intermediate frequencies for which the ABR relaxation is important. 

Therefore, for the set of parameters accepted in Eqs. (\ref{NR}), (\ref{abr}) and (\ref{br}), one can define three regimes of magnetic relaxation. For frequencies smaller than the spin-spin relaxation rate approximately from $10^{9}$ s$^{-1}$ to $3 \times 10^6$ s$^{-1}$, the Barnett relaxation dominates, and for the frequency range from $3 \times 10^6$ s$^{-1}$ to $5.2 \times 10^5$ the ABR relaxation dominates. For the frequencies lower than that, the nuclear relaxation is the dominant process. For grains rotating thermally, the Barnett relaxation dominates for grains smaller than $0.4\times 10^{-5}$ cm, while for grains in the range from $0.4\times 10^{-5}$ cm to $0.9 \times 10^{-5}$ cm the AMR relaxation is dominant. The grains larger than this latter size are dominated by the nuclear relaxation. In other words, the ABR relaxation introduced above is an important process that should be accounted within the quantitative theory of grain alignment. It is easy to see that the ABR-dominance range gets more extended if grains rotate slower. For grains rotating suprathermally, i.e., with the effective rotational temperature $T_{\rm rot}>3600$ K, the interval of ABR dominance disappears making nuclear and Barnett relaxation the only major players.

\begin{figure}
\includegraphics[width=0.5\textwidth]{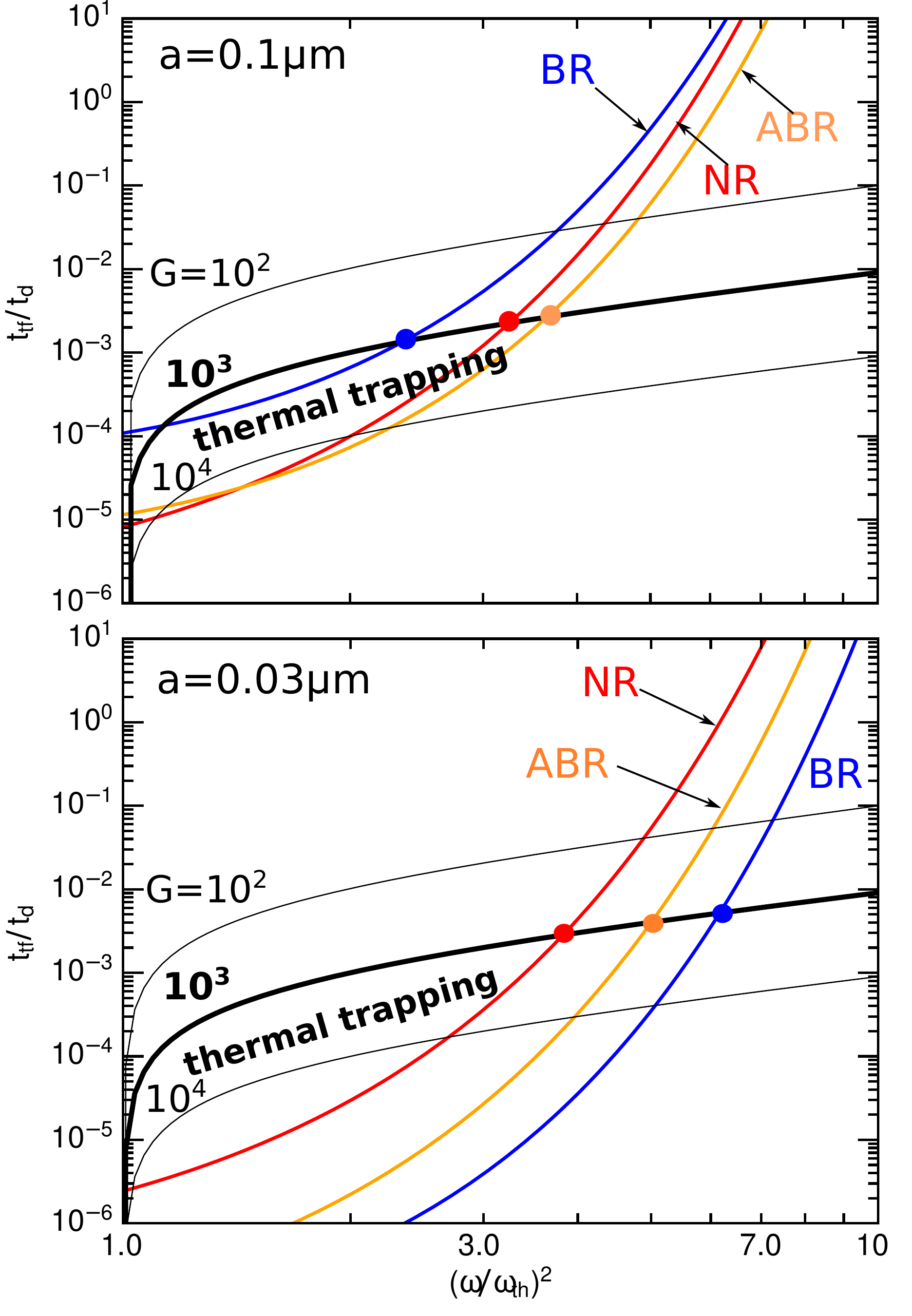}
\caption{Illustration of grain trapping with the rates from nuclear, ABR and the Barnett relaxation for $a=0.1\mu$m (upper panel) and $a=0.03\mu$m (lower panel) grains in the standard ISM of $T_{\rm gas}=100\K$ and $T_{d}=20\K$. Three values of the spin-up parameter $G$ is considered.}
\label{fig:flipping}
\end{figure}

The consequence of the Fluctuation Dissipation Theorem is that the larger the internal dissipation the larger are the thermal fluctuations of grain axes in terms of the direction of grain angular momentum (\citealt{Lazarian:1994}; \citealt{LazarianRoberge:1997}).  The fluctuations occasionally can be large and  \cite{1999ApJ...520L..67L} identified a new phenomenon  related to the internal relaxation, namely, the the thermal flipping. For the uncompensated torques (\citealt{Purcell:1979}), e.g. for the torques arising from H$_2$ formation over grain survace, the grain flips when its angular momentum is small. After flipping the Purcell's torques change their direction. If the flipping is fast enough, then yet another new phenomenon was identified in \cite{1999ApJ...520L..67L} takes place, namely, thermal trapping.\footnote{The ability of grains to undergo thermal flipping was questioned in \cite{2009ApJ...690..875W}. By considering the action of gaseous bombardment \cite{2009ApJ...695.1457H} showed that flipping does take place. Later \cite{2017MNRAS.471.1222K} showed that for triaxial grains the flipping happens even in the absence of gaseous bombardment.} 
As the uncompensated torques change their direction fast due to flipping, they cannot spin up the grain to high velocities and the grains stays in the state of thermal rotation. Thermal flipping and thermal trapping are important elements of grain dynamics that depend on rate of internal relaxation. Therefore in  Figure \ref{fig:flipping} we illustrate the modification of the thermal trapping process described in the presence of the three internal relaxation processes described above. For the typical grain size of $a=0.1\mu$m, thermal trapping is dominated by ABR for $(\omega/\omega_{\rm th})^{2}<4$.\footnote{We illustrate in Figure \ref{fig:Jmax_PSI} that the wide variety of rotational velocities is expected depending on the angle between the direction of radiation and the magnetic field.} Here $G$ is the strength of a systematic torque given by $(G-1)^{1/2}I_{\|}\omega_{\rm th}/\tau_{\rm damp}$. 

The effect of thermal trapping can act against the Purcell torques that act to increase grain angular momentum at the low-J attractor point (HL08).\footnote{The importance of Purcell torques was shown in an observational study by  \citeauthor{2013ApJ...775...84A}. The observed increase of the alignment associated with the action of H$_2$ formation torques testifies that the magnetic relaxation enhancement cannot be too large and at least some grains should increase their degree of alignment in the presence of H$_2$ torques.}  

\subsection{Internal relaxation within grains with magnetic inclusions}

Are there ways to increase the efficiency of internal relaxation? Our discussion in the previous section testifies that both Barnett and nuclear relaxation processes can be enhanced in the presence of magnetic inclusions. The efficiency of the two processes are enhanced the same way as the efficiency of magnetic relaxation of a grain rotating in the external magnetic field (see Appendix \ref{apdx:enhanced}). Indeed, the Barnett equivalent magnetization changes its direction in the grain body and the increase of magnetic response in the form of $K(\omega)$ induces higher rate of energy dissipation. Therefore, rather than repeating our arguments relevant to the enhance magnetic relaxation within the grains we refer our reader to the previous section of the magnetic dissipation in the external magnetic field. Our arguments there are valid if for the electron and nuclear systems of spins one uses
the relaxation that is induced by the Barnett-equivalent magnetic field or its nuclear spin analog.

The obvious outcome of silicate grains having magnetic inclusion is that both they are subject to MRAT and that their internal relaxation is enhanced. This is also applicable to the composite silicate/carbonaceous grains that we expect to be present in molecular clouds. 

\section{Randomization and RAT alignment}\label{sec:random}

\subsection{Classical randomization}

In the theory of grain alignment, the damping of grain rotation is traditionally related to the gaseous interactions. Therefore, the time scale of grain randomization is equal to the time for its colliding with its own mass of the gas. 

The rotational damping rate due to the dust-gas collisions is given by
\bea
\frac{\langle \Delta J\rangle}{\Delta t}=-\frac{J}{\tau_{\rm gas}},
\ena
where $\tau_{\rm gas}$ is the gaseous damping time:
\bea
\tau_{\rm gas}&=&\frac{3}{4\sqrt{\pi}}\frac{I_{\|}}{n_{\H}m_{\H}
v_{\rm th}a^{4}\Gamma_{\|}},\nonumber\\
&=&7.3\times 10^{4} \hat{\rho}\hat{s}a_{-5}\left(\frac{100\K}{T_{\gas}}\right)^{1/2}\left(\frac{30
\cm^{-3}}{n_{\H}}\right)\left(\frac{1}{\Gamma_{\|}}\right) \yr,~~~~\label{eq:taugas}
\ena
where $v_{\th}=\left(2k_{\B}T_{\gas}/m_{\H}\right)^{1/2}$ is the thermal velocity of a gas atom of mass $m_{\H}$ in a plasma with temperature $T_{\gas}$ and density $n_{\H}$. Above, $\Gamma_{\|}$ is a geometrical parameter, which is equal to unity for spherical grains (see Roberge \& Lazarian 1999). This timescale is comparable to the time required for the grain to collide with an amount of gas equal to its own mass.

In addition, IR photons emitted by the grain carry away some of the grain's angular momentum, resulting in the damping of the grain rotation. The rotational damping rate by IR emission can be written as
\bea
\tau_{\rm IR}^{-1}=F_{\rm IR} \tau_{\gas}^{-1},\label{eq:tauIR}
\ena
where $F_{\rm IR}$ is the rotational damping coefficient for a grain having an equilibrium temperature $T_{\rm d}$ (see \citealt{1998ApJ...508..157D}), which is given by
\bea
F_{\rm IR}\simeq \left(\frac{0.4}{a_{-5}}\right)\left(\frac{u_{\rm rad}}{u_{\rm ISRF}}\right)^{2/3}
\left(\frac{30 \cm^{-3}}{n_{\H}}\right)\left(\frac{100 \K}{T_{\gas}}\right)^{1/2},\label{eq:FIR}
\ena
where $u_{\rm rad}$ is the energy density of the interstellar radiation field (ISRF) and $u_{\rm ISRF}=8.64\times 10^{-13}\erg\cm^{-3}$ is the energy density of the local ISRF as given by \cite{1983A&A...128..212M}.

Other processes are also important for randomizing grains. Therefore, the total damping rate is described by
\begin{equation}
\tau_{\rm damp}^{-1}=\tau_{\rm gas}^{-1} + \tau_{\rm IR}^{-1} + \tau_{\rm other}^{-1},\label{eq:tdamp}
\end{equation}
where the time of the gaseous damping time $\tau_{\rm gas}$ is a quantity well defined for neutral gas from the beginning of the studies of the grain alignment (see \citealt{Purcell:1979}), $\tau_{\rm IR}$ is the damping due to IR emission, and the damping by other processes $\tau_{\rm other}$ has been a subject of more recent research. In particular, in the ionized gas, due to Coulomb focusing effect, grain-ion interactions correspond to the cross sections that are different from the geometric cross sections of grains, and the cross-section for positively charged grains is larger than that for negatively charged grains (see \citealt{1998ApJ...508..157D}). Moreover, grains having an electric dipole moment interact with plasma and this induces additional damping.  All these processes have been quantified and elaborated in the studies of emission of spinning dust for the case of the rotating nanoparticles producing anomalous microwave emission (see \citealt{1998ApJ...508..157D}; \citealt{Hoang:2010jy}). However, the obtained expressions can be also used for the classical $10^{-5}$~cm grains. \footnote{A possible exception is that the assumption that the electric dipole moment increases with the grain size in the random walk fashion adopted in \citealt{1998ApJ...508..157D}. We expect that if the random walk adding persists the electric forces would become more and more important and start to affect the grain structure. Therefore it is natural to assume that the dipole moment grow slower with size as grains get larger. Obtaining the corresponding law may require the detailed modeling of the grain growth.} As shown in \cite{2016ApJ...821...91H}, for grains larger than  $a\sim 0.1\mu$m, the damping is dominated by neutral-grain collisions and IR emission (see their Figure 2).

\subsection{Anomalous randomization due to fluctuations of electric and magnetic moments}

The interstellar medium is turbulent (see \citealt{Armstrong:1995}; \citealt{ChepurnovLazarian:2009}; \citealt{XuZhang:2016}; \citealt{XuZhang:2017}). In this situation the charged grains experience acceleration (\citealt{2002ApJ...566L.105L}; \citealt{2003ApJ...592L..33Y}; \citealt{Hoang:2012cx}) and get significant velocities. Rapidly moving grains experience the electric field $E_{induced}=V_{grain}/c\times B$ while moving with velocity $v$ in the magnetic field $B$. The electric field interacting with the grain electric dipole moment induces an additional precession. It was noted  by \cite{2006ApJ...647..390W} that if the electric dipole moment of a grain changes, this causes an additional randomization. This process that can be termed "anomalous randomization" was further elaborated in \cite{2009MNRAS.400..536J}(henceforth JW09). 

The grain with an electric dipole moment $\mu_{elecric}$ precesses in the field $E_{induced}$ about the axis perpendicular to the magnetic field direction. The latter precession is not regular if $\mu_{electric}$  is fluctuating. The fluctuations of the electric charge on the grain surface were identified in JW09 as the major source of electric moment variations. It was suggested by JW09 that the torques arising from the interaction of the stochastic fluctuations of the electric dipole moment  with the electric field should induce the random walk randomization of the grain angular momentum ${\bf J}$. The estimates in JW09 show that for typical interstellar conditions the rate of randomization arising from this process can be faster than $\tau_{\rm gas}^{-1}$.  

JW09 note the possible problems related to explaining the observed grain alignment. However, we cannot agree completely with their conclusions. JW09 base their reasoning on the grain alignment picture in \cite{2003ApJ...589..289W}. There it was claimed that grains tend to stay in the state of low subthermal angular momentum and therefore are easy to randomize. However, the dynamics of dust in the presence of RATs was revisited in LH07 and the subsequent studies. In particular, LH07 showed that the grains can have high-J attractor points and, in fact, LH08 showed that in the presence of enhanced magnetic response most of the grains have high-J attractor points. Even if  grains have only low-J attractor points (see Figure 24 in LH07) than, in the presence of randomization, grains spend a significant amount of time at high angular momentum, moving in the phase space around high-J repellor point (HL08). As a result, such grains have, on average, a significant $J$ and demonstrate a significant alignment (HL08). Last, but not the least, if irregular grains are moving in respect to the ambient gas they are subject to uncompensated mechanical torques. This effect was identified in \cite{LazarianHoang:2007b} and proven numerically in \cite{Hoangetal:2018}. Due to this torques the irregular grains moving in respect to the gas  can rotate suprathermally. According, however, to \cite{2009ApJ...695.1457H} the suprathermal rotation "lifts" the low-J attractor point. This is another factor in reducing the randomization of grain angular momentum. This effect acts to decrease the resulting randomization. All these effects that decrease the rate of anomalous randomization must be accounted for. 
  
 All the above was relevant to the B-RAT alignment, i.e., the alignment by RATs in respect to magnetic field. If we deal with the k-RAT alignment, an new effect of magnetic anomalous randomization must be accounted for. Indeed, consider a setting when both the magnetic field and the RATs induce the precession in the two perpendicular directions. If the RAT precession dominates, the irregularity of the Larmor precession induces a randomization similar to that induced by the irregularity of the electric dipole precession that we discussed above.  The fluctuations of the magnetic dipole moment of the grain due to thermal magnetic fluctuations can induce the randomization. This effect is significantly enhanced in the presence of  superparamagnetic inclusions. Naturally, a similar randomization can be considered for an arbitrary angle between the magnetic field and the direction of radiation anisotropy. This provides a new type of anomalous randomization that will be considered elsewhere.

\subsection{Complexity of RAT alignment}\label{sec:complex}

The description of the RAT alignment in LH07 and HL08 assumed the classical randomization by gas-grain collisions and disregarded paramagnetic relaxation. An additional strong magnetic relaxation introduced in LH08 and HL16 made the alignment simpler as it allowed to perfectly align grains irrespectively of the other factors, in particular the $q^{\rm max}$ factor, that enter the LH07 analytical theory. Numerical simulations for RAT alignment for grains with iron inclusions are performed in \cite{2016ApJ...831..159H} where the perfect alignment was demonstrated. 

The present paper demonstrates the potential complexity of the alignment. The complications arise both from (1) the potential dependence of the rate of magnetic relaxation and the internal relaxation on the rate of grain rotation, (2) the presence of anomalous randomization. These additional effects require detailed studies in order to formulate the comprehensive theory covering grains of arbitrary composition and being in an arbitrary dynamical state. In what follows, instead, we describe briefly what we believe is the consequence of these effects. An additional complication, namely, (3) the dependence of grain alignment direction on the grain composition and intensity of anisotropic radiation flux will be considered in more detail for the rest of the paper.  

The mechanisms of enhanced relaxation that we have explored show a dependence of the relaxation on the frequency of grain rotation. The range of frequencies at which a change of relaxation takes place depends on the mechanism and can be from $\sim 10^{3}$ s$^{-1}$ for carbonaceous grains which relaxation is enhanced by the Barnett effect as it discussed in \S 3, to $10^{6}$ s$^{-1}$ when inclusions within silicate grains are considered. "Classical" $10^{-5}$ cm grains have characteristic rates of thermal rotation $\sim 10^5$ s$^{-1}$. Under the influence of RATs, the rates can increase by a factor of 10 or more for the standard interstellar radiation field. The increase can be larger in the vicinity of stars and other luminous radiation sources. The high rotation and perfect alignment are associated with the alignment of grains at the high-J attractor point (see LH07). The existence of high-J attractor point for ordinary paramagnetic grain depends on the shape of the grain as it is reflected by its  q-factor\footnote{This factor depends on the ratio of the torques in the grain system of reference and depends on the shape and the composition of the grain (LH07).} the magnetic response and the angle between the magnetic field and the radiation direction. 

The anomalous randomization depends on the grain velocities in respect to the magnetized astrophysical plasmas. Stellar radiation was identified as the driver of grain velocities in the early literature on gas physics. The expected motions of grains were assumed to be parallel to magnetic field and therefore the radiation would not cause the anomalous 
randomization. Grain acceleration through the gyro-resonance mechanism (see \citealt{2003ApJ...592L..33Y}), on the contrary, induces the grain motion perpendicular to magnetic field and this induces the anomalous randomization as it was noted in \cite{2006ApJ...647..390W} and JW09. The randomization increases with the increase of the grain velocity component perpendicular to magnetic field, as well as the amplitude of the fluctuations of electric dipole moment of the grain. The latter arises from both the fluctuations of the grain charge distribution over grain surface and the grain flipping (\citealt{1999ApJ...516L..37L}; \citealt{HoangLazarian:2008}; \citealt{2017MNRAS.471.1222K}). Most processes involved in the anomalous randomization have not been thoroughly identified, unfortunately. This presents a problem for a rigorous study. We may make a few statements, nevertheless. 

The magnetic moment of carbonaceous grains is expected to be much smaller than that of silicate grains (see \citealt{2016ApJ...831..159H}). Therefore, as first noted in JW09, the anomalous randomization of carbonaceous grains in respect to magnetic field is expected to be significantly enhanced compared to that of silicate ones. In fact, we may expect only marginal  B-RAT alignment of carbonaceous grains. 

For B-RAT alignment of silicate grains, a further reduction of the anomalous randomization is expected for grains with magnetic inclusions. However, for k-RAT alignment, as we discussed earlier, the new type of anomalous randomization arising from the interactions of thermal fluctuations grain magnetization and the ambient magnetic field is expected. This randomization can be called B-type anomalous randomization to contract to the E-type anomalous randomization described in JW09.  Therefore the composition of grains significantly changes both types of RAT alignment. Studies of k-RAT alignment of silicate grains can provide the insight on the importance of B-type anomalous randomization and therefore on the magnetization of grains and their velocities. The B-type anomalous randomization is not expected for the carbonaceous grains. Therefore, one may expect higher k-alignment of carbonaceous grains compared to the silicate grains.

\section{Theoretical expectations}\label{sec:theory}

\subsection{Simple arguments}

{\bf Magnetic susceptibility of grains}. In LH07 it was explained that the alignment can happen both in terms of magnetic field and in terms of radiation depending on whether  the rate of precession induced by magnetic field or the rate of precession induced by the radiative torques is higher. The rate of precession in magnetic field depends on the magnetic susceptibility which is currently not constrained from observations. Indeed, superparamagnetic  inclusions can change the magnetic susceptibility by many orders of magnitude. In comparison to this, the rate of precession of grains is well constrained if the radiation sources, the grain sizes and the rate of grain rotation are known. The latter is not certain, but still its variations are better constrained compared to the grain magnetic susceptibilities.

Our suggestion is to constrain the magnetic susceptibility of dust grains observationally by identifying the point where the grain alignment changes from being in respect to radiation to being in respect to magnetic field. In many relevant settings the stars act as the source of anisotropic radiation. Therefore, by knowing the star luminosity and the distance at which the transition from one direction of alignment to the other, one can estimate the precession rate corresponding to the point of the transition. This precession rate corresponds to the precession of grains in magnetic field. This provides the following alternatives:
\begin{itemize}
\item If one has an estimate of magnetic field, this can constrain grain magnetic susceptibility. The uncertainties in the magnetic  field estimation usually do not exceed a factor of few, which is much smaller compared with the uncertainties in the dust magnetic susceptibilities. The latter estimates in the literature vary by a factor of $10^6$. 
\item Assuming that grains do not have magnetic inclusions, the transition point provides an upper limit of magnetic field strength.
\item If the magnetic susceptibility of grains is known, the transition point gives the actual value of magnetic field strength.
\end{itemize}

We note that resolving the issue of the actual  magnetic susceptibilities one can address a question of whether it is the RAT or the MRAT mechanism that is responsible for the grain alignment. The answer to this question is essential for making the grain alignment theory really predictive. 

{\bf Relative Magnetic strength}. Even if magnetic susceptibility is not precisely known, it is reasonable to assume that the grains of the same composition have the same magnetic susceptibility. As a result, the aforementioned effect of the change of the direction of the alignment provides a way to study of {\it relative} magnetic strength in different regions. The uncertainty in the absolute values of magnetic field would in this situation arise from the uncertainties of the dust magnetic susceptibilities. If magnetic field in any of these regions is measured using other techniques, e.g. Zeeman measurements, Chandrasekhar-Fermi technique (see \citealt{2010HiA....15..438C}), this will allow to improve the estimates of the magnetic susceptibilities and enable the technique to measure the absolute values of magnetic field in different part of the interstellar medium, molecular clouds and accretion disks. 

There are a number of ways to observe the transitions from B-RAT to k-RAT alignment. High resolution observations of the polarization arising from aligned grains is a natural way to find the transition point from the alignment in respect to radiation to the alignment in respect to magnetic field. ALMA naturally presents such a possibility for the studies of accretion disks. In addition, similar studies are also possible by measuring the point where circular polarized radiation is produced. Indeed, as we discuss in \S 9.4 single scattering from aligned grains produces circular polarization, but only in the situation when the direction of radiation and the direction of the alignment axis do not coincide (see \citealt{2007JQSRT.106..225L}). Therefore the grains aligned in respect to radiation are not
expected to produce circular polarization. As we discuss below, the latter effect is possible, however, if the alignment happens in respect to the ambient magnetic field. Finally, if the radiation source is variable, as this is the case of variable stars or novae, then one can also observe the variations of polarization (both linear and circular) induced by the change of the alignment direction.

\subsection{If anomalous randomization is present}

We mentioned earlier that the can be additional processes that increase grain randomization.  For instance, estimates in JW08 provide a randomization rate  larger than accepted rates of gaseous randomization for typical ISM parameters, i.e., $\tau_{\rm gas}^{-1}$. In this case the value of $\delta_m$ can be significantly reduced. For carbonaceous grains the randomization potentially may happen faster than the precession time, which will preclude carbonaceous grains from being aligned. Faster precessing silicate grains can still be aligned by the fast RAT alignment described in LH07. The latter type of alignment takes place over time scale much shorter than $\tau_{\rm gas}$. 

\begin{figure}
\includegraphics[width=0.5\textwidth]{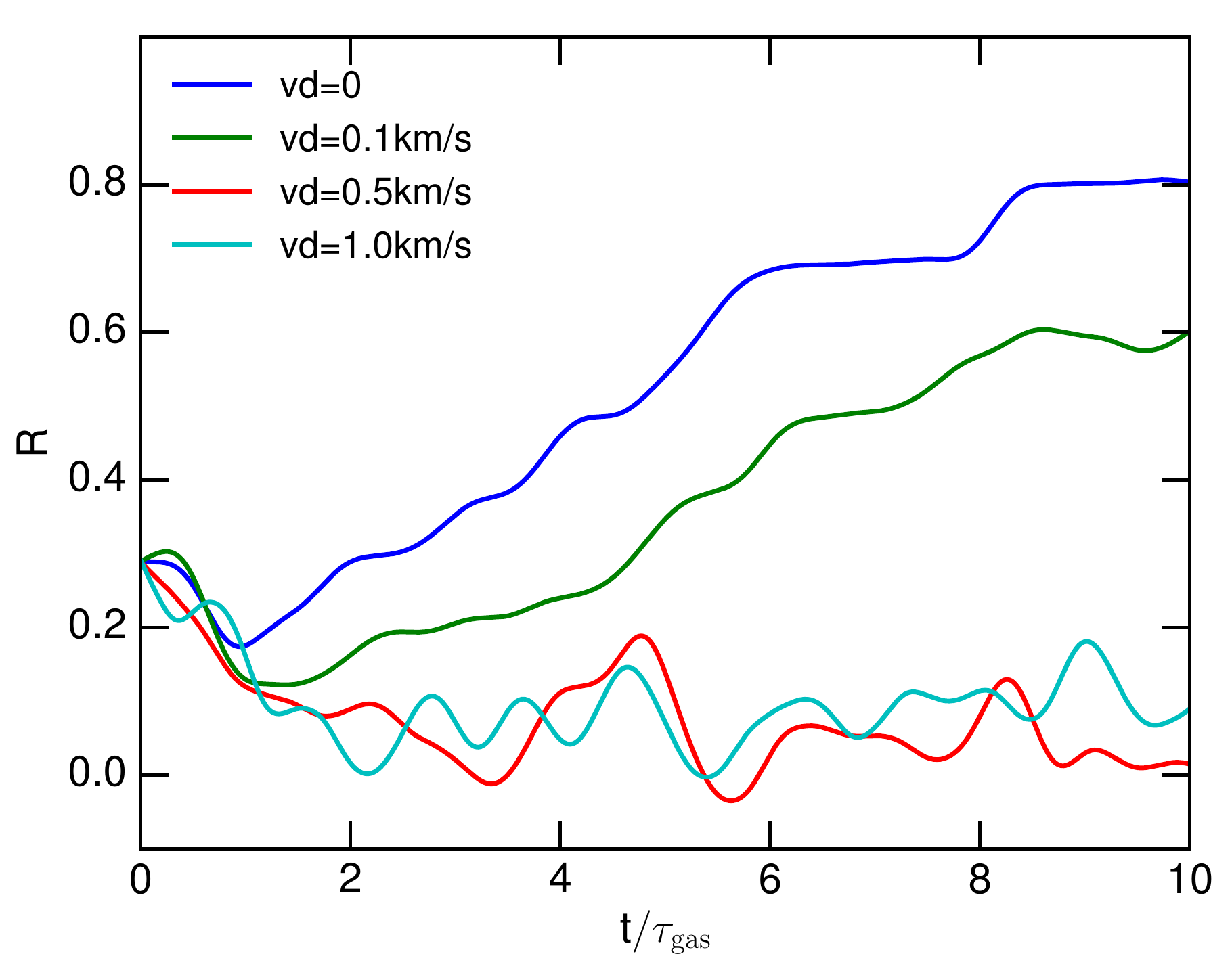}
\includegraphics[width=0.5\textwidth]{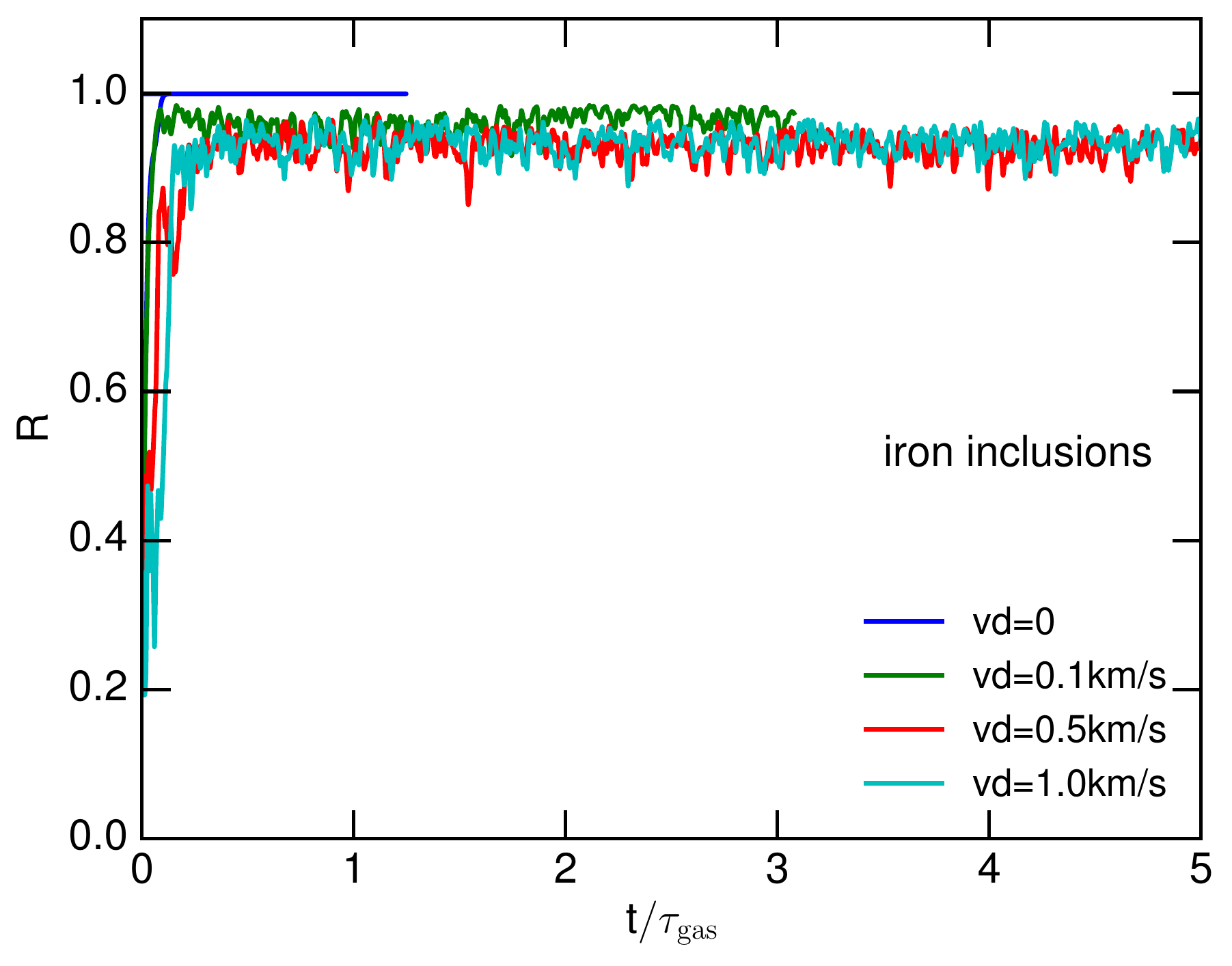}
\caption{Rayleigh reduction factor of grains aligned by RATs in the presence of anomalous randomization for paramagnetic grains (upper panel) and superparamagnetic grains of $\delta_{m}=60$ (lower panel). The grain size $a=0.1\mu m$ and the typical interstellar medium ($n_{\rm H}=30\cm^{-3}, T_{\rm gas}=100$K) are considered.}
\label{fig:R_vd}
\end{figure}

To investigate the effect of anomalous randomization on grain alignment, we perform numerical simulations for grain rotational dynamics subject to RATs, stochastic collisions, magnetic relaxation (see \citealt{2016ApJ...831..159H} for details). We also account for the effects of anomalous randomization by adding the corresponding random torques (see \S 6.2). To characterize grain alignment we calculate the Rayleigh reduction factor (Greenberg 1968):
\begin{equation}
    R=3/2 \langle \cos^2\theta -1/3\rangle
\end{equation}
where $\theta$ is an angle between the grain axis and the magnetic field. The initial rotation of grains is assumed suprathermal with $J=25J_{th}$ and we assume that the alignment of grain angular momentum in grain axes is perfect. The calculations are performed for a model grain with $q^{max}=0.78$, which ensures that the paramagnetic grain has the high-J attractor point (see LH07). We compute the evolution of the Rayleigh reduction factor for B-RAT alignment for an ensemble of grains subject to the interstellar radiation field (see LH07). Upper panel of Figure \ref{fig:R_vd} present the evolution of $R$ for a paramagnetic grain, while the lower panel of the figure depicts $R$ for a grain with enhanced magnetic response corresponding to $\delta_m=60$. The calculations were provided for different velocities of grains (see Hoang \& Lazarian 2019, for more details). 

Fig. \ref{fig:R_vd} testifies that the alignment of a paramagnetic grain is significantly altered by anomalous randomization for sufficiently large drift velocities. On the contrary,
the grain not subject to the anomalous randomization with $v_d=0$ gets well aligned. Its alignment is increasing in time and is expected to become a perfect alignment as more and more of grain in the ensemble end up rotating at high-J attractor point.\footnote{The grains experience thermal gas bombardment and diffuse in the phase space under the action of random torques, but when they get into high-J attractor points, the probability of their escape from these points decreases significantly.}  However, we see that for grains moving with velocities larger than $v_d=0.5$ km/s, the resulting alignment of interstellar grains is marginal. To allow ordinary paramagnetic grains to get aligned, grain velocities in respect to magnetic field should be significantly reduced. This is an important conclusion that can testify in favor of the necessity of grains to have enhanced magnetization as we discuss below. 

Grain velocities of the order of 1 km/s were reported for large grains (see Yan \& Lazarian 2003). One can question the estimates for the acceleration of charged grains by MHD turbulence and claim that the velocities obtained in the corresponding papers may overestimate the actual velocities. However, the lower limit estimate of the grain relative velocities to magnetic field can be obtained from the hydrodynamic approach in \cite{2002ApJ...566L.105L}. The calculations there testify that for grains larger than $10^{-5}$ cm $V_d\approx 0.1$ km/s and it increases with the increase of the grain size. Therefore, the effects of the anomalous randomization cannot be neglected. This makes it unlikely that for typical conditions of the interstellar medium paramagnetic grains can be aligned by the RATs to the degree that can explain the highest degree of observed polarized radiation. This can happen, however, at the regions of higher radiation fluxes and/or higher magnetic field strength.

On the other hand, Figure \ref{fig:R_vd} testifies that the alignment of superparamagnetic grains is efficient even at high drift velocities. 
This happens due to the fact that grains have magnetic inclusions the rate of precession of such magnetic grains increases significantly. In addition, grains at high-J attractor point are much less susceptible for randomization.  When $\delta_m\gg 1$ the RAT alignment transfers into the MRAT alignment \cite{LazarianHoang:2007b}, which is characterized by nearly $100\%$ degree of alignment \citep{2016ApJ...831..159H}. In other words, to reconcile the alignment theory predictions and the high degree of observed polarization of silicate grains, one should allow for the enhanced magnetic relaxation.

It is obvious that the B-RAT alignment of carbonaceous grains is significantly reduced compared to the alignment of paramagnetic silicate grains. This follows from the lower rate of precession of carbonaceous grains. Therefore we do not expect to observe the B-RAT alignment for such grains in normal interstellar conditions. The k-RAT alignment is expected to be present if the radiation field is strong enough. The distance over which such alignment takes place depends on the rate of randomization. Therefore the detailed high resolution observation of the polarization from carbonaceous grains can provide an insight into the efficiency of randomizing anomalous torques.

We note that the anomalous randomization depends on the grain environment. For instance, the dipole fluctuations depend on the grain charge, which in turn depends on the gas ionization fraction. In the circumstellar disk, the ionization is very small. Therefore, anomalous randomization can be significantly reduced there. This, for instance, can induce a more efficient k-RAT alignment of carbonaceous grains.

\section{k-RATs versus B-RATs}\label{sec:kBRAT}

Below we consider a few consequences that the interplay of , B-RAT and k-RAT alignment processes. The alignment via k-RATs is expected to be widely spread for grains with silicate and carbonaceous grains without inclusions, provided that the anomalous randomization is not significant. For grains with magnetic inclusions  MRAT is the major mechanism of alignment, and, as a result, the k-RAT can dominate only in the vicinity of strong radiation sources.

Table \ref{tab:notations} summarizes the abbreviation and meaning of various alignment mechanisms.

\begin{table}
\caption{Notation and Terminology}\label{tab:notations}
\begin{tabular}{l l } \hline\hline\\
{\it Abbreviation} & {Meaning}\\[1mm]
\hline\\
RAT & RAT alignment for ordinary paramagnetic grains\\[1mm]
MRAT & RAT alignment for grains with \\[1mm]
 & enhanced magnetic susceptibility \\[1mm]
SRAT & RAT alignment for superparamagnetic grains\\[1mm]
B-RAT & RAT alignment along B-field\\[1mm]
k-RAT & RAT alignment along k-field\\[1mm]
MAT & Alignment by mechanical torques\\[1mm]
B-MAT & Alignment by mechanical torques along B-field\\[1mm]
v-MAT & Alignment by mechanical torques along v-field\\[1mm]
\\[1mm]
\hline\hline\\
\end{tabular}
\end{table}

\subsection{Differences of alignment at low and high J attractors}

An interesting new effect emerges in the process of RAT alignment. It is known from the RAT theory that grains can be aligned at low and high-J attractor points. The existence of the high-J points depends on the direction of radiation and the grain parameter, which was termed $q-$factor in LH07. The q-factor depends on the ratio of the radiative torques calculated in the system of the grain (see LH07) and is the intrinsic property of a grain, similar to its axis ratio. The parameter space corresponding to the alignment with both high and low-J attractor points was discussed in LH07 for a typical paramagnetic grain. The parameter space for the high-J attractor alignment increases as the magnetic susceptibility of grains increases (HL16). The corresponding calculations were done for the magnetic or B-RAT alignment. 

\begin{figure}
\includegraphics[width=0.5\textwidth]{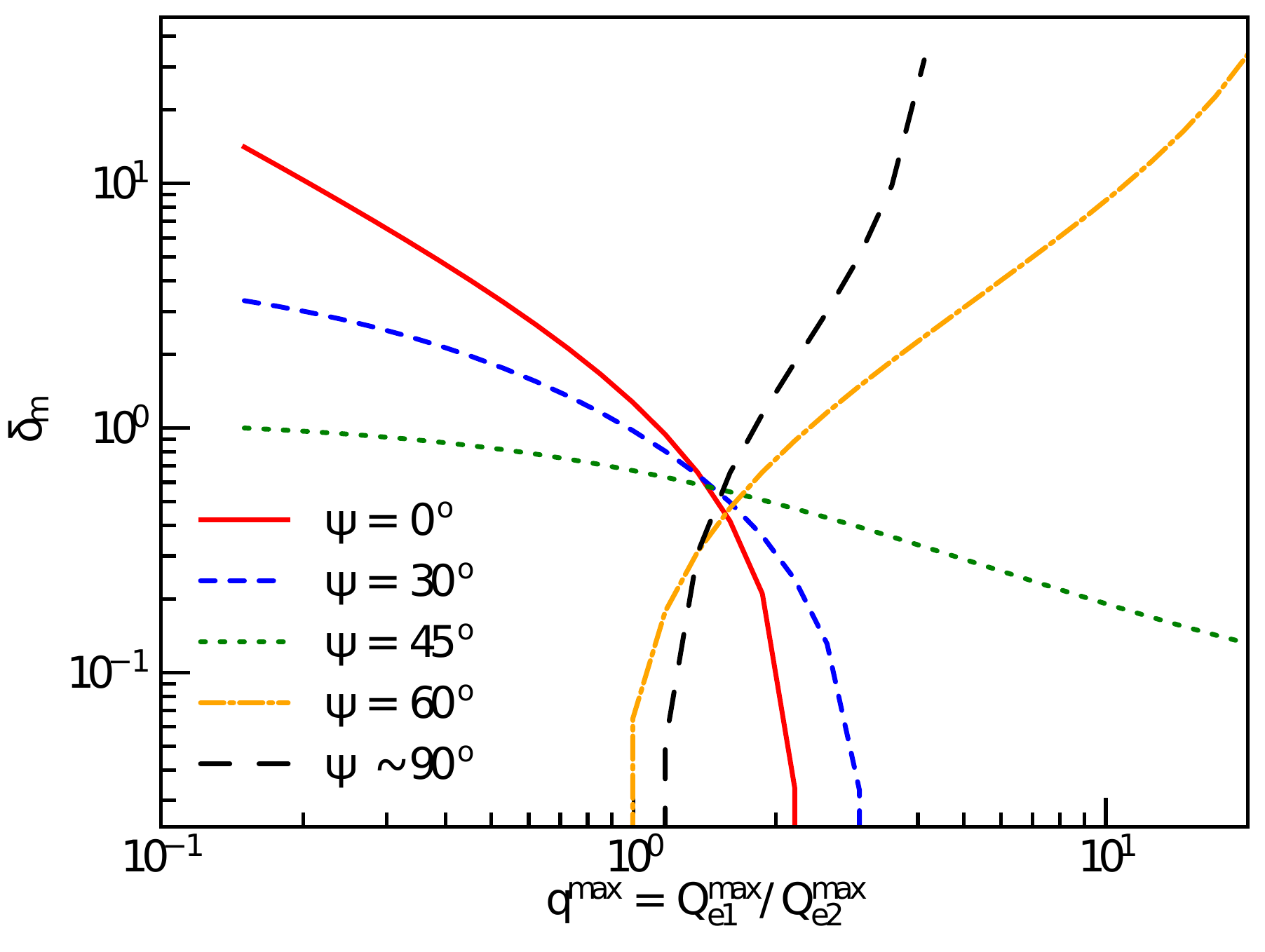}
\caption{Magnetic parameter $\delta_{m}$ required for the existence of high-J attractor as a function of the torque parameter computed for the different radiation direction in respect to magnetic field (angle $\psi$). Standard physical parameters of the CNM with gas density $n_{\H}=30 \cm^{-3}, T_{\rm gas}=100 \K, T_{d}=20 \K$, and the grain size $a=0.2\mu$m are assumed.}
\label{fig:highJ_qmax}
\end{figure}

 In Figure \ref{fig:highJ_qmax} we show the parameter space for the high-J attractors in coordinates of $\delta_m$ and $q-factor$ for the different values of the angle $\psi$ between the magnetic field and the direction of radiation. We observe that paramagnetic relaxation has some effect on creating high-J points even for $\delta_m<1$. At the same time, the importance of magnetic relaxation on stabilizing grain for high-J rotation significantly increases as $\delta_m$ increases values larger than 1. In fact, we see observe that for all explored $\psi$ the range of $q^{\rm max}$ from 1 to 2 the grains gain high-J points as $\delta_m\rightarrow 1$. This is the range of most probable $q^{\rm max}$ revealed a study of irregular grains in \cite{Herranen:2019kj}. 

If the alignment takes place in respect to radiation, i.e. it is k-RAT alignment, it corresponds to anisotropic radiation along the magnetic field, i.e., $\psi=0$. Indeed, in this situation both magnetic field and radiation induce the same type of precession.

To change the direction of the axis of alignment the anisotropic radiation should should induce the precession faster than the Larmor one. This is more difficult for fast rotating grains at high-J attractor. The parameter space for the high-J attractors depends on the angle that anisotropic radiation makes with the magnetic field direction. Figure \ref{fig:highJ_qmax} demonstrates the parameter $\delta_m$ versus $q^{\rm max}-$factor space (see LH07) for high-J attractors for the different angle $\psi$, including $\psi\approx \pi/2$.  Therefore one can potentially think of cyclic behavior of grain alignment moving from one alignment direction to another. For instance, consider  a grain that has a low-J attractor point for the B-RAT alignment in the vicinity of  $\psi=\pi/2$. This slowly rotating grains is subject to the RAT-induced precession (see \S 3.3) the rate of which is inversely proportional to grain angular velocity (see Eq. (eq:QRATs)). 
This can induce k-RAT alignment, i.e. the alignment in respect to the radiation direction. As we discussed above, the alignment when the axis of precession and the radiation direction coincide correspond formally to $\psi=0$. For this angle the grain is more likely to have an high-J attractor point (see Fig.\ref{fig:highJ_qmax}). If this the case, the grain rotational velocity may increase by a factor of $>10^2$ which would induce the decrease of the rate of the RAT-induced precession by a similar factor. As a result, the rate of Larmor precession that does not depend on the grain angular velocity can become larger that the rate of RAT- induced precession. This would revert the alignment to the B-RAT alignment. Their, as we assumed, there is no high-J attractor point and the grain is bound to slow down, increasing the rate of RAT-induced precession.    If this happens, the cycle of realignment repeats. In other words, the dependence of the appearance of the high-J attractors on $\psi$ as well as the value of $J$ as a function of $\psi$  may in some cases provide a complex dependence for $t_{rad, p}$. 

We note that the RAT component causing the precession, denoted by $Q_3$ in LH07 and described in detail there, gets to zero at $\psi$ equal $\pi/2$. This can modify the scenario of cyclic alignment suggested above for a range of $\psi$ close to $\pi/2$.\footnote{Cyclic alignment a possibility of which we discussed should not be confused with the cyclic phase trajectories discussed in Draine \& Weingartner (1997). In the latter work those trajectories were an artifact of the treatment that disregarded the dynamics of crossover events (see Spitzer \& McGlynn 1979).} However, the situation that the grains are perfectly aligned in respect to the magnetic field which is exactly at perpendicular to the incoming radiation is rather degenerate case.  

\begin{figure}
\includegraphics[width=0.45\textwidth]{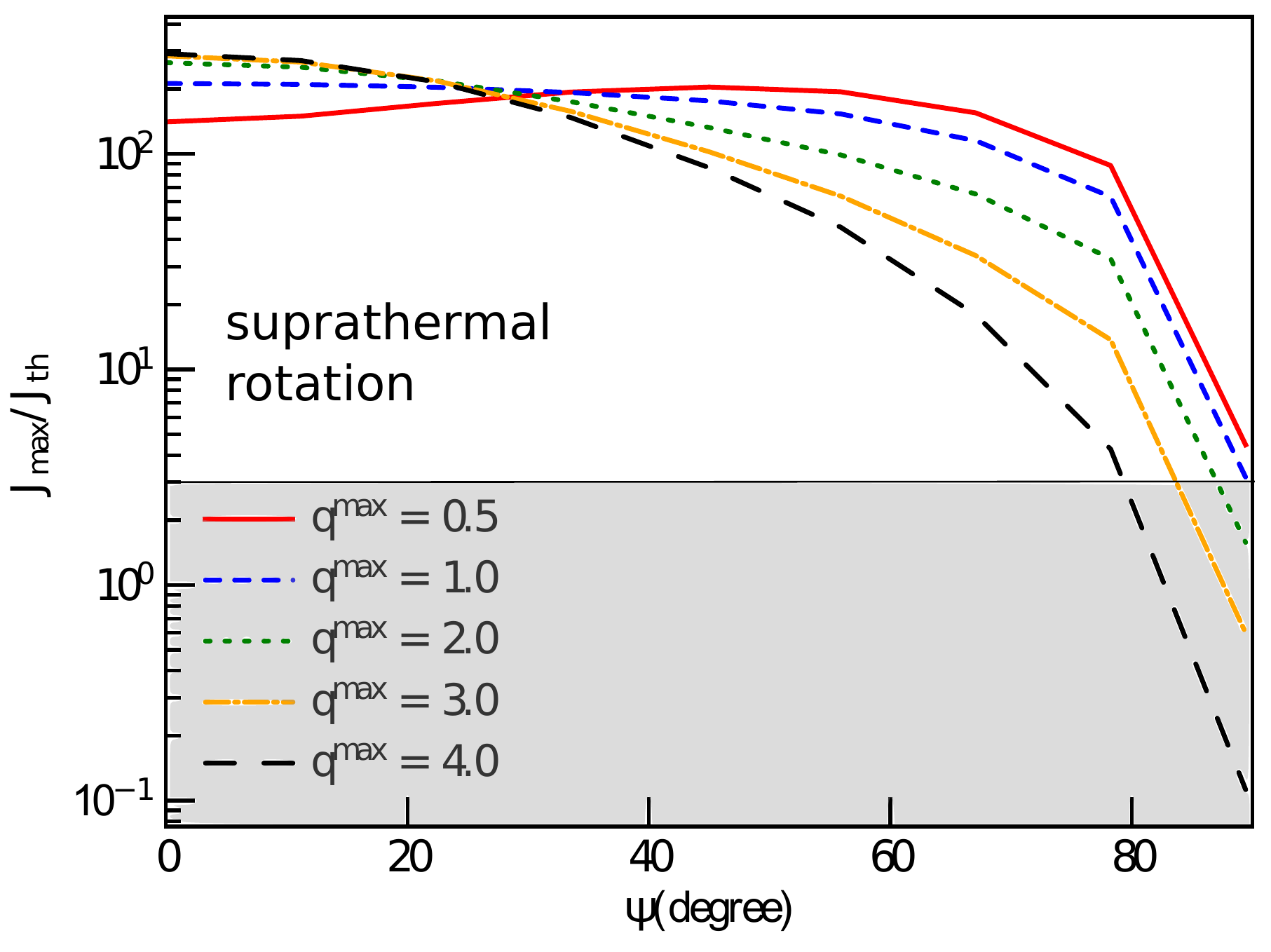}
\caption{The maximum angular momentum spun-up by RATs computed for the different values of $q^{\rm max}$. Gaseous density with $n=100$ cm$^{-3}$ and temperature $T_{\rm gas}=100$K are assumed for the gaseous damping calculations. The white area shows the suprathermal rotation region with $J_{\rm max}\gtrsim 3J_{\rm th}$. Decrease of $J_{\rm max}$ with increasing $\psi$ is observed.}
\label{fig:Jmax_PSI}
\end{figure}

We note that, to predict the actual dynamics  of grains is one should also take into account also the change of the amplitude of the RATs illustrated by Figure \ref{fig:Jmax_PSI}.This dynamics depend on $q^{max}$ value and for $q^{max}$ in the range of 0.5 to 2, which, in fact, covers most of the random shapes explored in Herranen et al. (2019), the rotational rate does not change significantly over $\psi$ in the range from 0 to 80 degrees. It is clear also clear as $\psi$ gets close to $\pi/2$ the efficiency of torques in spinning up grains decreases. The calculations predict the subthermal rotation, which, however, is an artifact of the absence of the gaseous bombardment. Another issue related to Figure \ref{fig:Jmax_PSI} is that the calculations do not take into account that for paramagnetic grains a range of $\psi$ the attractor points transfer into repellor points (see LH07). As a result, the results in Figure \ref{fig:Jmax_PSI} overestimate the rotational rate of paramagnetic grains, but are applicable to grains with strong magnetic relaxation that always have high-J attractor points.  

In fact, Figure \ref{fig:Jmax_PSI} demonstrates that even for grains with enhanced magnetic relaxation the alignment by the radiation direction is in the range of $85$ to $105$ degrees in respect to magnetic field can be significantly reduced.\footnote{Incidentally, this can be the reason for the inefficiency of mechanical torque, or MT alignment. MHD turbulence induces grain motions mostly perpendicular to the magnetic field lines. At our earlier discussion of precession rates in \S 3 suggests that the alignment is typically happens in respect to magnetic field. In this situation, the amplitude of MTs is expected to be reduced a similar way as the amplitude of RATs in the vicinity of $\psi=pi/2$.}    The RAT alignment gets inefficient if the $J_{max}$ gets comparable with $J_{th}$. The grains can still be aligned by Davis-Greenstein mechanism, but this mechanism fails when the gas and dust temperatures become comparable (Jones \& Spitzer 1967, Roberge \& Lazarian 1999, Lazarian 2003 and ref. therein).         

The suppression of magnetic response for the fast rotating grains is another interesting effect that can change their dynamics in very non-trivial way. Indeed, as we discussed in \S 4 and \S 5, both the rate of internal relaxation and magnetic relaxation in the external field depend on the grain rotation rate. Therefore very fast rotating magnetic grains lose their ability to dissipate fast and therefore stabilize their high-J attractor point. 

\begin{figure}
\includegraphics[width=0.45\textwidth]{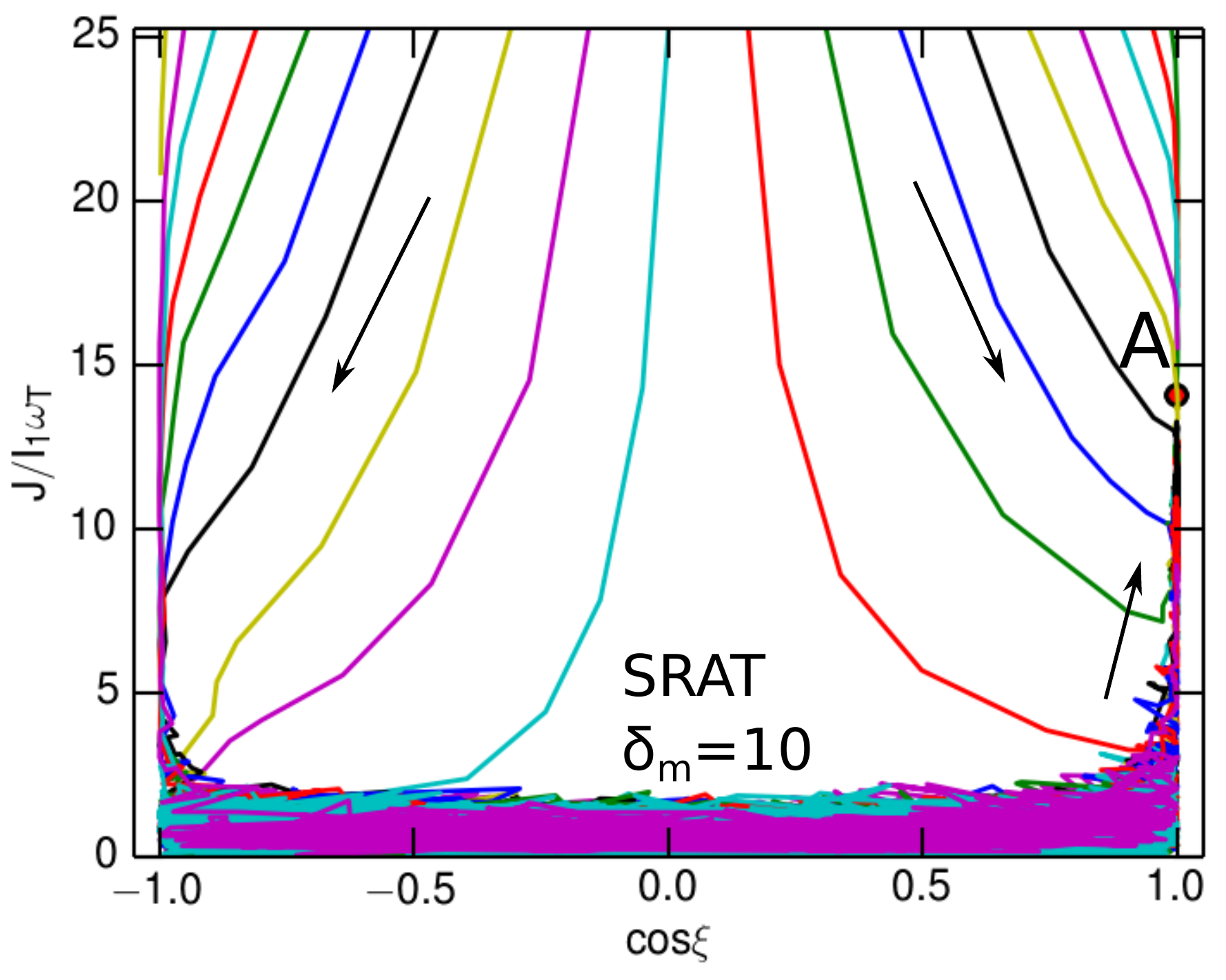}
\includegraphics[width=0.45\textwidth]{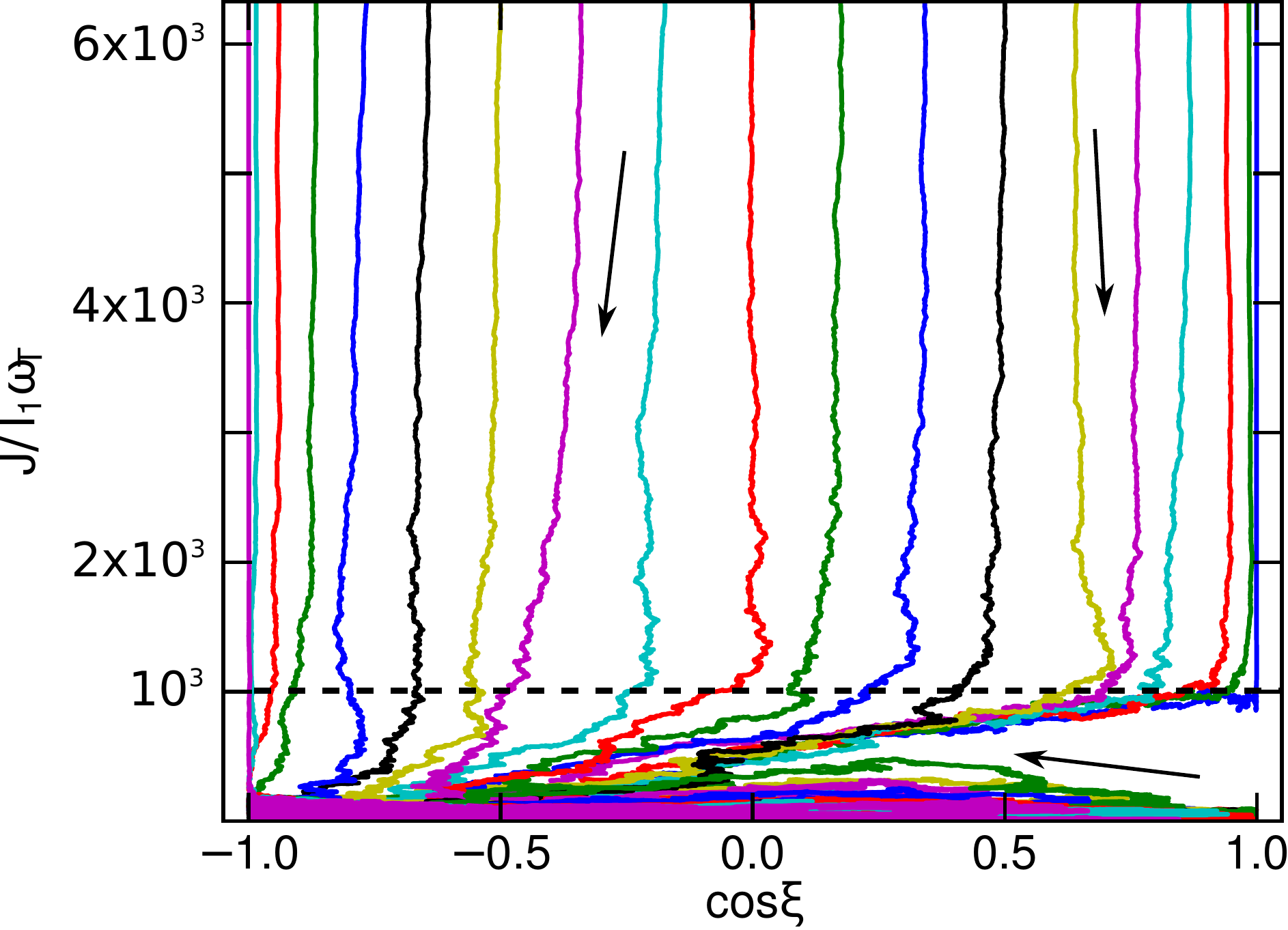}
\caption{Upper panel: Trajectory map of a superparamagnetic grain $a=0.1\mu m$ with high-J attractor point (filled circle). Lower panel: Schematic illustration of the phase map when the radiation intensity is significantly increased. High-J attractor points now becomes repellor point due to the decrease of magnetic relaxation at high frequencies. The dashed line marks the angular velocity above which the fluctuations destroy the high-J attractor point. The typical interstellar medium ($n_{\rm H}=30\cm^{-3}, T_{\rm gas}=100$ K) are considered. Only the randomization by gas collisions is considered.}
\label{fig:phase}
\end{figure}

In Figure Figure \ref{fig:phase} we illustrate the dynamics of a grains with enhanced magnetic response subject to RATs. It is well known from earlier studies (LH07, HL08) that in the presence of both low and high-J attractor points, the majority of the phase trajectories initially end up at the low-J attractor point. The angular momentum at the low-J attractor point is of the order of its thermal value corresponding to the grain temperature. As a result, grains at low-J attractor point are subject to the randomization by gaseous bombardment. This causes grain angular momentum to wander in the parameter space which eventually result in grains getting trapped at the high-J attractor point as shown in Figure \ref{fig:phase} (upper panel). For such a point, the randomization is significantly reduced. Indeed, due to the high value of $J$ the grain is not susceptible to collisional randomization and can stay there essentially forever, provided that the radiation field is sufficiently strong. 

We can show there is a new effect that arises in the situation when the radiation is so strong that it makes the grain rotate at the rate larger than the critical rate at which magnetic response drops (see \S 4). In this situation the additional magnetic relaxation that stabilizes the high-J attractor point disappears and the grain dynamics changes. If the grain is in the range of $q^{max}$ and $\psi$ for which only high-J repellor point is present, the grains move to low-J attractor point as illustrated by the lower panel of Figure \ref{fig:phase}. The dashed line corresponds to the rotation rates at which the enhanced magnetic relaxation fails and the high-J attractor point transfers to a repellor point. As a result the grain alignment {\it decreases} for grains with enhanced magnetic response after the radiation field increases over a particular value. This is a new interesting effect that deserves further studies.

\subsection{Measuring magnetic fields in the vicinity of radiation sources}

The flip of the direction of alignment from being in respect to magnetic field to being in respect to radiation defines the transition point from B-RAT to k-RAT alignment.   By equating Eq. (\ref{rad_p}) and Eq. (\ref{larmor_p}) we get the expression:
\begin{equation}
\chi B = \frac{2\pi I_{\|} g \mu_B}{\xi V \hbar t_{rad, p}},\label{chiB}
\end{equation}
which shows that the measurements of the change of the direction of alignment delivers the product of the grain magnetic susceptibility and the magnetic field strength. In many cases the quoted in the literature uncertainty of $\chi$ are significantly larger than those of $B$.  

The most natural way of determining the $\chi B$ is by observing the distance at which the alignment changes in the vicinity of a star from being in respect to magnetic field to being in respect to radiation. The intensity of the stellar flux changes as $1/R^2$. Therefore radiation precession time in  Eq. (\ref{chiB}) increases as $R^2$. If the star luminosity and the distance at which the alignment change occurs are know Eq. (\ref{chiB}) provides the required $\chi B$ product.

We note that Eq. (\ref{chiB}) provides the estimate that does not depend on the value of anomalous randomization. The obtained values of $\chi B$ put observational constrains on the efficiency of the anomalous randomization.  

As we discuss in \S 9.3 the changes of the direction of alignment can also take place in the presence of a light flash related to a nova flash or a supernovae explosion (\citealt{2017ApJ...836...13H}). We expect to see the change of alignment from being in respect to magnetic field to getting in respect to the radiation. The fast RAT alignment which happens on the time scale much less than $\tau_{\rm gas}$ is described in LH07. Potentially observations of such realignment in our galaxy and other galaxies can provide a measurement of the $\chi B$ for a significant part of a galaxy. For such measurements one should provide calculations of $t_{rad, p}$ as a function of time and compare those with $\tau_L$. A detailed modeling of this process goes beyond our present paper, however. 

\subsection{Differential alignment of carbonaceous and silicate grains}

Silicate and carbonaceous grains are known to show the difference in the polarization that they produce.  The difference in the Larmor precession rate of the two species is the most likely reason for this. Indeed, in the absence of anomalous randomization,  the RAT-induced precession is most likely to dominate for carbonaceous grains as discussed earlier in \cite{LAH15}. Therefore in most of the ISM the alignment of the carbonaceous grains will happen in respect to radiation, while silicate grains in respect to magnetic field. This effect of k-RATs should be measurable. If the anomalous randomization is present, then carbonaceous grains can be aligned only very close to the stars. Measuring this distance one can observationally evaluate the efficiency of  this type of randomization.

As we mentioned above, the existence of the separate carbonaceous and silicate grain in diffuse media populations can be due to centrifugal destruction of fast rotating composite grains subject to RATs (see Hoang 2019). We expect these composite grains to mostly survive in molecular clouds, which is one of the consequence of the decrease of radiation field there. The laboratory modeling of the mechanical properties of the composite grains would be useful to address this issue.

\section{Observational probing of magnetic susceptibilities}\label{sec:obs}

\subsection{Direction of alignment and rate of RAT-alignment}

As we discussed earlier, the interpretation of the observational results on the assumption that alignment must necessarily happen in respect to magnetic field is wrong in the vicinity of sources of strong radiation. Another fallacy is that the alignment must necessarily take at least one gaseous damping time. In fact, LH07 discussed the so-called fast RAT alignment happens on the time scale comparable with the RAT-induced precession time\footnote{The longer estimate of the fast alignment in \cite{2017ApJ...839...56T} resulted from assuming that grains are already rotating with $\omega\gg \omega_{thermal}$.} (see \S 3.2). Our present paper elaborates the relevant processes which let us provide a more reliable interpretation of the polarimetric observations. 

\subsection{Analysis of the ALMA polarimetry}

 To date, observational evidence for alignment in the vicinity of stars is increasingly available. The ALMA observations for HL Tau disk by \cite{Kataoka:2017fq}  show the azimuthal polarization from dust. The authors interpreted this as the evidence of the dust scattering. However, a similar pattern is expected for the RAT alignment in respect to radiation, i.e. for the k-RAT alignment.  The interpretation that at least part of the polarization arises from the aligned grains is also supported  by ALMA observations at 875$\mu$m and 1.3 mm by \cite{Stephens:2017ik} where the transition of the polarization pattern, from being along short disk axis at 870$\mu$m to azimuthal pattern at 3.1 mm.

Very recently, \cite{Alves:2018vl} found the similar azimuthal polarization patters at three different wavelengths and claimed to detect the  magnetic field in disks. However, this azimuthal pattern is expected from the k-RAT alignment for  which the polarization pattern is also perpendicular to the radiation direction and is independent of wavelengths. \cite{2018ApJ...854...56L} presents ALMA observations for an edge-on protoplanetary disk and found a complex polarization vectors, which include the polarization vectors parallel to the short axis and azimuthal polarization pattern in the far-sided region and is explained by poloidal field and the  RAT alignment. \cite{Sadavoy:2018vj} observed azimuthal polarization in the outer rings around the inner disks, which is explained by magnetic alignment along the poloidal field, but the radiation alignment is not ruled out. ALMA observations at 1.15 mm of a massive protostar HH 80-81 by \cite{2018ApJ...856L..27G} show a sharp change in the polarization vectors from radial directions in the inner region to azimuthal pattern in the outer region, and the transition occurs at $\sim$ 156 AU. In a massive protostar, hot inner regions can produce strong radiation of long wavelengths which can align large grains along the radiation direction in the outer region when gas collisions are reduced significantly, producing azimuthal polarization patterns.

While the available data is not conclusive and can have different interpretations, below we show how the measurements of the change of the alignment pattern in some disks can provide us with  the constraints on the grain magnetic susceptibility.  
From radiative transfer calculations in \cite{2017ApJ...839...56T}, one obtains the radiation density at distance of 150 AU in the disk plane is $u_{\rm rad}=10^{-10} \erg~ \cm^{-3}$ and the mean wavelength of radiation is $\bar{\lambda}\sim 100\mu$m. Plugging $u_{\rm rad}$ and $\bar{\lambda}$ into the precession time $\tau_{rad,p}$, one obtains the magnetic susceptibility normalized over the standard value (see \citealt{2014MNRAS.438..680H})
\bea
\frac{\chi(0)_{\rm max}}{10^{-4}}&\simeq& 800\left(\frac{a}{100\mu m}\right)^{3/2}\left(\frac{10 \rm mG}{B}\right)\nonumber\\
&&\times \left(\frac{\bar{\lambda}}{100\mu m}\right)\left(\frac{u_{\rm rad}}{10^{-10}\erg \cm^{-3}}\right) 
\label{response}
\ena
where the typical magnetic field strength $B\sim 10 m$G is taken. The particular estimate is motivated by the interpretation of data in \cite{Kataoka:2017fq} in terms of the RAT alignment. This reveals that the maximum magnetic susceptibility should be $\sim 10^{3}$ times larger than ordinary paramagnetic material, assuming that the grain size is $a\sim 100\mu$m. Comparing with our earlier discussion of the superparamagnetic effects we conclude that the magnetic response suggested by  Eq. (\ref{response}) requires inclusions that are much larger than those responsible for the super paramagnetic behavior for rapidly rotating astrophysical dust grains. These larger magnetic inclusions, however, can induce enhanced magnetic relaxation due to the spin-lattice relaxation as we discussed in Appendix A. 

Our analysis above indicates that the grains have large magnetic inclusions that induce enhanced magnetic relaxation enabling the MRAT alignment. Much smaller inclusions are efficient for producing the MRAT alignment, but via super paramagnetic response. Whether the size of inclusions in large grains in the disk and in the smaller in the ISM grains is an important question. Assuming that the disk dust grains have a power-law size distribution of magnetic inclusions $\sim l^{\alpha}$, $\alpha<1$ it is easy to show (see \citealt{1986ApJ...308..281M}) that, if large grains arise from the coagulation of smaller grains, the probability of grains to have larger size magnetic inclusions is increasing with the grain size. As a result, the majority of "classical" $\sim 10^{-5}$ cm  interstellar grains can have super paramagnetic or even paramagnetic response, while circumstellar disk grains may have larger magnetic inclusions that induce spin-lattice relaxation.   

The grain size in disks is uncertain. Theoretical coagulation model predicts $a\sim 10\mu$m for the density of $n\sim 10^{10}\cm^{-3}$. Assuming that self-scattering is dominant for ALMA polarization at mm wavelength, then, the grain size must be $a_{\rm max}\sim \lambda/2\pi\sim 150(\lambda/1\rm mm) ~ \mu$m. This size is considered upper limit for grain growth if the polarization from aligned grains is ignored. In reality, the grain size could be smaller. 

Observations also reveal the increase of grain size with decreasing the disk radius (\citealt{2015ApJ...813...41P}). If self-scattering is a dominant mechanism, then, one expects the variation of the polarization pattern with the stellar radius. It is not the case of HL Tau. \cite{Ohashi:2018th} suggested that  the polarization from southern par of HD 142527 measured by ALMA arises from "small" grains of $100\mu m$ that are aligned with the magnetic field, i.e. the grains aligned by B-RATs. The authors concluded that the polarization in the northern part 
is due to scattering of radiation by larger grains. We believe that the observed polarization pattern can be explained by the grains aligned with the radiation direction, i.e., by k-RATs. If this is the case, the HD142527 present an interesting case for future detailed modeling in order to test the idea of the alignment direction switch. Such detailed modeling that constrain the magnetic properties of grains is beyond the scope of the present paper. We note that the polarization pattern in the northern part of HD 142527 is consistent with the synthetic observations in \cite{2017MNRAS.464L..61B} that employed results of MHD simulations of circumstellar accretion disks with the toroidal magnetic field.  These simulations intended to reproduce the actual ALMA observations assumed a perfect grain alignment with grains aligned with long axes perpendicular to magnetic field. As we discussed in the paper above, this is not the only possible outcome for the grain alignment in accretion disks. 


More polarimetry observations are required in order to clarify the relative importance of B- and k-RAT alignment as compared to polarization arising from scattering. Our discussion above has a preliminary character and it alerts the community to the striking possibilities in terms of obtaining the information about grain magnetic properties via high resolution studies of polarization pattern in accretion disks. It is important that the RAT theory provides the predictions on the distance for the transition from the k-RAT to B-RAT alignment that depends on the grain size. Thus it is possible to test the theory by measuring the polarization at different wavelengths. 

A recent study performed using CanariCam at 10.4m Gran Telescopio Canarias mapped out a disc at mid-infrared polarization  (see \cite{2016ApJ...832...18L}). It proves that ALMA is not the only instrument for disk magnetic field exploration. The study shows indicates the B-alignment of grains at distances larger than 80 AU from the star.  This, combined with modeling of the radiation transfer, can also be used for constraining the magnetic properties of grains.

\subsection{Measuring the polarization in the vicinity of novae and supernovae}

Testing of the dust grain magnetic properties can be done if the radiation source is changing its intensity. This can induce two effects: (1) change of the polarization arising from the change from the k-RAT to B-RAT alignment with the increase of intensity and the opposite transition for the decrease of intensity; (2) alignment of smaller size grains as the intensity of radiation increases. The latter is the direct consequence of the RAT theory (see LH07). 

The second effect, i.e. the time-dependent change of the small grain alignment as a result of intense irradiation arising from supernovae, was discussed in \cite{2017ApJ...836...13H}. There it was shown that the polarization is changing as grains of smaller size get aligned as a result of the time-dependent flux of intensive radiation. The characteristic time of such alignment is roughly the precession time of the grain in the radiation field. This corresponds to the time of "fast alignment" that was introduced in LH07.\footnote{In LH07 the problem of alignment of grains already rotating suprathermally was considered. The initial suprathermal rate was $60\omega_{th}$. Thus the obtained rates of fast alignment were $\approx 60 t_k$, where $t_k$ is the precession time for a thermally rotating grain. For a thermally rotating grain the "fast alignment" time is of the order of $t_k$.}  

We note, that apart from aligning the small grains the intensive radiation from supernovae torn apart larger grains due to centrifugal stress that fast rotation entails (\citealt{Hoang:2019}). The destruction of large grains will happen in two stages. First, approximately half of grains will be destroyed as their angular velocity will be increasing, while for half of the grains it will be decreasing (see the phase trajectories in Figure 28 of LH07). This is true both for initially aligned, e.g. silicate, and initially not aligned, e.g. carbonaceous grains. However, the strength of the spinning up RATs can depend on the initial direction of the alignment of grains and the radiation (see Figure 8). As a result, the grain centrifugal destruction will depend on the angle between the direction toward the supernovae and the direction of the magnetic field. This is an interesting new effect that can be observationally tested. For instance, for magnetic field and radiation direction being perpendicular to each other we expect the minimal centrifugal destruction. At this stage we also expect to see the polarization from the aligned carbonaceous grains. Staying at the low attractor point, the aligned grains will not be torn apart due to their fast rotation. However, due to the randomization processes, mostly from IR radiation of grains will move the grains out of the low-J attractor points (see HL08) and get significant rotational rate. In this case, the large grains can be destroyed. The time of the second stage is determined by the time scale of grain randomization through IR emissivity which varies with the grain size. It also varies on the ratio of the q-factor of grains (see \S 8.1).\footnote{Grains with q-factor larger than approximately 2 (see Figure 24 in LH07) have high-J attractor point and will be preferentially destroyed. Numerical experiments with arbitrary distributions of grain shapes (\citealt{Herranen:2019kj}) are suggesting that it is the minority of grains that have q-factor larger than 2.}   
 
In this paper we augment the arguments in \cite{2017ApJ...836...13H} by discussing the change of the grain alignment orientation that arises from the change of from B-RAT to k-RAT alignment. Indeed, while generically paramagnetic grains are expected in the ISM to exhibit the B-RAT alignment,  a strong increase of the radiation field is expected to change this. 

The interstellar grain alignment involves the grains larger than $\sim 10^{-5}$ cm. In the presence of the increase of the radiation field due to the supernovae explosion the alignment of smaller grains takes at the regions illuminated by the supernovae flush. The corresponding alignment of smaller grains happens in a limited illuminated region and therefore the corresponding polarization can be distinguished from that arising from the other line of sight material. This allowed \cite{2017ApJ...836...13H} to evaluate the distance to the regions where the supernovae-induced alignment takes place and the time over which the grains were exposed to the radiation. The latter times were found sufficiently long to cause the "fast" RAT alignment that we mentioned above (see also LH07).

Polarimetric observations of four SNe Ia by \cite{Patat:2015bb} do not show the variability of the polarization vectors. In particular, they found that the polarization vectors are closely aligned with the local spiral structures which are also the large scale interstellar magnetic fields. This suggests the grain alignment along the magnetic field instead of along the radiation direction. \cite{2017ApJ...836...13H} show that small grains of $a\sim 0.01\mu m$ can be aligned by UV-optical radiation of the average wavelength $\bar{\lambda}\sim 0.35\mu m$. Thus, using Equation (\ref{chiB}) one can estimate the magnetic susceptibility as
\bea
\frac{\chi(0)_{\rm max}}{10^{-4}}&\simeq& 5.6\left(\frac{a}{0.01\mum}\right)^{3/2}\left(\frac{5 \mu G}{B}\right)\nonumber\\
&&\times \left(\frac{\bar{\lambda}}{0.35\mum}\right)\left(\frac{u_{\rm rad}}{10^{-7}\erg \cm^{-3}}\right),
\label{response_SN}
\ena
which implies a factor 5 enhancement of the magnetic susceptibility from the ordinary material. This enhancement provides the lower limit on grain $\chi(0)$ and it is much smaller than it is required to explain the magnetic alignment of grains in accretion disks given by Eq. (\ref{response}). Note, however, that the smaller grains have lower probability of having larger magnetic inclusions (\citealt{1986ApJ...308..281M}). 

The alternative explanation of magnetic field being 5 times larger brings a worrisome suggestion of $B\approx 25$ $\mu$G. While not impossible, this value of magnetic field seems excessively large. Probing the same region with Zeeman measurement would be most advantageous. Other types of measurements, e.g. using Chandrasekhar-Fermi technique can be also helpful in clarifying the issue.

Assuming that it is the magnetic susceptibility that is responsible for the B-RAT alignment enhancement, we conclude that the analysis in \cite{2017ApJ...836...13H}  presents the first evidence of the enhanced magnetic susceptibility response in the ISM. Eq. (\ref{response_SN}) provides a lower limit of the magnetic susceptibilities. At the same time, according to HL16, an increase of magnetic susceptibilities by a factor of 10 is sufficient to provide the perfect alignment of interstellar grains via the MRAT mechanism. This is suggestive of the universality of the MRAT alignment process in the ISM. Other time-dependent variations of radiation field, e.g., those associated with novae and Gamma-ray burst afterglows (Hoang and Giang, to be submitted) can be also used to study grain magnetic susceptibilities. 

One important feature of the RAT alignment is the spin-up time (\citealt{2017ApJ...836...13H}):
\bea
\tau_{\rm spin-up}\simeq 0.5 \left(\frac{d_{pc}^{2}}{L_{8}e^{-\tau}} \right)\left(\frac{\bar{\lambda}}{1.2\mu m}\right)^{1.7}a_{-5}^{-2.2} {~\rm day},
\ena
which is the same order of magnitudes as the radiative precession time (Eq. \ref{rad_p}).
This RAT alignment time is another important parameter that should be considered for the measurements of the variations of polarization from the material surrounding a supernovae. From \cite{2017ApJ...836...13H} it is shown that for a cloud at distance $d\sim 1-10$ pc (i.e., well beyond the sublimation radius, see \citealt{Hoang:2015bn}), the $a\sim 0.03\mum$ grains can be radiatively aligned on $t_{\rm spin-up}\sim 0.3-30$ days for $L_{\rm SN}=10^{8}L_{\odot}$. 

It is noted that polarimetric observations of SN 2008fp were performed on -2, 3, 9 and 31 days relative to the maximum epoch \citep{Cox:2014fq}. For SN 2014J, the observations were carried out on three epochs (Jan 28, Feb 3, Mar 8, 2014) with the maximum brightness in the first week of February (\citealt{Patat:2015bb}). As a result, RAT alignment by the supernova radiation requires the cloud be within 10 pc such that $t_{\rm obs}>\tau_{\rm spin-up}$. In contrast, if clouds are within 1 pc from the supernova, the alignment can achieve its maximum level before the maximum brightness epoch as shown in \cite{Giang:2019tb}.

Talking about the time-dependent polarization from the sources with changing intensity of radiation one should not necessarily assume that it is only the alignment of grains that is changing. We mentioned earlier that the RATs can spin up grains so fast that the grains can be destroyed by centrifugal forces. This can also decrease the polarization of the starlight as soon as rotational disruption occurs. For instance, in \cite{Giang:2019tb} it was shown that the degree of polarization can drop as large grains that are affected by RATs are destroyed, which occurs during the stage following the enhanced polarization by increased alignment by RATs. 

\subsection{Measuring circular polarization}

Circular polarization is most frequently discuss in terms of the multiple scattering of the polarized radiation. However, if grain aligned it can arise via a single scattering of unpolarized light. In fact, the circular polarization arising from single scattering by aligned grains can be another tool that can pinpoint to the place where the alignment direction changes. 

The physics of producing circular polarization via scattering by an aligned grains is rather straightforward. An electromagnetic wave interacting with a single grain coherently excites dipoles parallel and perpendicular to the grain's long axis. In the presence of adsorption, these dipoles get phase shift, thus giving rise to circular polarization. This polarization can be observed from an ensemble of grains if these are aligned. The intensity of circularly polarized component of radiation emerging via scattering of
radiation with ${\bf k}$ wavenumber on small ($a\ll \lambda$) spheroidal particles is (\citealt{Staude:1972ud})
\begin{equation}
V( {\bf e}, {\bf e}_0, {\bf e}_1)=\frac{I_0 k^4}{2 r^2}i(\alpha_{\|}
\alpha^{\ast}_{\bot}-\alpha^{\ast}_{\|}\alpha_{\bot})\left([{\bf e_0}\times
{\bf e}_1] {\bf e}\right)({\bf e}_0 {\bf e})\;\;\;,
\label{circular1}
\end{equation}
where ${\bf e}_0$ and ${\bf e}_1$ are the unit vectors in the directions of incident and scattered radiation, ${\bf e}$ is the direction along the aligned axes of spheroids; $\alpha_{\bot}$ and $\alpha_{\|}$ are the particle polarizabilities along ${\bf e}$ and perpendicular to it.

The intensity of the circularly polarized radiation scattered in the volume
$\Delta \Gamma({\bf d}, {\bf r})$ at $|{\bf d}|$ from the star at a distance
$|{\bf r}|$ from the observer is (\citealt{1976Ap&SS..43..291D})
\begin{equation}
\Delta V ({\bf d}, {\bf r})=\frac{L_{\star} n_{\rm dust}\sigma_{V}}{6\pi |{\bf d}|^4
|{\bf r}||{\bf d}-{\bf r}|^2}R \left([{\bf d}\times {\bf r}] h\right)
({\bf d}{\bf r})\Delta \Gamma({\bf d}, {\bf r})~~~,
\label{circular}
\end{equation}
where $L_{\star}$ is the stellar luminosity, $n_{\rm dust}$ is the number of
dust grains per a unit volume, and $\sigma_V$ is the cross section for
producing circular polarization, which for small grains is
$\sigma_V=i/(2k^4)(\alpha_{\|}\alpha^{\ast}_{\bot}-\alpha^{\ast}_{\|}\alpha_{\bot})$.
The circular polarization arising from
single scattering on aligned grains can be high, e.g., $10\%$.

Circular polarization has been discussed in the literature studying magnetic field (\citealt{2007JQSRT.106..225L}; \citealt{2014MNRAS.438..680H}). In particular, circular polarization reveals the alignment of the dust in comet coma as well as the Zodiacal dust. Incidentally, the fluctuations of circular polarization can be used to study temporal variations of magnetic field arising from turbulence. This was suggested as a cost effective way to study interplanetary turbulence studying the variations of circular polarization arising from the Zodiacal dust (see \citealt{2007JQSRT.106..225L})  

In the present paper we focus on another useful property of circular polarization, namely, it cannot arise if the direction of alignment and the direction of the scattered radiation coincide.  This property is obvious from the analysis of Eq. (\ref{circular1}). 
Thus if the grains are aligned by the k-RATs of the radiation star, the circular polarization of the scattered light is zero. For grains that are aligned in respect to magnetic field which symmetry deviates from one of radiation, the circular polarization is expected. 
Therefore the circular polarization from circumstellar regions is indicative of the B-RAT alignment of dust. Moreover, measuring the distance at which circular polarization emerges one can determine the transition from k-RAT to B-RAT alignment. Such studies can be complementary to the ALMA observations of linear polarized light that we discussed earlier. 

The time dependent changes of in circular polarization is also expected in the presence of time dependent change of radiation intensity which takes place near time variable sources, in particular in the vicinity of novae and supernovae.  The scattering on aligned grains is expected to induce circular polarization, provided that the grains aligned in respect to B-field. As in higher radiation field grains get aligned by the k-RAT mechanism, the circular polarization is expected to decrease. The B-alignment of small grains in the high intensity radiation is expected to induce circular polarization from those grains. Both effects can be used to study magnetic susceptibilities of grains.

\subsection{Alignment of grains with enhanced magnetic response in cometary coma}
Grains with enhanced magnetic response, e.g. superparamangetic grains, get their high-J attractor stabilized and therefore demonstrate higher degree of polarization (\citealt{Lazarian:2008fw}; HL16). Within our present study it is important that the enhanced magnetic susceptibility increases the magnetic moment of the rotating grain. As a result, the rate of Larmor precession increases. This allows grain align in respect to magnetic field even in the presence of the strong nearby sources of radiation. For instance, in \cite{Kolokolova:2016cd} the circular polarization from comet 67P/Churyumov-Gerasimenko was reported and it was claimed that the origin of it is the single scattering from aligned grains.

From Equation (\ref{circular}) it is clear that if the alignment happens in respect to the radiation, the circular polarization is not expected to arise. Therefore in \cite{2007JQSRT.106..225L} and \cite{2014MNRAS.438..680H} it was suggested that the axis of alignment is determined by the electric field in the comet atmosphere. However, \cite{Kolokolova:2016cd} provided the evidence of the magnetic field presence in the region and suggested that the alignment should be happen in respect to magnetic field. 

As shown in \cite{Kolokolova:2016cd}, the magnetic field strength in the coma is $B\sim 800\mu$G. Using the typical parameters of the coma for Equation (\ref{chiB}), one obtains the grain susceptibility required for the alignment with the magnetic field as 
\bea
\frac{\chi(0)}{10^{-4}}&\gtrsim& 718\left(\frac{a}{0.1\mum}\right)^{3/2}\left(\frac{800\mu G}{B}\right)\nonumber\\
&&\times \left(\frac{\bar{\lambda}}{0.5\mum}\right)\left(\frac{d}{1\rm AU}\right)^{2},
\ena
where $d$ is the heliodistance of the comet. This is fulfilled with the high level of iron inclusions of $N_{\rm cl}\sim 1000$. This is consistent with the properties of grains discussed by \cite{1995ApJ...455L.181G}. 

\section{Discussion}\label{sec:discuss}

Grain alignment is an astrophysical problem of a very long standing (see \citealt{2003JQSRT..79..881L}).  While the understanding of grain dynamics has been significantly advanced recently there are still outstanding issues. Below we overview the current state of the 
field and point out to a few existing uncertainties.

\subsection{Grain rotation rate} 

The difference between the Larmor precession and that induced by RATs is related to the dependence on the rate of grain rotation. Unlike the Larmor precession, the faster grain rotates, the smaller is the RAT-induced precession rate. This provides a way to distinguish between fast and slow rotating grains. In terms of our defining the grain magnetic susceptibility, this is a limiting factor as grain rotation can vary significantly, e.g., by a factor of 100. 

The rotation rate is different for grains aligned at high-J and low-J attractor point (LH07). Ordinary paramagnetic grains get aligned with high or low angular momentum depending on the several factors which include the grain shape and the direction of radiation. For instance, for grain shapes which alignment was studied in LH07 and  more recently in \cite{Herranen:2019kj}  for typical ISM radiation field have the high-J attractor points for a limited range of angles (in the vicinity of 90 degrees) between the direction of radiation and magnetic field. In terms of grain rotation at high-J attractor points, this is a serious limitation. However, it is important to keep in mind that the grains which have only low-J attractor point are not staying at this point all the time. As we discussed in the paper, in the presence of gaseous bombardment the phase trajectories of such grains are wandering in the vicinity of high-J point, which in this case a repeller point \cite{HoangLazarian:2008} and \cite{2016ApJ...831..159H}.  In terms of the statistical averages that are important for our study, this reduces the difference of velocities of grains with high and low-J attractor points. For a typical grain of $0.1\mu m$ in the ISM, the difference in $J$ at high-J and low-J can be around $\sim $30 (\citealt{2009ApJ...695.1457H}). Correspondingly, the uncertainties in magnetic susceptibilities that arise from the difference in the rate of rotation of grains at low and high-J attractor points decreases. 

In the presence of magnetic inclusions most grains get into the high attractor points (see \citealt{Lazarian:2008fw}; \citealt{HoangLazarian:2008}; \citealt{2016ApJ...831..159H}) and this results in all grains rotating at the maximal rate. In view the transition between $k-$ and $B-$ alignments the increase of grain rotation makes $B-$ alignment preferable. Some uncertainty arises if grains rotate at high-J too fast and this decreases the magnetic dissipation (see Figure \ref{fig:kappa_omega}). However, this uncertainty is not expected to affect the grain alignment radically as the grains will stay then in the "critical" state at the boundary of losing their high-J attractor. 

\subsection{Grain acceleration and anomalous randomization}

The anomalous randomization of grains that we have briefly discussed  presents an interesting problem. Apart from a number of physical uncertainties, it also depends on the grain velocities. The calculations of grain acceleration are expected to vary with the level of turbulence and grain size and charge. However, the quantitative calculations are  based on a number of assumptions, which are not easy to test at present. For instance, fast modes are found to dominate the acceleration (see \citealt{Hoang:2012cx}). Currently, it is difficult to determine observationally what percentage of energy is in different modes, although the first promising steps in this direction have been done in  (\citealt{2012ApJ...747....5L}; \citealt{2016MNRAS.461.1227K}; \citealt{Chepurnov:2019}).

In addition, the efficiency of the anomalous randomization depends on the grain rotational state. Grains at high-J attractor points are less subjected to randomization. It was shown in \cite{2009ApJ...695.1457H} that the low-J attractor points can be also modified in the presence of uncompensated torques. According to \cite{Purcell:1979} these torques can arise due to the variation of gas atom accommodation coefficient, emission of electrons and, importantly, formation of H$_2$ molecules at a catalytic sites on the grain. The efficiency of these processes requires further studies.   

The observed polarization in diffuse media and molecular clouds suggests that the alignment of silicate grains is an efficient process for  typical ISM conditions. Our calculations indicate that this corresponds to the alignment with the high angular momentum. This naturally happens for dust having enhanced magnetic susceptibilities significantly or/and subject to uncompensated mechanical torques, e.g. Purcell's torques (\cite{Purcell:1979}) or the "helicity-induced" torques arising from the relative motion of grain in respect to the ambient media (\citealt{LazarianHoang:2007b}).  

If we deal with the carbonaceous dust, the enhanced magnetization is not likely. The spin-up by the uncompensated mechanical torques is a possible way of enhancing the alignment of carbonaceous grains in respect to radiation.  

In circumstellar disks, the level of turbulence may be significantly reduced compared to the ISM and, as a result, the anomalous randomization arising from grain re-charging 
is expected to be lower there compared to one in the diffuse ISM. At the same time the anomalous randomization arising from grain flipping (see \citealt{2006ApJ...647..390W}) can still be important. However, the flipping rate of grains
(\citealt{1999ApJ...516L..37L}; \citealt{2009ApJ...695.1457H}; \citealt{2017MNRAS.471.1222K}) require more work to be quantified for realistically irregular grains.

 Additional randomization is possible in the case of the k-RAT alignment. As we discussed above, magnetic thermal fluctuations of the grain material interact with the external magnetic field and induce an additional randomization. This randomization increases as the magnetic susceptibility of grains increases. We discussed in the paper that the alignment of grains with superparamagnetic inclusions in respect to radiation is requires much stronger radiation flux compared to that of paramagnetic grains.  Naturally, superparamagnetic grains that are being aligned via k-RATs are expected to experience the magnetic version of the anomalous randomization that we described in this paper. 

\subsection{Differences in the alignment of carbonaceous and silicate grains}

In the paper we provided arguments why carbonaceous grains should be less efficient in producing polarization compare to the silicate ones. Several processes that disfavor the alignment of carbonaceous grains are at work. First of all, the small magnetic moment favors k-RAT compared to B-RAT alignment. Thus we expect the direction of the alignment to vary in space depending on the distribution of the radiation sources. This decreases the observed polarization. The rate of precession for carbonaceous grains may be comparable to the rate of randomization via traditional, i.e. gaseous bombardment, randomization as well as anomalous randomization. If any of the latter rates are larger than grain precession, the alignment of carbonaceous grains is suppressed (see \citealt{2006ApJ...647..390W}).  

In addition, a less efficient internal relaxation of energy in carbonaceous grains, may make the RAT alignment less deterministic. Indeed, it was shown in \cite{2009ApJ...697.1316H} that grains without internal relaxation can align with their axes both parallel and perpendicular to the alignment axis.\footnote{In \cite{2009ApJ...697.1316H} the discussion was performed in terms of B-alignment, but the same arguments are applicable to the k-alignment.} Appealing to Fig. 24 in LH07, we can suggest that the alignment at the low-J attractor point is the most likely one. For the k-RAT alignment the alignment with long grain axes parallel to the radiation direction is expected. 

If carbonaceous grains are aligned, e.g. in respect to radiation, the degree of alignment of their alignment  is expected to be reduced by the anomalous randomization effects that we discussed. Studies of q-factors for a variety of random grain shapes in \cite{Herranen:2019kj} shows that most probable value of $q$ for the studied grains is around 1. Such grains are expected to get k-RAT alignment with low-J attractor point. If the internal relaxation within grains is large, such grains are expected to align with with long axes perpendicular to the radiation direction with the alignment degree of $\sim 30$\% in the presence of typical ISM gaseous bombardment (see HL08). This degree is expected to decrease in the presence of the anomalous randomization. An additional complication that potentially can take place is that the internal relaxation via inelastic relaxation (see \citealt{LazEfroim:1999}; \citealt{Efroimsky:2002p6086}) can depend on the size and structure, e.g. porosity, of grains. In this situation some carbonaceous grains may be aligned parallel to the radiation flux and some can be aligned perpendicular to this flux. 

The relative motion of grains in respect to gas induces both the alignment of irregular grains and the precession of grain of any shape about the direction of the flow (\citealt{LazarianHoang:2007b}; \citealt{Hoangetal:2018}). Both carbonaceous and silicate grains are subject to this effect. However, the lower degree of Larmor precession for carbonaceous grains makes the alignment in respect to the gaseous flow more probable. Historically, the grain acceleration by radiation was considered (see Purcell 1964). However, MHD turbulence is a more likely cause for the grain acceleration (\citealt{2003ApJ...592L..33Y}; \citealt{Hoang:2012cx}). Fast precession induced by mechanical torques can induce a new type of alignment by radiative torques, i.e. the alignment for which the RATs do the alignment, but the direction of alignment is determined by the direction of the relative velocity of grains in respect to the ambient media. This type of alignment can be termed  the {\it v-RAT} alignment. The direction of grain velocity is expected to be along the magnetic field for grains accelerated by radiation. In this case, the v-RATs and B-RATs work in unison. In the case of MHD turbulence the acceleration direction depends on the modes that dominate the motions as well as on the mechanism of acceleration. The Alfvenic modes induce the relative grain-gas motions mostly perpendicular to magnetic field (see \citealt{Lazarian:1994}; \citealt{2003ApJ...592L..33Y}). As we discussed in \S 3 the precession induced by mechanical torques is expected to be slower than the Larmor precession. In this situation the mechanical torques still induce alignment with long grain axes perpendicular to B. In special situations when we the precession in respect to grain velocity is faster,  carbonaceous grains can be aligned with long axes {\it parallel to the magnetic field}.  If, however, the dust is accelerated by compressible modes via the Transit Time Damping mechanism (\citealt{Hoang:2012cx}; \citealt{Xu:2018um}), grain velocities are parallel to B and the traditional alignment with long grain axes perpendicular to magnetic field dominates. The quantitative description of these processes requires more observational studies of turbulence as well as theoretical studies of grain dynamics.

As we mentioned earlier, we expect composite the composite grains containing carbonaceous and silicate fragments to exist  in molecular clouds. These composite grains are expected to have fast Larmor precession, fast internal and paramagnetic relaxation especially if paramagnetic fragments contain magnetic inclusions.  The carbonaceous material would supply the mass, while the magnetic response would be due to the component with higher magnetic susceptibility. A search of the spectral features arising form the alignment of carbonaceous material can clarify the nature of the grain composition and structure. This could explain the studies (e.g. Giles Novak 2018, private communication, \citealt{2019arXiv190500705S}) that testify in favor of the alignment of the carbonaceous grains in molecular clouds.  As the radiation field in molecular clouds is weaker, the RATs there are less likely to destroy the composite silicate-carbonaceous grains. Future observations should test this hypothesis.  

\subsection{Linear and circular polarization}

For poorly resolved sources the study of circular polarization provides a way to study grain alignment. If the alignment is the outcome of the k-RAT action, one does not expect to see circular polarization as a result of scattering from aligned grains. If, however, the alignment takes place in terms of magnetic field, then the circular polarization may be expected. From the circular polarization pattern it is possible to get insight into the magnetic field structure of the disks. 

Circular polarization can also arise from the multiple scattering from grains. However, the probability of multiple scattering is low at long wavelength. In fact the observations in \cite{2016ApJ...820...54K} are consistent with single scattering supports the model of single scattering for the accretion disks observed by ALMA. Therefore the detection of circular polarization will be the evidence of grains aligned in respect to magnetic field. It is important to note that if a fraction of grains has enhanced magnetization, this fraction will dominate the circular polarization signal, while its contribution may be subdominant in terms of linear polarization. 

In general, not only magnetic field and radiation can define the directions of the alignment axes. As we discuss in \S 10.3, in exceptional cases the grain motion relatively to gas may make the grain velocity direction the alignment axes. It was speculated in Lazarian (2007) that the direction of electric field in the comet coma can also define the alignment axes. Therefore, circular polarization provides a unique way of identifying the change of the direction of the alignment axis. 

\subsection{Radiative and Mechanical torques}

Our study above was focused on grain alignment by RATs. Apart from RATs, mechanical torques of irregular grains can produce both grain precession (LH07) and the alignment (\citealt{LazarianHoang:2007b}; \citealt{2016MNRAS.457.1958D}; \citealt{Hoangetal:2018}). Consider first grain precession. The precession of grains in mechanical flows is similar to that induced by radiation flows. 

We discussed in \S 5.2 that the k-RAT alignment of superparamagnetic grains can be affected by magnetic anomalous randomization. A similar effect is present for the mechanical alignment of irregular grains if mechanical torque rotation rate is larger than the Larmor rotation rate. 

The mechanism of the MT alignment  is similar to the RATs and is based on the helicity. 
However, the difference between the two mechanisms stems from the fact that as an irregular grain exposes its different facets to the flow, the helicity of the grain changes. This is different from the RATs, for which the radiation samples the entire grain. As a result, the functional form of RATs can be well approximated by the Analytical Model (AMO) suggested in LH07. At the same time, the mechanical torques are more chaotic and, in general, do not follow a simple analytical model (see \citealt{Hoangetal:2018}). Due to this variations of grain helicity that is sampled by the flow, the effective helicity of a grain decreases and this decreases the efficiency of the MTs in terms of grain spin up. Therefore MATs are expected to render grains rotational rate that are smaller than the maximal rotational rate that can be obtained on the basis of the AMO. Nevertheless, numerical simulations show that the irregular grains exhibit the averaged net helicity. Therefore grains get aligned due to the relative motion of the dust and gas.

The MT alignment dominates the Gold one (\citealt{1952MNRAS.112..215G}; \citealt{Lazarian:1995p3034}; \citealt{1997ApJ...483..296L}; \citealt{Lazarian:1998p3032}) as well as other types of mechanical alignment (see \citealt{1997ApJ...483..296L}). This is due to the fact that the grain helicity allows the regular increase of angular momentum as opposed to the random walk increase that is present in the Gold and similar stochastic mechanisms of grain alignment. 

The relative role of the alignment of irregular grains by MTs and by the RATs requires further studies. The motion in respect to the gas can be caused by different factors with turbulence being the most likely driver (see \citealt{2002ApJ...566L.105L}; \citealt{Hoang:2012cx}). Note, that, unlike RATs, the MTs do not have a well defined cut-off size for the alignment. This can induce situations that grains for a range of sizes is aligned by the RATs, while smaller and larger grains can be aligned by the MTs. The differences of the RAT and MT alignments of grains of different sizes can be a source of the variations of observed polarization with frequency (see \S 10.4). 

The importance of MTs in molecular clouds can be tested using the effect predicted recently in \cite{HoangTram:2019} and \cite{HoangTung:2019}. There it was shown that the desorption of molecules from dust grains can strongly be increased by grain suprathermal rotation. As a result, the grain chemistry can be an indicator of the grain rotation. By detecting the variations of chemistry not compatible with that expected in the presence of RATs and correlated with the local level of turbulence can provide a way of observationally constrain the importance of MTs.\footnote{Compared to the grain disruption by the centrifugal stress (see \citealt{Hoang:2019}) grain chemistry provides a less ambiguous measure. Indeed, in the regions of higher turbulence and shocks not only MTs but also grain-grain collisions are responsible for the change of the grain sizes.} 

In terms of the present study the MTs can produce additional effects. In the presence of a significant grain drift, the drift direction can become the alignment axis.\footnote{In shocks both the alignment of grains by the MTs and the breaking of grains by centrifugal stress arising from fast MT-induced grain rotation are important. In fact, the latter process may be more efficient compared to the traditionally considered grain shuttering by grain-grain collisions. Therefore observations of shocks in interstellar media can present a laboratory for testing the efficiency of the MTs in terms of both alignment and grain spin-up.} This, similar to the cases discussed in \S 9.3, can result in circular polarization. 

One may expect that the accretion disks present a case where the interplay of MT alignment and B- and k-RAT alignment may be present. Therefore the multifrequency linear and circular polarization observational studies of such regions are essential.

\subsection{Variations of polarization with frequency}

For many astrophysical settings it is essential to determine the magnetic response of the dust. Only if this response is known, it is possible to reliably trace magnetic fields in the star formation regions. 

It is important to mention that magneto-dipole emission from dust (\citealt{2000ApJ...536L..15L}; \citealt{2013ApJ...765..159D}; \citealt{2016ApJ...821...91H}) is potentially an important component of the microwave foregrounds. To determine the importance of this emission one should know the magnetic susceptibility of the dust grains. Note, that the polarization from the magneto-dipole and electric-dipole emissions are orthogonal. Therefore the combination of dust components
with distinctly different magnetic susceptibilities is expected to result in non-trivial dependence of the polarization degree and direction on frequency. The search of the effect of magneto-dipole emission can be complicated by the effect of flipping polarization direction when the wavelength is comparable with the grain size as discussed, e.g., in  \cite{2019MNRAS.tmp.1705K}. Therefore more detailed studies of the variations of the polarization with the wavelength are necessary to distinguish the two effects. 

Incidentally, there are other factors that induce the variations of polarization. For instance, if carbonaceous grains are aligned by k-RAT alignment while silicate ones are aligned by B-RAT, this  can also result in the frequency-dependent polarization. Locally, in the presence of sources of radiation, the direction of alignment can change from B- to k-alignment for both species of grains. We discussed in \S 10.3 that in the presence of mechanical uncompensated torques acting on irregular grains the direction of polarization can change for grains of different sizes. In addition, the variations of polarization with frequency can arise if the the anisotropy of external radiation is different at different frequencies. Indeed, as the grains of different sizes $a$ experience RATs from the parts of the spectrum with $\lambda \sim a$ the grains of different sizes may have both different temperature and different degree of alignment.

In addition, it was suggested by Mathis (1986) that the magnetic properties of grains of different sizes are different as larger grains are more likely to have magnetic inclusions. With grains of different sizes having different temperature this already induces a non-trivial spectrum of polarized radiation both in terms of electro-dipole emission and magneto-dipole emission. On the top of this, grains with and without inclusions can be differentially aligned through the B-RAT and k-RAT mechanism, making the interpretation of the polarization measured at a single frequency even more complicated. 

In view of the existence of all these different ways that grain emission and grain emission and grain alignment physics can affect both degree and the direction of polarization at different frequencies the multi-frequency observations of polarization are essential. Unlike synchrotron polarization measurements, a single frequency measurement of polarization may  sometimes be misleading in terms of identifying the direction of magnetic field. Moreover, the removal of the CMB polarization foregrounds in order to search for enigmatic cosmological B-modes do require a very detailed modeling of the variations of grain polarization with the wavelength. 

\subsection{Importance of polarization from dust: synergy with other techniques}

For years polarization from dust has been used as a major source of exploring magnetic fields in diffuse media and molecular clouds. The assumption that the direction of polarization necessarily reflects the line-of-sight averaged direction of the underlying magnetic field is frequently made. This, however, is not necessarily true and the present paper is provides clear evidence for this. First of all, the alignment is going to be suppressed for regions of high density and lower illumination of grains,\footnote{Note, that we are talking about mostly by the illumination by an anisotropic part of the radiation flux, the spin-up and the resulting alignment from the isotropic part of the radiation is negligible in most settings.} e.g., in the dark cores of molecular clouds. Second, in the vicinity of radiation sources the alignment can happen not in respect to magnetic field, but in respect of the radiation anisotropy direction. The range at which the transfer from one type to another type of alignment happens depends on grain magnetic susceptibilities, the latter being different for grains with and without magnetic inclusions as well as being different for carbonaceous and silicate grains. This requires caution in interpreting polarization in terms of magnetic fields.

The good news, however, that by now we have a quantitative theory of RAT alignment and thus we can potentially identify the variations of both grain alignment degree and its direction with the physical conditions in the grain environments as well as with the grain composition. The latter dependencies are important as they allow extracting more information from polarimetry that it was thought possible earlier. In other words, the progress in the grain alignment theory provides a big change compared to earlier studies that were done when grain alignment efficiency was not possible to quantify. Nevertheless, as we discussed in this paper, some additional issues, e.g. the influence of mechanical torques on irregular grains and the quantitative theory of anomalous randomization require further studies.

The importance of quantitative understanding of dust polarization has increased recently due to the big challenge of detecting of B-modes originating from gravitational wave background in the early Universe, the B-modes. This requires the increase of the accuracy of accounting of foreground subtraction by, at least, a factor of 100. This is a big challenge taking into account the non-trivial wavelength behavior of dust polarization that is discussed in this paper. It is clear that the simple power-law interpolations that have been successfully used for removing emission contributions arising from most of the major foregrounds\footnote{It is known that unlike synchrotron, thermal dust emission or free-emission, the so-called anomalous microwave emission (AME) is not a power-law (see \citealt{Dickinsonetal:2018}, for a review).} are unlikely to work for removing the polarization from galactic dust. As a result, the progress of this branch of cosmological studies will have to go hand by hand with the progress of understanding grain alignment in galactic environments.

We should also add that in terms of magnetic field studies, the dust polarimetry can be used in synergy with the some more recent techniques. For instance, both the accepted technique of magnetic field based on the Goldreich-Kylafis (GK) effect (\citealt{{Goldreich:1981},{Goldreich:1982}}; \citealt{Crutcher:2008}) and a newer tracing Ground State Alignment (GSA)\footnote{The term is not exact, as \cite{{YanLaz:2006},{YanLaz:2007}} explain that the effect is important for both atoms in the ground state and on metastable levels. However, instead of suggesting a new term, e.g. Ground-Metastable State Alignment (G-MSA), we prefer to use the GSA, as it was used in the recent papers, e.g. \cite{Zhangetal:2019}. In earlier papers (see \citealt{YanLaz:2015} for a review) the same effect was termed "atomic/ionic alignment", but this term has its own deficiencies.} technique (\citealt{{YanLaz:2006},{YanLaz:2007},{YanLaz:2008},{YanLaz:2012}}, see also more recent studies in \citealt{{ZhangYan:2018},{Zhangetal:2019}}), based on the alignment of atoms and ions in ground and metastable states can also provide the direction of magnetic field. The GK effect and the GSA obtained with measuring the polarization of emission or absorption line at a single frequency provide the direction of magnetic field with the 90 degrees ambiguity. Therefore the dust polarization measurements are valuable in removing this ambiguity. At the same time, in view of the potential issues related with the B- and k-alignment, as well as for the polarization arising from dust scattering (see \S 9.1) both GK and GSA provide a way of distinguishing different causes of polarization and therefore remove the ambiguity in interpreting the dust polarization data. We would like to add the GSA polarization at multiple frequencies allows studying the 3D orientation of magnetic field, providing the new dimension for studying magnetic field with polarization observations. 

In fact, recent develpments show that the magnetic field in diffuse ISM and molecular clouds can be also traced without observations of polarization, but using spectroscopic information. A new Velocity Gradient Technique (VGT) use Velocity Centroid Gradients (VCGs) (\citealt{CasLaz:2017}) or Velocity Channel Gradients (VChGs) (\citealt{LazYuen:2018}) to successfully trace magnetic fields. The VGT employes the tested properties of MHD turbulence/turbulent reconnection (see \citealt{GoldSri:1995}; \citealt{LazVish:1999}) to trace magnetic field and it has been successfully proven to map magnetic fields both in diffuse media and molecular clouds (\citealt{YuenLaz:2017}; \citealt{LazYuen:2018}). The VGT does not depend on the intensity of the radiation field and therefore it can map magnetic fields in the situations when the interpretation of the dust polarimetry is ambigous. It can also provide the 3D information on magnetic fields using either galactic rotation curve (\cite{CasLaz:2017}) or a variety of molecular spectral lines (\citealt{Huetal:2019}). In the former case, the testing of the technique ability of producing 3D magnetic field maps was performed using polarization of starlight from stars with known distances. In the presence of graviational collapse, the velocity gradients flip their direction by 90 degrees and, combining this with far infrared polarization observations, this allowed \cite{Huetal:2019} to identify such regions in molecular clouds. 

The dependence of the aforementioned techniques on the ambient density is different from that grain alignment. For instance, the VGT does not depend on the density of the gas, while, on the contrary, GST can be suppressed for high densities, unless the radiation field is sufficiently strong. This provides additional synergy in terms of tomographic studies of galactic 
magnetic fields.

\section{Summary}\label{sec:summary}

In this paper we have explored the consequences of the grain alignment in respect to the magnetic field, i.e., B-RAT alignment, as well as in respect to the radiation flux anisotropy, i.e., k-RAT alignment. We considered the effects of magnetic relaxation and discussed their influence on grain alignment. In our treatment we assumed that in the centrifugal stress produced by RATs can prevent the formation of composite carbonaceous-silicate grains in diffuse media. We consider carbonaceous and silicate grains to exist as separate entities. We claim that in molecular clouds, far from the embedded luminous stars, the centrifugal stress is reduced and carbonaceous and silicate dust particles aggregate composite grains. RATs may not be strong for destroying such grains, but can be still strong enough to align them producing in polarization the signatures of aligned carbonaceous material. Within this model, our results can be summarized as follows:
\begin{itemize}

\item For typical conditions of diffuse media we expect to observe the B-RAT alignment of silicate grains and k-RAT alignment, if any, of carbonaceous grains. We expect the B-RAT and k-RAT alignment of composite grains containing both silicates and carbonaceous fragments in molecular clouds and accretion disks. 
\item For B-RAT alignment, the enhanced magnetic response of grains is important both for increasing the degree of grain alignment as well as mean rotational speed of an ensemble of grains subjected to the RATs. The former is important for magnetic field tracing with polarimetry, the latter is important for rotational disruption of grains by centrifugal force. 

\item Grain magnetization arising from the Barnett effect induces additional effect of spin-lattice relaxation. This effect can be important for the external alignment of angular momentum $\bf J$ of carbonaceous grains in respect to the magnetic field and also in terms of the internal alignment of $\bf J$ in respect to grain axes.

\item The value of the magnetization induced by the Barnett effect within triaxial grains changes as they perform their complex rotation. This induces additional relaxation process that we termed Amplitude Barnett Relaxation (ABR). We identified a range of rotation frequencies in which the ABR dominates the nuclear and Barnett relaxation.

\item Whether the grain alignment happens in respect to magnetic field (B-RATs) or in respect to radiation (k-RATs) depends mostly on the grain magnetic properties, the magnetic field strength, and the radiation intensity. The existing observations of k-RATs dominance at an appreciable distance from the radiation source indicate that grains do not have large magnetic susceptibilities. Both high resolution linear polarization measurements and circular polarization measurements provide ways to determine the transfer from one type alignment to another.  

\item Due to the expected presence of magnetic inclusions, the B-RAT alignment is preferable for silicate grains, but, as we described in this paper, this induces additional B-type anomalous randomization of grains with enhanced magnetic susceptibilities in the case of k-RAT alignment. This potentially makes k-RAT alignment preferable for carbonaceous grains that exhibit weak magnetic response.

\item Our estimates of magnetic susceptibilities for dust grains provide the lower limit of the enhancement of $10$, relative to the ordinary paramagnetic material, for sub-micron sized grains in the supernovae environment and cometary coma where the alignment is found to be along the magnetic field. We also find an upper limit of 1000 for mm-sized grains in protoplanetary disks where there is observational evidence for grains being aligned with the radiation direction. With the better constrained knowledge of the magnetic susceptibilities the transfer from B- to k-alignment provides a new way of measuring the strength of magnetic field. 

\item The alignment of carbonaceous and silicate grains by B-RATs and k-RATs is different, and such difference in the alignment direction of these two types of grains can shed light on the nature of the anomalous randomization of grains that is related to the interaction of grain electric moment with electric field arising from the grain motion in magnetic field. 
\end{itemize} 

 \acknowledgments
We thank the anonymous referee for useful comments. AL acknowledges the NSF grant AST 1715754. T.H. acknowledges the support from the Basic Science Research Program through the National Research Foundation of Korea (NRF), funded by the Ministry of Education (2017R1D1A1B03035359). The authors express their gratitude to B-G Andersson for his stimulating discussions and for his testing their theoretical predictions which proved that the theory of grain alignment is a testable and reliable theory.  
 
 \appendix
 \section{Enhancement of paramagnetic relaxation}\label{apdx:enhanced}
 \subsection{Superparamagnetic and Ferromagnetic Inclusions}
 
 The paramagnetic dissipation in the rotating grains has negligible effect on the RAT alignment. However, the enhancement of the paramagnetic relaxation can increase the alignment efficiency (LH08 and HL16). Therefore below we summarize what is known about the enhancement of the magnetic response for grains with magnetic inclusions. 
 
For paramagnetic materials, i.e. the materials with unpaired electrons in the partially filled electron shells the static magnetic susceptibility corresponding to $\omega=0$ is 
\begin{equation}
\chi (0)=\frac{n_p \mu_B^2}{3kT},
\label{chi_p}
\end{equation}
where $n_p=f_{p}n$ is the density of paramagnetic species, $\mu_B= e\hbar/2m_e c$ is the Bohr magneton, $T$ is the grain temperature. The fraction of paramagnetic atoms/ions in the grain, $f_p$, changes with the grain composition. The interstellar depletion suggests that most Fe atoms are in grains. Assuming a uniform mixing of heavy elements in dust the Fe impurity would produce $f_p$ is at least 0.1 (\citealt{Draine:1996p6977}). The uniform mixing of iron may not be a good approximation, however. One may expect the variations of $f_p$ to happen within the grains. 

The two relaxation processes are applicable to a rotating grain. The one is the spin-spin relaxation that happens as electron spins interact with the magnetic field induced by their neighbors. The latter is 
\begin{equation}
H_i= 3.8 n_p p\mu_B,
\end{equation}
where $p\approx 5.5$ for iron ions.
Therefore the spin-spin coupling can be associated with the spin precession time in the field $H_i$:
\begin{equation}
\tau_{\rm SS}= \frac{\hbar }{g_e \mu_B H_i} \approx \frac{\hbar}{3.8 n_p p \mu_B^2} .
\label{eq:SS}
\end{equation}
To estimate the value of spin-spin relaxation for silicate grains one can assume that iron ions are the major paramagnetic species. Then $\tau_{\rm SS} = 2.9 \times 10^{-12} f_p$ s. 

At frequencies higher than $\tau_{\rm SS}^{-1}$ the paramagnetic relaxation gets suppressed and
the expression 
\begin{equation}
K(\omega)=\frac{\chi (0) \tau_{\rm SS}}{1+(\omega \tau_{\rm SS})^2} \approx  1. 2 \times 10^{-13} \left(\frac{15~{\rm K}}{T}\right) \frac{1}{1+(\omega \tau_{\rm SS})^2}, 
\label{K_par}
\end{equation}
are commonly used (see \citealt{Draine:1996p6977}). The high frequency response of materials is not known, but recent studies by Draine \& Hensley (2012)
are supportive of the form suggested by Eq. (\ref{K_par}).  

For paramagnetic relaxation Eq. (\ref{tau_mag}) $\delta_m<1$. The suggestion that grains can have superparamenetic inclusions in  \cite{Jones:1967p2924} allows to increase the $\delta_m$. Indeed, if iron is clustered into aggregates of $N_{\rm incl}$, the corresponding clusters then behave as if they were single entities having very large magnetic moments. If the volume of the cluster is small enough its magnetic moment flips due to thermal fluctuation and the cluster does not have a spontaneous magnetic moment. This grain with such inclusions exhibit superparamagnetic response. As a result, its magnetic susceptibility is significantly enhanced.  The zero-frequency susceptibility for a grain with superparamagnetic inclusions \citep{Morrish:2001vp},
\begin{equation}
 \chi_{sp} (0) = 1.2 \times 10^{-2} N_{\rm incl} f_{sp} \left(\frac{15~{\rm K}}{T}\right),
 \label{eq:chi_sp}
 \end{equation}
$f_{\rm sp}$ is the fraction of atoms that are within the superparamagnetic clusters, $N_{\rm incl}$ is the number of atoms per cluster  that can vary from 10 to $10^6$ (\citealt{Kneller:1963p6410}). 

We have evidence for the existence of grains with superparamagnetic inclusions. For instance, the clusters discovered in 
interplanetary grains suggests $f_{sp}=0.03$ \citep{{Bradley:1994p6379},{1995ApJ...445L..63M}}, and $N_{\rm incl}$ is expected to have $N_{\rm cl}=10^{3}-10^{5}$ \citep{Jones:1967p2924}. For this sort of inclusions we can expect to have the enhancement of the magnetic susceptibilities by a factor of a thousand. For smaller values of $N_{\rm incl}$, e.g.  $\sim 10^3$, the increase by a factor of 30 is expected. For the relaxation the value of that is $K(\omega)$ given by Eq. (\ref{K_par}) is important. The change of $N_{\rm incl}$ changes also the effective value of $\tau_{\rm SS}$ as we discuss below. 

The inclusions are superparamagnetic as their magnetic moment oscillates due to thermal fluctuations. The characteristic rate is
 \begin{equation}
 \tau_{fluct}^{-1}= \nu_0 \exp\left(\frac{-N_{\rm incl} \aleph}{T}\right) , 
 \label{tau_fluc}
 \end{equation}
  where $\nu_0\approx 10^9$ s$^{-1}$ and $\aleph\approx 0.1$ K for metallic iron inclusions (see \citealt{Morrish:2001vp}). For large inclusions with $N_{\rm incl}\sim 10^5$ and the dust temperature of 10 K, the fluctuating frequency $\tau_{fluct}$ is significantly reduced due to the exponential dependence of the frequency in Eq. (\ref{tau_fluc}). 
  In terms of magnetic relaxation $\tau_{fluct}$  acts as $\tau_{\rm SS}$ for paramagnetic materials (see Eq. (\ref{K_par})). As a result, inclusions larger than $10^5$ cannot provide the super paramagnetic response for rapidly rotating interstellar grains.  The increase of the magnetic relaxation $K(\omega)$ is proportional to $N_{\rm incl}(\tau_{fluct}/\tau_{\rm SS})$ and it can be of the order $3\times 10^4$ if $\sim 3 \%$ of iron atoms are concentrated in inclusions  $N_{\rm incl}=10^3$. This is a significant increase of relaxation, but it requires a particular inclusion sizes. Potentially, to make the magnetically enhanced alignment by RATs, i.e. MRAT alignment, significantly more efficient in the interstellar environments, an increase of the  by a factor of $\sim 10 $ can be sufficient (HL16). Therefore, in spite of the aforementioned limitations of the required size of the super paramagnetic inclusions, we believe that the superparamagnetic inclusions present the best possibility for enabling the MRAT alignment. 
 
 Large inclusions preserve their magnetization. This is related to the single domain inclusions which $\tau_{fluct} \gg \omega$ and larger ferromagnetic/ferrimagnetic inclusions. In the presence of larger inclusions an additional way of enhancing magnetic relaxation was pointed out by \cite{1978ApJ...219L.129D}. He noticed that the constant magnetic field of the ferromagnetic or ferrimagnetic inclusions  induces an additional energy dissipation via spin-lattice relaxation which has $\tau_{\rm SL}\approx 10^{-6}$ s. The latter 
 is significantly larger than the spin-spin relaxation time $\tau_{\rm SS}$. In fact, the suggestion by \cite{1978ApJ...219L.129D} amounts to modifying the magnetic response of the grains to
 \begin{equation}
 K(\omega) = X \frac{\chi (0) \tau_{\rm SL}}{1+(\omega \tau_{\rm SL})^2} +(1-X) \frac{\xi (0) \tau_{\rm SS}}{1+(\omega \tau_{\rm SS})^2},
 \label{K_sl}
 \end{equation}
 where the field $H_{\rm incl}$ induced by the inclusion in the surrounding material entails the change of the coefficient
 \begin{equation}
 X=\frac{H_{\rm incl}^2}{H_{\rm incl}^2 + 0.5 H_i^2},
 \end{equation}
 where $H_i$ is the magnetic field within the body without inclusions.  The estimates of $H_i$ depend on the material adopted. \cite{1978ApJ...219L.129D} provided $H_i\approx 10$ Oe.\footnote{A higher value of $10^3$ Oe was quoted by \cite{Draine:1996p6977} in relation to paramagnetic salts. We believe that the silicate dust grains are different from such salts and use the estimate in \cite{1978ApJ...219L.129D}.}  We will use this as a reference number, but we are aware of its uncertainty. Naturally, for grain rotation frequencies $\omega<\tau_{\rm SL}^{-1}$ a significant increase of magnetic relaxation is expected.

 For a fraction $f_{in}$ of the total number of atoms $n_{tot}$ being in the form of single domain clusters, the rms magnetic field $H_{\rm incl}$ induced by the inclusions within the material inclusions  can be estimated \cite{Draine:1996p6977}
 \begin{equation}
  H_{incl}\approx 3.8 f_{in} n_{tot} p \mu_p/\sqrt{3}=300 (f_{in}/0.03)~~ Oe
 \end{equation}
 where the approximate character of this relation is obvious. Indeed, in the close vicinity of inclusions the magnetic field is expected to be significantly larger, although the volume with larger magnetic field decreases. This correction is not important as soon as $H_{incl}>H_i$.
 
\subsection{Relaxation via nuclear spins}
  
 We note, that nuclear moments of dust also induce magnetic relaxation. The effect is noticed in \cite{Purcell:1969p3641}. Indeed, for low rates $\omega$, $K(\omega) \approx \chi (0) \tau_{\rm SS}$ for the spin-spin relaxation. The paramagnetic susceptibility $\chi (0)$ is proportional to the product of density of species $n$ and the square of their magnetic moment $\mu$ (see Eq. (\ref{chi_p}), i.e. $\chi (0)\sim n \mu^2$. At the same time according to Eq. (\ref{eq:SS}) $\tau_{\rm SS}\sim (n \mu^2)^{-1}$. Therefore the resulting $K(\omega)$ {\it paradoxically} does not depend either on the number of paramagnetic species or their magnetic moments. This is indeed true for $\omega <\tau_{\rm SS}^{-1}$ and for very slowly rotating grains the relaxation arising from nuclear spins is as efficient as the relaxation arising from electron spins. Naturally the relaxation gets suppressed for $\omega>\tau_{\rm SS}^{-1}$. The corresponding spin spin relaxation times of nuclear spins and electron spins is proportional to $n_{n} \mu_{n}/(n \mu)$, which for most cases $\gg 1$  and therefore the velocities for which nuclear relaxation is efficient are significantly smaller that those for electron relaxation.  
 
 In the presence of ferromagnetic inclusions we also expect to have spin-lattice relaxation
 effect that we discuss above to be present for the nuclear spins. As discussed in LD00 the transition of energy from the nuclear spin system happens on the time $\tau_{ne}$ of interaction of nuclear and electron spins. The corresponding time should be used in Eq. (\ref{K_sl}) instead of $\tau_{\rm SL}$. Estimates in LD00 provide $\tau_{ne}$ to be of the order of nuclear spin-spin relaxation time for silicate grains. As a result, we may expect an increase of relaxation by a factor of unity, which does not change the conclusion that the nuclear relaxation is marginally important for increasing $\delta_m$ for silicate grains. At the same time, for large carbonaceous grains that slowly rotate and also have larger $\tau_{ne}$, the corresponding increase may be important.

\section{Grain alignment in the presence of the Barnett magnetization}\label{apdx:alignment}

According to our discussion in the Appendix \ref{apdx:enhanced}, this induces additional spin-lattice relaxation with 
\begin{equation}
K(\omega)= \frac{H_{\rm Barnett}^2}{H_{\rm Barnett}^2 + 0.5 H_i^2} \frac{\chi (0) \tau_{\rm SL}}{1+(\omega \tau_{\rm SL})^2},
\end{equation}
where $\tau_{\rm SL}$ is the nuclear spin-lattice relaxation rate, while $H_i$ is the field produced by the paramagnetic atoms. 
We adopt, as earlier,  $H_i \sim 10$ Oe for paramagnetic materials (see \citealt{1978ApJ...219L.129D}).
For paramagnetic grains the typical spin-spin relaxation time $\tau_{\rm SS} \sim 10^{-10}$ s, while the spin-lattice relaxation times are $\tau_{\rm SL} \sim 10^{-6}$ s. Therefore for thermally rotating, i.e. $\omega \sim 10^{5}$ s$^{-1}$ , silicate grains the
effect of the additional spin-lattice relaxation is negligible. For carbonaceous grains $\tau_{\rm SL}$ can
be $\sim 10^{-6}$ s, while the spin-spin relaxation time $\sim 10^{-9}$ s. If the intrinsic magnetic field decreases to $\sim 10^{-1}$ Oe, the spin-lattice relaxation induced by the Barnett-equivalent field becomes dominant for thermally rotating grains. Its importance increases as $\omega$ increases to $10^6$ s$^{-1}$ and then starts decreasing. The values of the $\tau_{\rm SL}$ and $\tau_{\rm SS}$ for carbonaceous grains are not certain. However, our rough estimate above suggests that this new effect of spin-lattice relaxation in carbonaceous grains should be accounted for the calculations of $\delta_m$. 

It is interesting that the nuclear Barnett-equivalent magnetic field is significantly stronger than the magnetic field induced by nuclear spins. It can be obtained from Eq. (\ref{barn_equiv}) by changing the electron g-factor and the value of the electron magneton to the nuclear g-factor and the nuclear magneton respectively. For protons that are an important component of carbonaceous grains, the expression for the equivalent field is
\begin{equation}
H_{Barnett, nucl}= \frac{\hbar \omega}{g_{n} \mu_{n}},
\label{B_nucl}
\end{equation}
where the nuclear magneton $\mu_{n}\equiv e\hbar/2m_p c$, where proton mass $m_p$ is the much larger than the electron mass $m_e$. Therefore in the idealized case of no electron spins, the Barnett-equivalent field that is more than thousand times stronger than the one arising from electron spins dominate by more than a thousand times the grain material intrinsic magnetic field $H_i$. As a result a slowly rotating hypothetical grain  with no paramagnetic electron can demonstrate the relaxation arising exclusively from nuclear spins
\begin{equation}
K_{n}(\omega)= \frac{\chi_n (0) \tau_{n, SL}}{1+(\omega \tau_{n, SL})^2} + \frac{\chi_n (0) \tau_{n,SS}}{1+(\omega \tau_{n,SS})^2}
\label{Knucl}
\end{equation} 
where the $\chi_n (0)$ is the nuclear zero frequency susceptibility, $\tau_{n, SL}$ is spin-lattice nuclear relaxation time and $\tau_{n,SS}$ is nuclear spin-spin relaxation time. The latter consists of the interaction between nuclear spins of time scale $\tau_{nn}$ and the interaction between the nuclear spin with electron spin of timescale $\tau_{ne}$. The total nuclear spin-spin relaxation rate is $\tau_{n,SS}^{-1}=\tau_{ne}^{-1}+\tau_{nn}^{-1}$.

In a hypothetical setting of no electron spins $\tau_{n, SL}$ may be significantly larger than then the nuclear spin-spin relaxation time:
\begin{equation}
\tau_{nn}=\frac{\hbar}{3.8 g_{n} n_{n} \mu_{n}^2}.
\label{nucl_ss}
\end{equation}

In the presence of electrons, the role of nuclear spin-lattice exchange is played by the exchange of momentum between electron and nuclear spins that take place over the time scale $\tau_{ne}$. Later the momentum is transferred to the lattice via electron spin-lattice exchange, which is much faster than the direct nuclear one. The comparison between the nuclear spin-spin relaxation to the nuclear spin to electron spin relaxation  times is provided in \cite{1999ApJ...520L..67L} which gives 
\begin{equation}
\frac{\tau_{nn}}{\tau_{ne}}=\left(\frac{g_{n}}{g_{e}}\right)\left(\frac{\mu_{N}}{\mu_{n}}\right)^{2}\left(\frac{n_{e}}{n_{n}}\right)\simeq 0.185 \left(\frac{2.7}{g_{n}}\right)\left(\frac{n_e}{n_{n}}\right).
\end{equation}
The ratio above can be $\ll 1$ for carbonaceous grains with $n_e/n_{n}\ll 1$. For silicate grains, $g_{n}\ll 2.7$ (see Table 1 in \citealt{1999ApJ...520L..67L}). Thus, $\tau_{nn}\sim \tau_{ne}$. 

As we discussed earlier, for slow rates of grain rotation the nuclear spin-spin relaxation and the electron spin-spin relaxation of the same order, which means that the second term in Eq. (\ref{Knucl}) is the of the same order as the low frequency limit of the relaxation coefficient for paramagnetic grains given by Eq. (\ref{K_par}). The fact that in such low frequencies the first term in Eq. (\ref{Knucl}) is much larger than the second one means that the new mechanism of magnetic relaxation that we introduced above is orders of magnitude more efficient than the ordinary paramagnetic relaxation. The physical reason for this is much more slower response to external magnetization of the nuclear spins in the Barnett-equivalent field that aligns nuclear spins.\footnote{A similar effect of the enhanced relaxation associated with the nuclear magnetism is the effect of nuclear relaxation of internal energy of wobbling grains introduced in \cite{1999ApJ...520L..67L}). There also the larger value of the Barnett-equivalent magnetic field with the slow response for reorientation of nuclear spins resulted in the $\sim 10^6$ increase of the nuclear relaxation compared to the electron spin based Barnett relaxation introduced by \cite{Purcell:1979}. } In other words, the nuclear relaxation rate can significantly exceed the rate of paramagnetic relaxation if $\omega$ does not exceed much $\tau_{n, e}^{-1}$. The latter according to \cite{1999ApJ...520L..67L} is
\bea
\tau_{ne}&=& \frac{\hbar g_e}{3.8 n_e g_{n}^2 \mu_{n}^2} \nonumber\\
&=& 3\times 10^{-3} \left(\frac{2.7}{g_{n}}\right) \left(\frac{10^{21} {\rm cm}^{-3}}{n_e}\right) s
\label{nucl_electr}
\ena
which means that the effect is very important for large slowly rotating grains. In carbonaceous grains with very few unpaired electrons the Barnett-induced nuclear relaxation can make $\delta_m\gg 1$ for slowly rotating grains. This is an interesting that the decrease of the concentration of unpaired electrons increases the nuclear relaxation through the decrease of the effective spin-lattice rates. If $n_e\sim n_{n}$ the both terms in Eq. (\ref{Knucl}) are of the same order and the enhancement of the relaxation is an effect of order unity. Thus we do not expect any significant increase of $\delta_m$ for silicate grains.

The effect of Barnett magnetization on the relaxation of grain rotational energy is not limited to the inducing additional spin-lattice relaxation. The spin-spin relaxation is also being modified. 
In LD00 the Barnett magnetization was identified as a source of enhanced relaxation that was termed "resonance relaxation" (see also HL16). The boundary between the normal paramagnetic and resonance relaxation is related to the ratio of the internal magnetic field and the Barnett-equivalent magnetic field given by Eq. (\ref{barn_equiv}). If $H_{\rm Barnett}/H_i \gg 1$ the resonance relaxation dominates. The corresponding expression for the resonance relaxation was derived in LD00
\begin{equation}
K_{res} (\omega)=\chi (0) \frac{\tau_{\rm SS}}{1+\gamma^2 g_e^2 \tau_{\rm SS} \tau_{\rm SL} H_{\rm ext}^2 \sin^2 \theta}
\label{K_res}
\end{equation}
where $\gamma\equiv e/2mc = 8.8 \times 10^6$ s$^{-1}$ G$^{-1}$, $H_{\rm ext}$ is the interstellar magnetic field acting upon the grain, $\theta$ is an angle between the direction of ${\bf H}_{\rm ext}$ and $\omega$. The difference between the Eq. (\ref{K_res}) and the classical expression given by Eq. (\ref{K_par}) is that there is no suppression of the relaxation at high frequencies. This is the consequence of the Barnett magnetization that makes electrons resonate with the changing external field. Instead the relaxation in Eq. (\ref{K_res}) depends on the transferred energy $\sim H_{\rm ext}^2$ and the efficiency of energy transfer through spin-spin and spin-lattice interactions, i.e. on characteristic times $\tau_{\rm SS}$ and $\tau_{\rm SL}$. 

For very small fast rotating grains the Barnett field is dominant, but the issue of the possible increase of $\tau_{\rm SL}$ becomes an issue. LD00 calculated that Raman scattering of phonons can make the relaxation still efficient even for ultra small grains responsible for the microwave emission.\footnote{An experimental proof would be very helpful, as this type of analytical calculations are subject to uncertainties.} It is also clear that for carbonaceous grains the resonant relaxation gets more important for lower rotation speeds and therefore is relevant to larger slower rotating grains. For such grains $\tau_{\rm SL}$ takes its ordinary value $\sim 10^{-6}$ s.  

Interestingly enough, the resonant relaxation can take place in the system of nuclear spins for lower $\omega$ compared to the rotation rate required for the electron resonance relaxation. This arises from the Barnett equivalent magnetic field given by Eq. (\ref{B_nucl}) is significantly stronger than its electron counterpart. Therefore $H_{\rm Barnett, nucl}/H_i\gg 1$ is achieved for slower rotational rate. It is important as the paramagnetic relaxation due to nuclear spins effect is usually considered only up to $\omega$ less than the nuclear spin-spin relaxation rate $\tau_{nn}^{-1}$,  the latter being $\sim 3 \times 10^3$ s$^{-1}$ (\citealt{1999ApJ...520L..67L}). The arguments above are suggestive that the paramagnetic relaxation involving nuclear spins can be important for faster rotating grains if the rate of spin-lattice relaxation is fast. Our discussion above suggests that the role of $\tau_{\rm SL}$ for nuclear spins can be played by $\tau_{ne}$ given by Eq. (\ref{nucl_electr}).  However, in the presence of high density of electron spins the intrinsic magnetic field of the sample $H_i$ increases and the Barnett magnetization stops being the dominant. This makes resonance relaxation impossible unless grains rotate very fast. At the same time, if the density of electron spins is low, the $\tau_{ne}$ becomes long suppressing the effect of resonance nuclear relaxation. Thus the nuclear resonance relaxation marginally affects the values of $\delta_m$.

To increase the efficiency of relaxation arising from nuclear spins can be expected in the presence of superparamagnetic inclusions. It was noticed in
\cite{Lazarian:2008fw} that in the presence of the superparamagnetic inclusions the range of frequencies over which the relaxation related to nuclear spins can be extended.  This happens if the rate of thermal magnetization fluctuations $\tau_{fluct}^{-1}$ given by Eq. (\ref{tau_fluc}) is faster than the rate of the fluctuations of the internal magnetic field arising from the electron spin precession $\tau_{\rm SS}^{-1}$.  This is a common case, as the rate of nuclear spin precession arising from the interactions with other nuclear spins $\tau_{n, SS}^{-1}$ given by Eq. (\ref{nucl_ss}) is smaller than both $\tau_{fluct}^{-1}$ and $\tau_{\rm SS}^{-1}$. 

The arguments why the external fluctuating field can induce faster relaxation within the system of nuclear spins are similar to those in LD00. The external fluctuating magnetic field $H_i$ induces the random work variations of the nuclear spin direction with the step $\delta \theta \sim \tau_{f} \tau_{ne}^{-1}$, where 
$\tau_{f}^{-1}=max[\tau_{\rm SS}^{-1}, \tau_{fluct}^{-1}]$. The latter estimate assumes that the fluctuating magnetic field $H_i$ is the same for paramagnetic and superparamagnetic grain. While formally the mean field within the grain material $\langle H_i \rangle$ does not change whether the atoms are clustered or not,  the local value of $H_i$ near magnetic inclusions is $N_{\rm incl}$ times larger. Therefore for the spins in the vicinity of the inclusion $\tau_{ne}^{-1}$ can be up to $N_{\rm incl}$ times faster. The decorrelation time is the time over which the initial phase relationship of is lost, i.e. after $(\delta \theta)^{-2}$ steps, which provides
\begin{equation}
\tau_{n, super}\approx \frac{\tau_{ne}^2}{\tau_{fluct}}.
\end{equation}
Our considerations above suggest that the range of the nuclear relaxation can be extended to grain rotation with $\omega \approx \tau_{n, super}^{-1}$.   

In other words, if only nuclear spins are involved the low frequency response is 
\begin{equation}
K_{nn} (\omega) = \frac{\hbar}{11.4 g_{n} kT_{d}}.
\end{equation}which is similar to the low frequency dissipative response arising from the electron system.
In the presence of electron spins, the low frequency ($\omega< \tau_{ne}$ ) response changes to 
\begin{equation}
K_{ne} (\omega) = \frac{\hbar}{20 kT_{d}}\left(\frac{n_{n}}{n_e}\right)
\end{equation}
where the ratio of the $n_{n}/n_e$ can be significantly larger than 1 for carbonaceous grains making the relaxation faster.
For superparamagnetic grains, the low frequency ($\omega< \tau_{flip}$) response arising from the nuclear moments is approximately
\begin{equation}
K_{n, super} (\omega)= \frac{\hbar}{20 kT_{d}} \left(\frac{\tau_{\rm SS}}{\tau_{flip}}\right),
\end{equation}
where $\tau_{\rm SS}$ is the spin-spin relaxation time. For $\tau_{\rm SS}/\tau_{flip}>1$ the relaxation of nuclear moments is subdominant for superparamagnetic grains. 

For larger $\omega$, e.g. for $\omega> max[\tau_{ne}, \tau_{flip}]$, the relaxation decreases for the the process of nuclear dissipation due to the inability of nuclear spins  to response to the frequently changing external field. Therefore we expect the effect of paramagnetic relaxation within the nuclear system of a grain to be important only for slowly rotating carbonaceous grains. Due to grain growth such grains should be present in molecular clouds and accretion disks.  For the RAT alignment with the low-J attractor point (see LH07a) the increase of the nuclear-induced paramagnetic dissipation can be important. However, the increase of is mitigated as the frequency of grain rotation increase as the grains approach the high-J attractor/repellor point. Note that the grain dynamics in the vicinity of the high-J repellor point is important for increasing the alignment of grains that have only low-J attractors. This constraints the work of the mechanism of RATs with enhanced magnetic response (LH07b) that was termed MRAT in HL16. 

Our present work shows that the processes of magnetic relaxation that can affect the RAT alignment are more complex that it has been realized. For instance, our results suggest that the MRAT mechanism can be important for carbonaceous grains as the presence of the Barnett-equivalent field makes the spin-lattice relaxation within such grains dominant. Earlier on such effect was considered only in the presence of the ferromagnetic inclusions (see \citealt{1978ApJ...219L.129D}). If RATs are very strong the increase of the Barnett-equivalent field can make the resonance relaxation (LD00) possible. Our work calls for more detailed modeling of magnetic relaxation in order to compare the results with high precision observations of aligned dust, e.g. those possible with ALMA. 

\section{Effect of magnetic inclusions on the relaxation of nuclear spin system}

Magnetic inclusions can change the dynamics of the relaxation processes taking place with nuclear spins. The corresponding effects are most important
for carbonaceous grains as the enhanced magnetic relaxation within such grains is problematic. First of all, sufficiently large ferromagnetic inclusions can
induce spin-lattice relaxation within the nuclear system. This effect can be estimated using Eq. (\ref{K_sl_nucl}) with the modification that in Eq. (\ref{X})
one should use the magnetic field of the inclusions instead of the Barnett-equivalent magnetic field. Using earlier arguments one should substitute $\tau_{ne}$ given by Eq. (\ref{nucl_electr1}) for the nuclear spin-lattice relaxation time $\tau_{\rm SL}$ in Eq. (\ref{K_sl_nucl}). The potential enhancement of the magnetic relaxation will be proportional then to the ratio $\tau_{SL}/\tau_{SS}$. This ratio according to estimates in \cite{1999ApJ...520L..67L}  is  proportional to the ratio of the densities of nuclear and electron spins, more precisely to $1.7 n_{n}/n_{e}$ . The latter ratio can be large for
carbonaceous grains, and this is expected to enhance the relaxation. This approach assumes that $\omega$ which is less than $\tau_{ne}$, which, as we discussed earlier, means that it is applicable only for grains $>10^{-4}$ cm. Moreover, the effect is subdominant to the effect of the spin-lattice relaxation within the electron spin system. The latter is expected to be present within realistic carbonaceous grains \citep{1999ApJ...520L..67L}. 

If carbonaceous grains have super-paramagnetic inclusions the response time of the nuclear system can be shortened and the spin-lattice relaxation
for nuclear spins in the Barnett-equivalent magnetic field given by Eq. (\ref{B_nucl}) can be reduced. This means that the effect of nuclear spin-lattice
relaxation can be applicable to smaller carbonaceous grains that rotate faster. We describe this effect in Appendix \ref{apdx:alignment}. However, in the setting that we
already have superparamagnetic inclusions in a grain, the superparamagnetic relaxation is expected to dominate the energy dissipation within a rotating
grain. 


\end{document}